\documentclass{aastex63}

\submitjournal{ApJ}
\shorttitle{Solar Vortex Tubes}
\shortauthors{Silva et al.}

\begin{document}

\title{Solar Vortex Tubes: Vortex Dynamics in the Solar Atmosphere}
\correspondingauthor{Suzana S. A. Silva}
\email{suzana.seas@gmail.com}

\author[0000-0002-0786-7307]{Suzana S. A. Silva}
\affiliation{Department of Physics, Aeronautics Institute of Technology, São José dos Campos, Brazil \\
}

\author{Viktor Fedun}
\affiliation{Plasma Dynamics Group, Department of Automatic Control and Systems Engineering, University of Sheffield, Sheffield, UK
 \\}


\author{Gary Verth}
\affiliation{Plasma Dynamics Group, School of Mathematics and Statistics, University of Sheffield, Sheffield, UK
 \\}

\author{Erico L. Rempel}
\affiliation{Department of Mathematics, Aeronautics Institute of Technology, São José dos Campos, Brazil \\
}

\author{ Sergiy Shelyag}
\affiliation{School of Information Technology, Deakin University,
Geelong, Australia}

\begin{abstract}
In this work, a state-of-the-art vortex detection method, Instantaneous Vorticity Deviation, is applied to locate three-dimensional vortex tube boundaries in numerical simulations of solar photospheric magnetoconvection performed by the MURaM code. We detected three-dimensional vortices distributed along intergranular regions and displaying coned shapes that extend from the photosphere to the low chromosphere. Based on a well-defined vortex center and boundary, we were able to determine averaged radial profiles and thereby investigate the dynamics across the vortical flows at different height levels. The solar vortex tubes present nonuniform angular rotational velocity, and, at all height levels, there are eddy viscosity effects within the vortices, which slow down the plasma as it moves toward the center.  The vortices impact the magnetic field as they help to intensify the magnetic field at the sinking points, and in turn, the magnetic field ends up playing an essential role in the vortex dynamics. The magnetic field was found to be especially important to the vorticity evolution. On the other hand, it is shown that, in general, kinematic vortices do not give rise to magnetic vortices unless their tangential velocities at different height levels are high enough to overcome the magnetic tension.
\end{abstract}

\keywords{convection--Sun: granulation -- Sun: atmosphere---Sun: photosphere  
magnetohydrodynamics (MHD)}

\section{Introduction} \label{sec:intro}

The solar surface is covered by granules that constitute the tops of small convective cells. They appear when the hot plasma coming from the solar interior rises into the solar atmosphere and radiatively cools down. The study of photospheric flows is based on observational data for velocity fields derived from FLCT applied to intensity maps, \textit{e.g.} \citet{Fisher2008}, and velocity fields obtained from numerical simulations based on magnetohydrodynamics (MHD) equations.  Strong negative divergence regions of velocity field intensity tends to  behave as sinks and concentrate the magnetic field \citep{Balmaceda2010} and the observational data also suggest that downdraft centers may display vortical motions \citet{Bonet2010}. The main theory concerning the creation of the observed intergranular vortices describes the vortical dynamics to be originated at those downdraft centers as the plasma diverging from the granule centers has an angular momentum in relation to the sink, leading the elements of fluid to rotate as they approach the downdraft center. This process is also known as the ``bathtub effect'' and it is related to a free, i.e. not forced vortex.  The
``bathtub effect'' was suggested as a photospheric vortex creation mechanism by \citet{Nordlund85} based on simulations of convective motions. Not all downdraft centers present vortex dynamics, however. One necessary condition for their formation is the existence of vorticity in the sink region \citep{Simon_1997}. In the downdraft centers, the concentration of magnetic flux leads to the generation of vorticity by the magnetic tension term, which dominates the vorticity evolution in the solar atmosphere \citep{Shelyag2011}. Other mechanisms have also been shown to lead to vortices in the solar atmosphere, e.g., based on radiative hydrodynamic simulations, \citet{Kitiashvili2012a} found that both horizontal and vertical vortex tubes at the intergranular lanes can be generated by Kelvin–Helmholtz instability of shearing flows. 

Vortical flows  in the photosphere have been investigated using velocity fields derived from both observations \citep{Bonet2010, Attie08,Bonet2008,Balmaceda2010, Giagkiozis2017,Requerey2017, Tziotziou2018, Tziotziou2019, Shetye2019} and magnetohydrodynamics (MHD) simulations \citep[see e.g.][]{Kitiashvili2012, moll12, Wedemeyer2012, Wedemeyer2014, kato17}. The vortices observed in the solar atmosphere present a radius ranging from 0.1 to around 2 Mm \citep{Bonet2010,Silva_2018,Giagkiozis2017} and have an average lifetime of around  0.29 minutes \citep{Giagkiozis2017}, but in supergranular convection it is possible to find vortices that last for hours \citep{Requerey2017,Chian2019}.  Swirling motions have also been detected in the chromosphere based on Ca II
8542 \AA \hspace{1mm} and H$\alpha$ line observations \citep{Wedemeyer09, Shetye2019}. Those observations indicate that chromospheric swirls last for 10 minutes, and their vortical motions extend to around 1 Mm from the center. These swirls also present different shapes and seem to be correlated to magnetic concentrations at downdraft centers and can be observed in different line emissions \citep{Shetye2019}. \cite{Wedemeyer2012} showed observational signatures of vortical motions at different height levels that are spatially correlated, which suggests that the solar photospheric vortices are most likely the lower part of solar atmospheric vortex tubes that extends up to the solar corona.  Nevertherless, \cite{Shetye2019} could not determined whether the observed swirls correspond to  motions in the photosphere or a propagating Alfvén wave.

The magnetic field in the intergranular lanes also interacts with vortices, affecting their dynamics.  Magnetohydrodynamical simulations show that the presence of the magnetic field intensifies the vortex tube effects in the chromosphere in weak magnetic fields \citep{Kitiashvili2012}. On the other hand, the vortical flow gives stability to the magnetic fluxtube \citep{Requerey2017} and drags the magnetic field. \cite{Wedemeyer2012} and \cite{Wedemeyer2014} suggest that, as the magnetic field is twisted, it drives the vortical motion of the plasma in the chromosphere, which is, in turn, observed as swirl signatures in different line emission observations. Based on MHD modeling, \cite{moll12} and \cite{shelyag2013} found that the magnetic field lines are not considerably twisted, whereas \cite{Wedemeyer2014} and \cite{Rappazzo2019} simulation results display a rotating magnetic field coexisting with a kinematic vortex.

In this paper, we apply a state-the-art vortex detection method, Instantaneous Vorticity Deviation (IVD), to precisely define vortex tubes in the solar atmosphere. We investigate the dynamics across the vortical flows at different height levels and their impact on the magnetic field. The paper is organized as follows. First, we introduce the IVD technique and the construction of three-dimensional vortices in section \ref{sec:Methodology}. We then proceed in section \ref{sec:results} to describe the detected vortices and show radial profiles from selected vortices for velocity and magnetic field related variables. Section \ref{sec:discussions} deepens the analysis of the relationship between the magnetic field and the vortex dynamics. Lastly, conclusions are presented in section \ref{sec:conclusions}.

\section{Methodology} \label{sec:Methodology}
We analyze the data from the radiative MHD simulations of magnetoconvection in the solar photosphere and upper convection zone obtained with the MuRAM code \citep{voegler2005}. The rectangular domain we use has $960 \times 960 \times 160$ grid cells, which cover a region of 24 Mm in the $x-$ an $y$-directions and 1.6 Mm in the vertical $z$-direction. The model realistically simulates a solar plage region with the net vertical magnetic field of $200~\mathrm{G}$. The visible solar surface (Rosseland optical depth $\tau=1$) is located at $ H= 0.0~\mathrm{Mm}$, 600 km below the upper boundary. The upper boundary of the simulation domain is located in the temperature minimum. The simulation region size and resolution are chosen such that it covers horizontal and vertical convective spatial scales in the solar photosphere. The simulation box is positioned in the solar atmosphere so that it covers the region where the radiation comes from and where the transition from magnetically dominated (atmospheric part of the domain) to fluid-dominated (interior part) dynamics occurs, which are specifically of interest for this study. A standard gravitationally stratified radiative resistive magnetohydrodynamic model is used in the simulations, which are self-consistent with only a small number of parameters, such as solar gravity acceleration, the average outward radiative flux, initial vertical uniform magnetic field strength, and solar photospheric chemical composition. Further information regarding how those terms were implemented and the values used can be found in \cite{voegler2005}. The system of equations solved is essentially nonadiabatic. The equation of state is used in a very general form with a tabulated functional dependence of pressure and temperature on density and internal energy per unit volume. The nonideal MHD terms are ohmic resistive.
The average parameters of the modeled atmosphere were checked to make sure that it reached a quasi-stationary state. It was found that the total box mass and the net radiative flux oscillated around their required constant values. As small scales are of interest for this study, the phase of these 5 minute oscillations are of no importance. Partial ionization in the solar interior and photosphere is taken into account through the nonideal equation of state as explained by \cite{voegler2005}. 

\begin{figure}[ht!]
\plotone{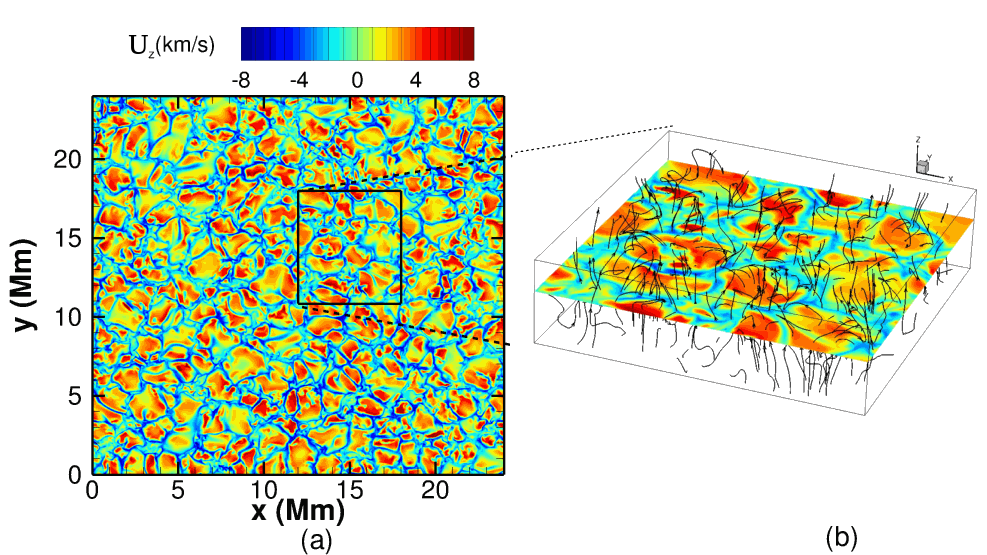}
\caption{Simulation domain at $t=0$. (a) 2D view: The $xy$-plane of the whole domain at $H=0.0~\mathrm{Mm}$ colored by the z-component of the velocity. The part of the domain investigated in this paper is delimited by a black square. (b) The 3D view of the region within the black square in (a). The $xy$-plane at $ H=0.0~\mathrm{Mm}$ ($z =1.0 \mathrm{Mm}$) and it is colored by the $z$-component of the velocity. The black lines are the magnetic field lines. \label{fig:domain}}
\end{figure}

In this paper, we focus on a fraction of the whole domain, with $240 \times 240$ grid points. In Fig. \ref{fig:domain} we display both the 2D view of the whole $xy$-plane from the original domain (left panel) and the 3D view of the partition used in our investigations. Our studies then concern a domain of size 6 Mm $\times$ 6 Mm $\times$ 1.6 Mm, which is large enough to cover multiple granules and their intergranular regions. 

 We applied the IVD technique \citep{Haller2016} to find vortices in the upper part of our domain, $z\geqslant 1.0$ Mm, shown in Fig. \ref{fig:domain}(b). The IVD field is computed by the following expression
\begin{equation}
\label{Eq:3}
\mbox{IVD}(\mathbf{x},t):= |\mathbf{\omega}(\mathbf{x},t) -\langle \mathbf{\omega}(t) \rangle|,
\end{equation}
where $\mathbf{x}$ is a position vector, $\mathbf{\omega} = \nabla \times \mathbf{u}$ is the vorticity, and $\langle \cdot \rangle$ denotes the instantaneous spatial average. 
\cite{Haller2016} establishes the boundary of a given vortex in 2D as the outermost convex closed contour of the IVD scalar field around the vortex center, which in turn is defined by a local maximum of the IVD field. Physically, this contour provides a locus of particles with the same intrinsic rotation rates \citep{Haller2016}.  In other words, IVD defines the vortex boundary using an intuitive notion where the particles have a coherent rotation along an approximately elliptical curve. In 3D flows, the method can be applied in successive 2D planes and interpolating among the closed contours found in each plane, forming a vortex tube. 

A significant advantage of employing IVD is the fact that the only parameter that needs to be chosen is the maximum amount of deviation from convexity a curve may have to describe a vortex. This parameter is also called convexity deficiency, $\epsilon$, and it is defined as:
\begin{equation}
    \epsilon = \frac{A_c - A_{ch}}{A_c},
\end{equation}
where $A_c$ is the area which is enclosed by the extracted contour whereas the term $A_{ch}$ stands for the area enclosed by its convexhull.

IVD is the instantaneous (Eulerian) version of the  Lagrangian Average Vorticity Deviation(LAVD) method, which has been successfully employed in a number of hydrodynamic and plasma problems and shown to perform better than other available vortex detection methods (see, e.g., \citet{Hadjighasem2017, Silva_2018}). 
Nonetheless, IVD and LAVD have some limitations when the velocity field has convex regions with strong shear leading to high vorticity even if no coherent swirling motion is seen, as reported by \citet{Silva_2018}. This may also cause a difference between the position of a local maximum of IVD and a true vortex center. 
To avoid such problems, the $d$-parameter was proposed by \citet{Silva_2018} to filter out false vortex detections by IVD/LAVD. The $d$-parameter first detects vortex centers as points in the flow surrounded by fluid particles that undergo circular motions during a certain time interval. The circular motion is determined by checking the relative positions of displacement vectors obtained by integrating four particles surrounding each grid point.  Once the vortex centers are found with the $d$-parameter, the IVD operator is employed to detect the vortex boundaries surrounding each vortex center.  

\subsection{Three-dimensional vortices}
\begin{figure}[ht!]
\plotone{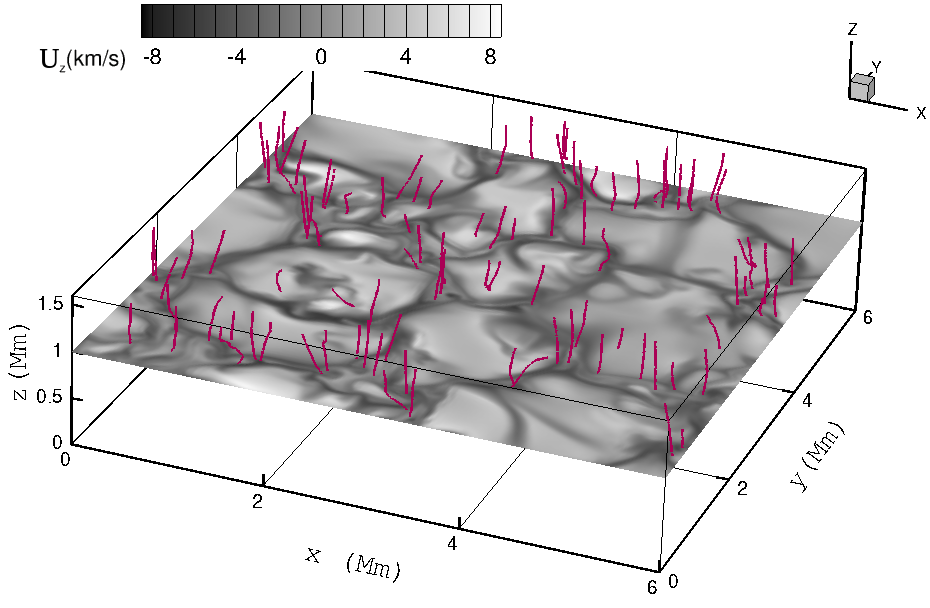}
    \caption{3D simulation domain at $t= 0$ with $xy$-plane at $H=0.0 \mathrm{Mm}$ ($z =1.0 \mathrm{Mm}$) colored by the $z$-component of the velocity. The core lines of  vertical vortices as computed by the d-criterion are shown in purple. }
    \label{fig:corelines}
\end{figure}

\begin{figure}[ht!]
    \plotone{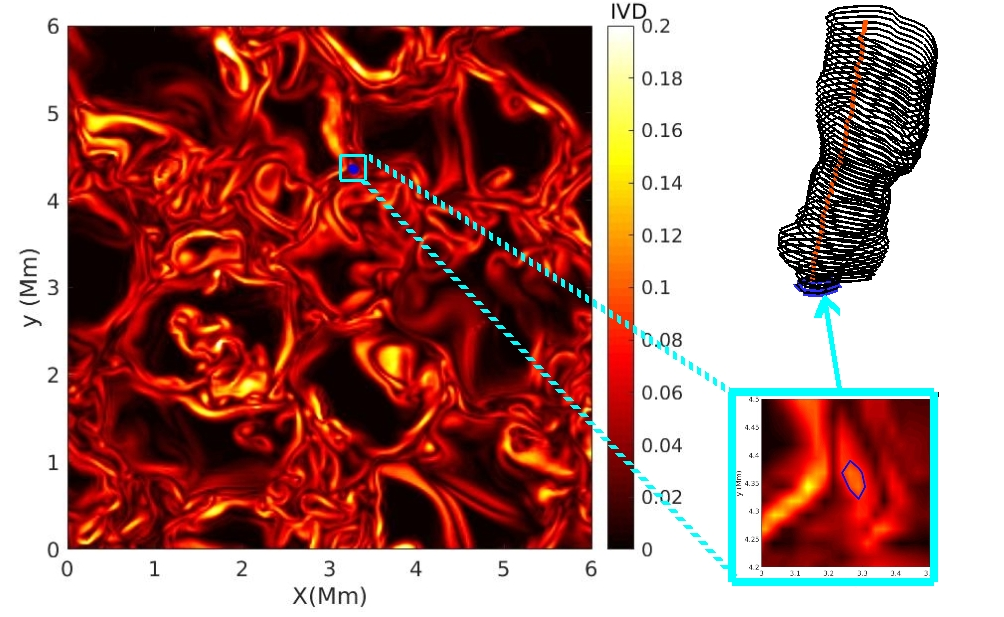}
    \caption{3D Vortex construction from the IVD field. The left panel shows the IVD field for the $xy$-plane at $H=0.0 $ Mm (or $z =1.0 \mathrm{Mm}$) with the dark blue line denoting the vortex boundary. The bottom-right panel is a closer view of the vortex boundary and the upper-right panel shows the complete vortex from IVD contours in several planes and its core line in orange.}
    \label{fig:3Dvortex}
\end{figure}

The construction of three-dimensional vortices performed in this work is based on the analysis of a series of two-dimensional IVD fields, which were computed for  $xy$-planes within a range of $z$ above the solar surface. The choice of using horizontal planes is based on previous works, e. g., \cite{Kitiashvili2012a} and \cite{ Wedemeyer2012}, that show that vortical motions are mainly in the horizontal direction. As we are mostly interested in solar atmospheric vortices, we use the range from $H = 0$ to $H=0.5$ Mm which corresponds to $z =1.0 \mathrm{Mm}$ to $z =1.5 \mathrm{Mm}$. The time evolution of the vortices was determined by computing the IVD field at all time frames within a time interval of 50 s, starting at $t=0$, which is labeled $t_0$ from now on. To avoid most of the abovementioned problems due to the presence of shear in the flow,  we first apply the d-criterion to determine possible vortex centers and also dismiss any false detection. As the d-criterion is based on the particle displacement after a time interval and IVD is an instantaneous method, we compute the d-criterion in the $xy$-plane within the minimal time interval of $\Delta t=2.5~\mathrm{s}$ between frames. Then, for each time frame, we pick the points that obey the d-criterion in different horizontal layers that are sufficiently close to determine a vortex core line. Thereby, we are able to compute the core lines of possible vertical vortices, as shown for $t=0$ in Fig. \ref{fig:corelines}. 
After that, we use each core line to compute the vortex boundary in all $xy$-planes from $H=0.0$ to $H=0.5$ Mm. The vortex boundary around the points given by the core line was defined as the outermost convex contour of the IVD field in each $xy$-plane.  In order to obtain this contour, we apply a convex deficiency of $c=0.03$.  The 3D vortex is then obtained by the group of all contours of a given vortex core line, as illustrated in Fig. \ref{fig:3Dvortex}.
One can see that even though the IVD field was computed separately for each height level, the final vortex boundary varies smoothly and shows a spatial coherence.

\section{Results} \label{sec:results}
\begin{figure*}
\gridline{\fig{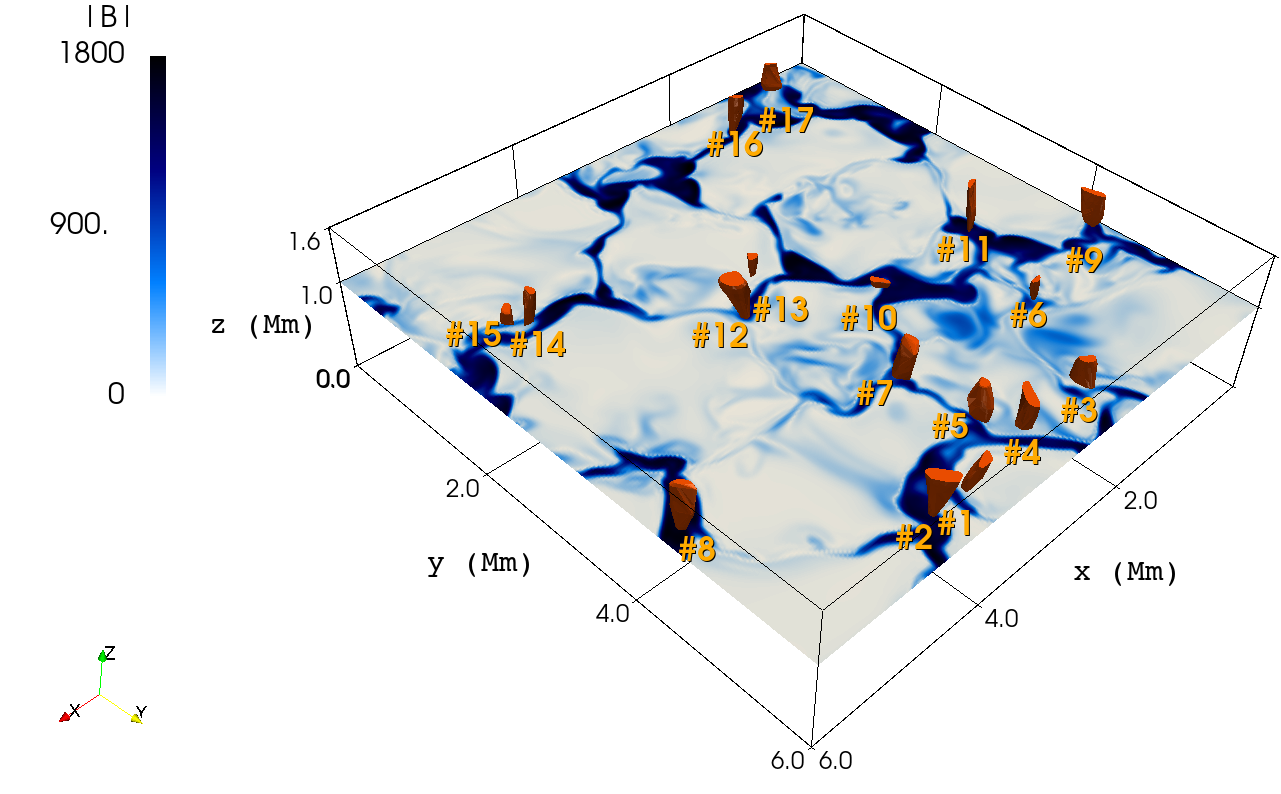}{0.8\textwidth}{}
          }

\caption{3D simulation domain at $t=25 s$. Label numbers identify the vortices detected by IVD. (a)  $xy$-plane at $z=1.0 $ Mm or $H=0.0 $ Mm colored by the intensity of the magnetic field and the vortex boundaries computed by IVD for  $t=25 s$ are shown in orange.\label{fig:allvortices1}}
\end{figure*}

\begin{figure*}
\gridline{
          \fig{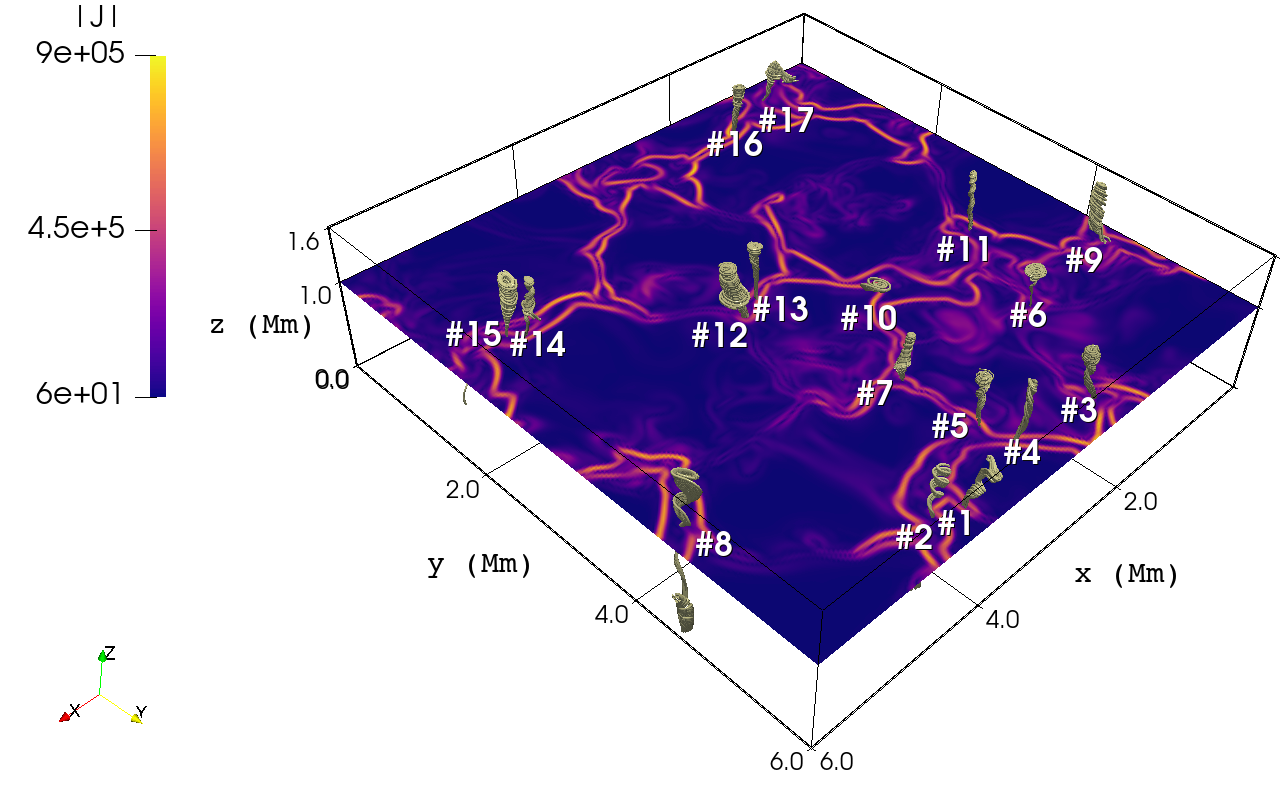}{0.85\textwidth}{(a)}
          }
\gridline{\fig{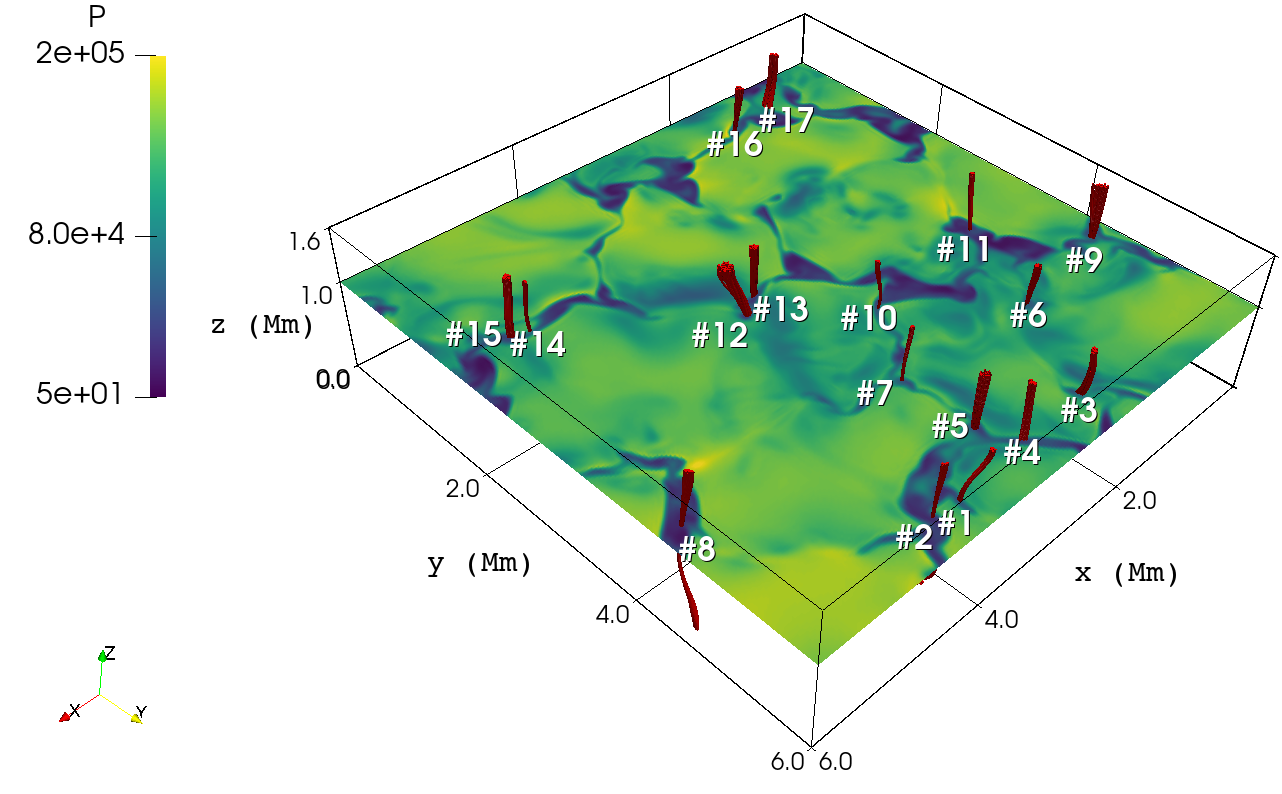}{0.85\textwidth}{(b)}
          }

\caption{3D simulation domain at $t=25 s$. Label numbers identify the vortices detected by IVD. (a) $xy$-plane at $H =0.0 $ Mm ( or $z =1.0 \mathrm{Mm}$) colored by the magnitude of current density and the instantaneous streamlines are for the velocity field, traced from points within each vortex boundary. (b) The magnetic field lines are shown in red, traced from points within each vortex boundary\label{fig:allvortices}, and the $xy$-plane at $H =0.0 $Mm ($z =1.0 \mathrm{Mm}$) is colored by the pressure.}
\end{figure*}

We detect a total of 17 vortices that persist for the whole time interval considered in this analysis, 50 s. Eleven vortices rotate counterclockwise, and six rotate clockwise. The vortices present different shapes and sizes, as illustrated in Fig.\ \ref{fig:allvortices1}(a), which shows those 17 detected vortex boundaries colored in orange for $t=25~\mathrm{s}$ and the $xy$-plane at $H=0.0~\mathrm{Mm}$ colored by the intensity of the magnetic field.  Another common feature in the geometry of the vortices is the widening up of the vortex boundaries in the upper part of the domain. The vortex radius is, on average, around 40 km at $H=0$ and up to approximately 80 km at $H =0.5$ Mm. Therefore, that vortex radius at $H= 0.0$ Mm is one order of magnitude smaller than the vortices obtained using the two-dimensional velocity field derived from observational data \citep{Bonet2008,Silva_2018,Giagkiozis2017}.
In Fig.\ \ref{fig:allvortices}(a) we see the instantaneous spiraling velocity streamlines, which were traced from points withing the vortex boundary. The vortices are located between regions of strong current density  as indicated in Fig.\ \ref{fig:allvortices}(a), and they tend to appear in low-pressure regions, as displayed in Fig.\ \ref{fig:allvortices}(b). The magnetic field lines traced from points encompassed by the vortex boundaries are shown in red in Fig, \ref{fig:allvortices}(b), they are mostly vertical lines and organized in tube structures. 

\begin{figure}[ht!]
\includegraphics[width=\textwidth]{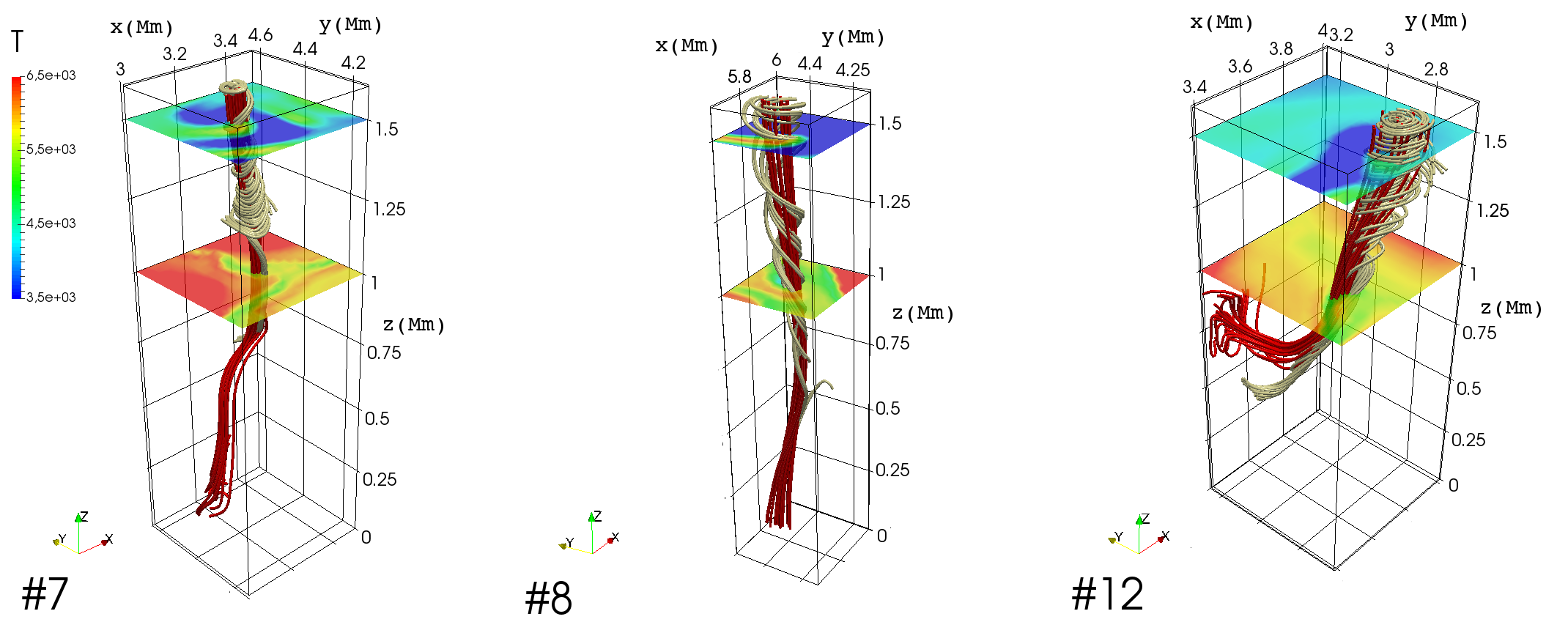}
    \caption{Magnetic and velocity field lines  traced from random points within the vortex boundary are show in red and dark khaki colours correspondingly for vortices \#7, \#8 and \#12 for $t=t_0$. The $xy$-planes are placed at $H= 0$ Mm ( or $z =1.0 \mathrm{Mm}$) and $H=0.5$ Mm ( or $z =1.5 \mathrm{Mm}$)}  and are colored by the plasma temperature. 
    \label{fig:selected}
\end{figure}

For our analysis, we select three vortices, \# 7 , \# 8,    and \#12, which are located in different parts of the domain. This choice was based on the fact that those vortices' boundaries were detected in the photosphere and in the upper part of the domain. In addition, they also give a good representation of the dynamics found for the detected solar vortices. 
In Fig.~\ref{fig:selected}, we show field lines in red for the magnetic field and in dark khaki for the velocity field for those selected vortices at the initial time instant, $t=t_0$.   They were traced from random points within the vortex boundary and the colors in $xy$-planes in the figure correspond to the local temperatures at the constant heights of $H=0$ and $H=0.5~\mathrm{Mm}.$ The vortices seem to encompass regions of different temperature ranges in the photosphere and in the chromosphere. We see that the vortical dynamics imposed on the plasma also seem to influence the temperature distribution, dragging, and mixing the hot and cold plasma. 

\subsection{Radial profiles}
\begin{figure*}
\gridline{\fig{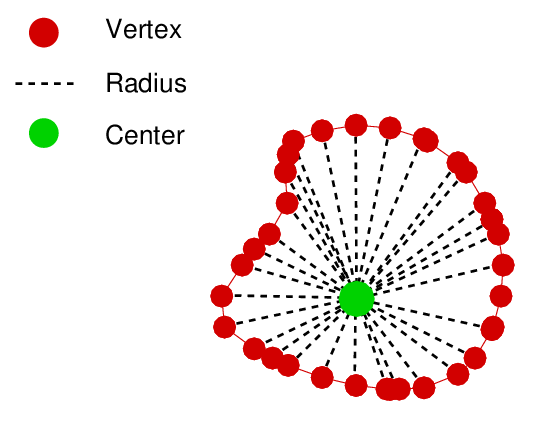}{0.3\textwidth}{}
          }

\caption{Contour of the vortex \# 12 at H=0.5 Mm for $t=t_0$ given by a red line. The vertices of the vortex are indicate by red circles and the vortex center is given by the green circle. The black dashed lines represents the radius of the vortex for each vertex. \label{fig:vertices12}}
\end{figure*}

The plasma dynamics across the vortex flow can be studied using the radial profile, which describes the changes in the plasma as one moves away from the center of the vortex toward its boundary.  Now, each vortex boundary at a given $xy$-plane is formed by a group of vertices that are not necessarily within the same distance from the vortex center as illustrated in Fig.\ \ref{fig:vertices12} for vortex \#12 at $H = 0.5$ Mm. For each vertex, we set a grid with 20 equally spaced points along the line  segment from the identified vortex center to the vertex (a vortex “radius”). The distance from those grid points to the vortex center, $r$, is normalized by the distance from the given boundary vertex to the vortex center, $R$. The physical quantities at the grid points are obtained by linear interpolation. Also, at each height level, we obtain the general tendency of the vortex radial profile  by averaging the radial profiles, obtained for each vertex, along the angular direction.

All the figures of the radial profile plots show the variable average distribution along the vortex radius from the center, $r=0$, to the vertex $r=R$ for different times $t_0= 0$ (red line), $t_1=25s$ (green line) and $t_2=50s$ (blue line) and different heights (a)-(c) $H = 0.1$ Mm, (d)-(f) $H = 0.3$ Mm,(g)-(i) $H = 0.5$ Mm. The left $y$-axis displays the values for the averaged radial profile given by the solid lines, and the right $y$-axis is for the averaged radial profile shown in dashed lines and all the values are in cgs units. The radial profile for vortex \#7 is on the first column of the figures, the panels in the middle column are for the vortex \#8, and the last column depicts the radial profiles of vortex \#12.

\begin{figure*}[htp!]
\gridline{\fig{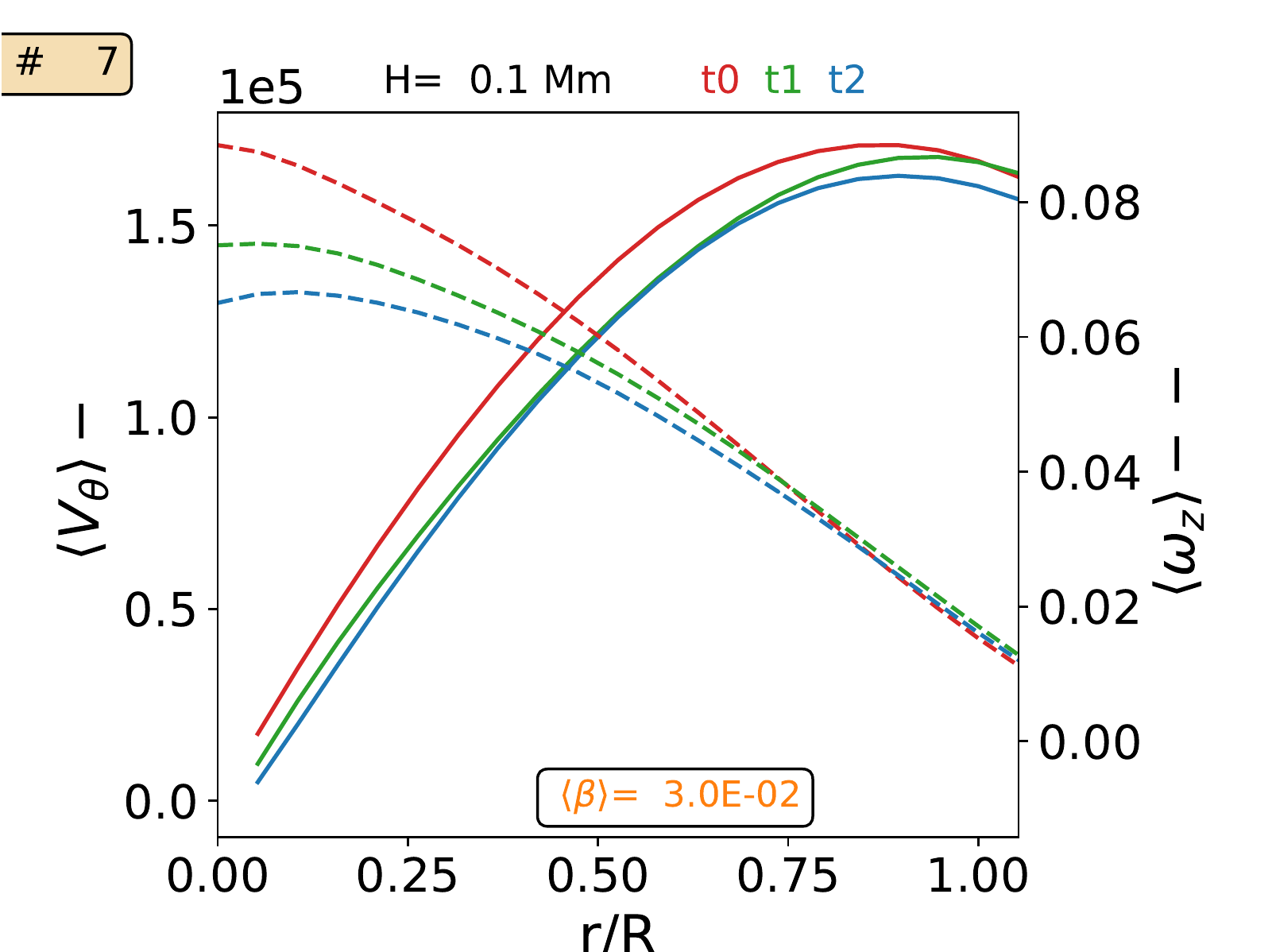}{0.3\textwidth}{(a)V7}
          \fig{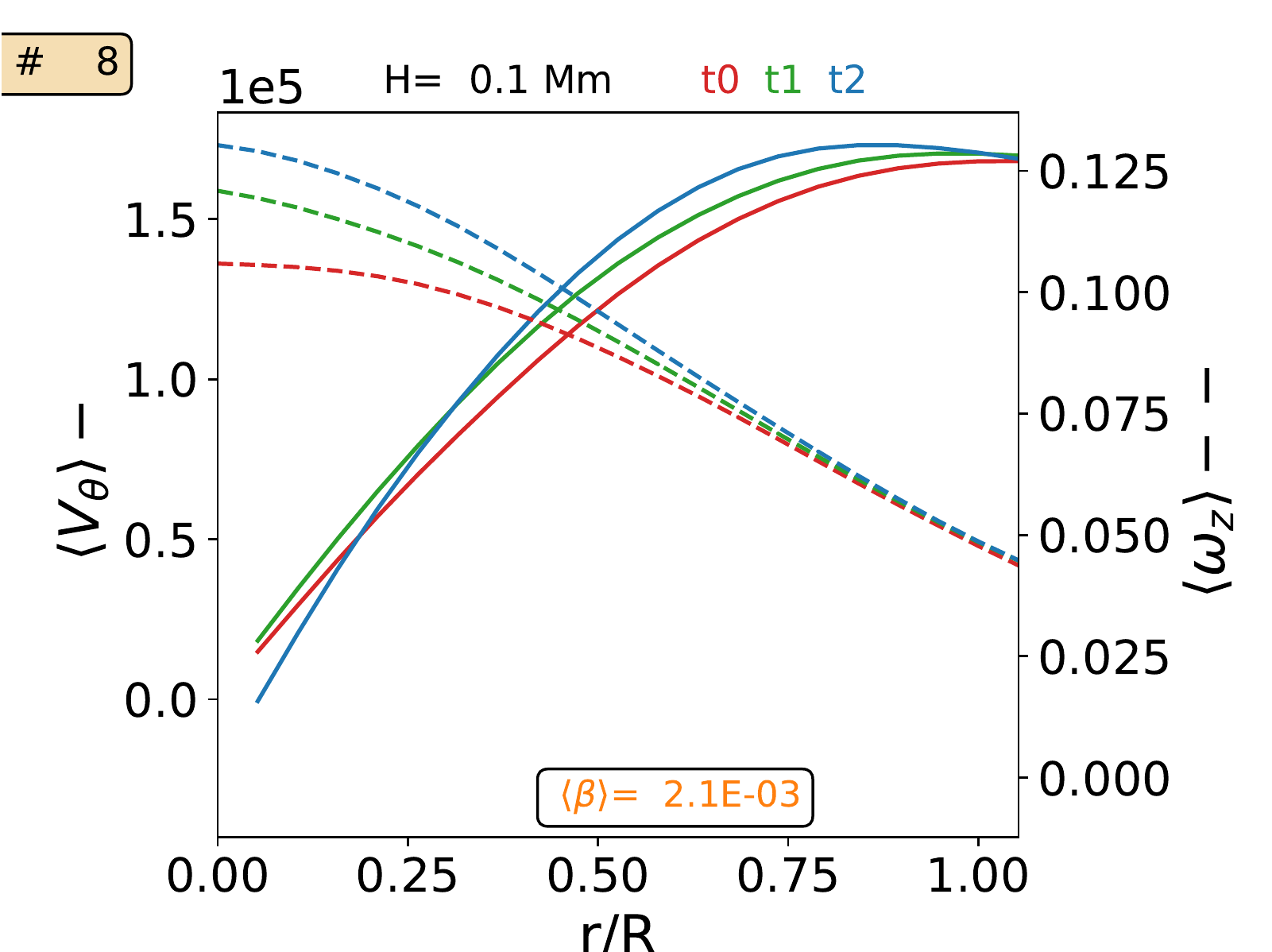}{0.3\textwidth}{(b)V8}
          \fig{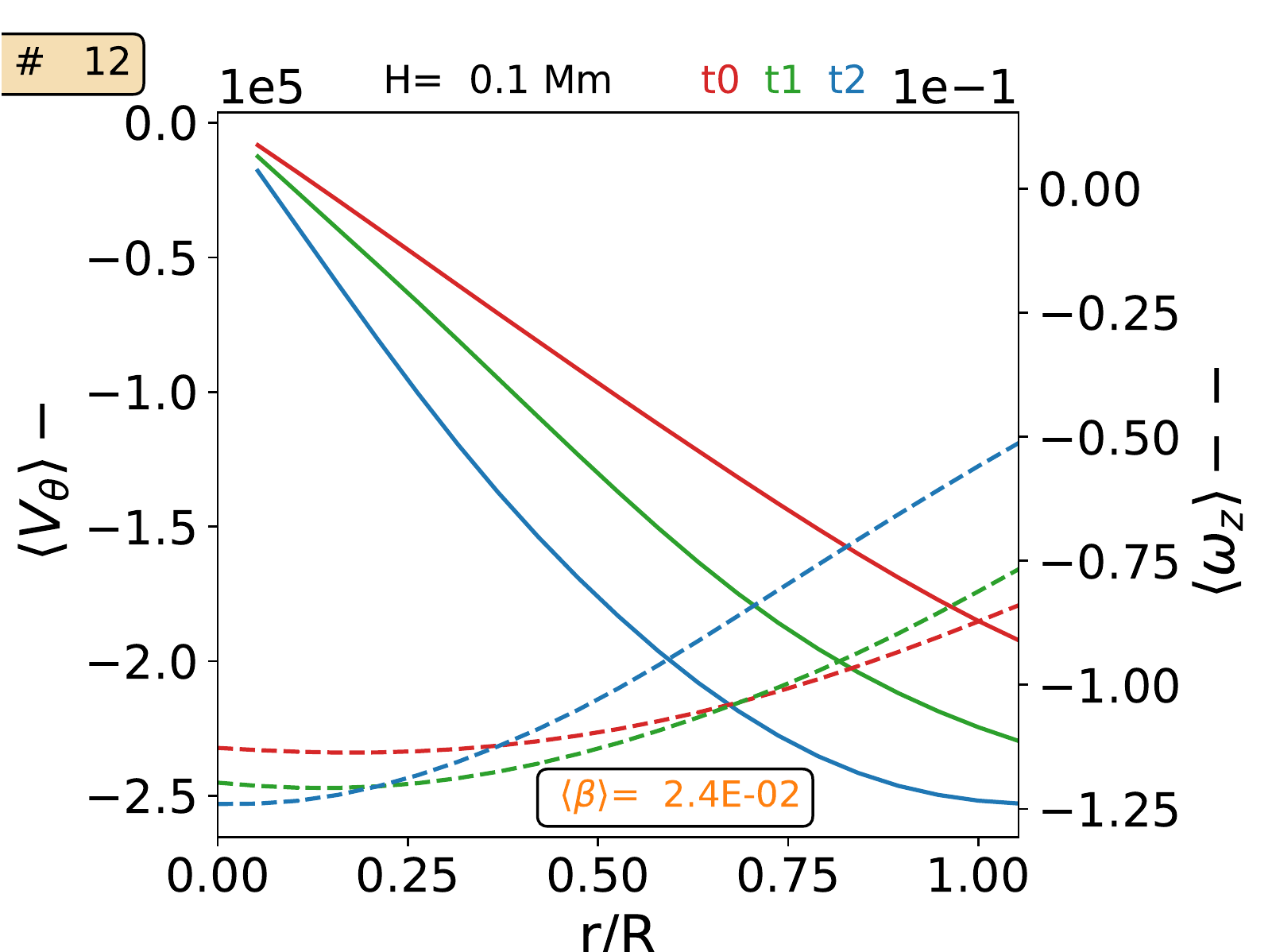}{0.3\textwidth}{(c)V12}
          }
\gridline{\fig{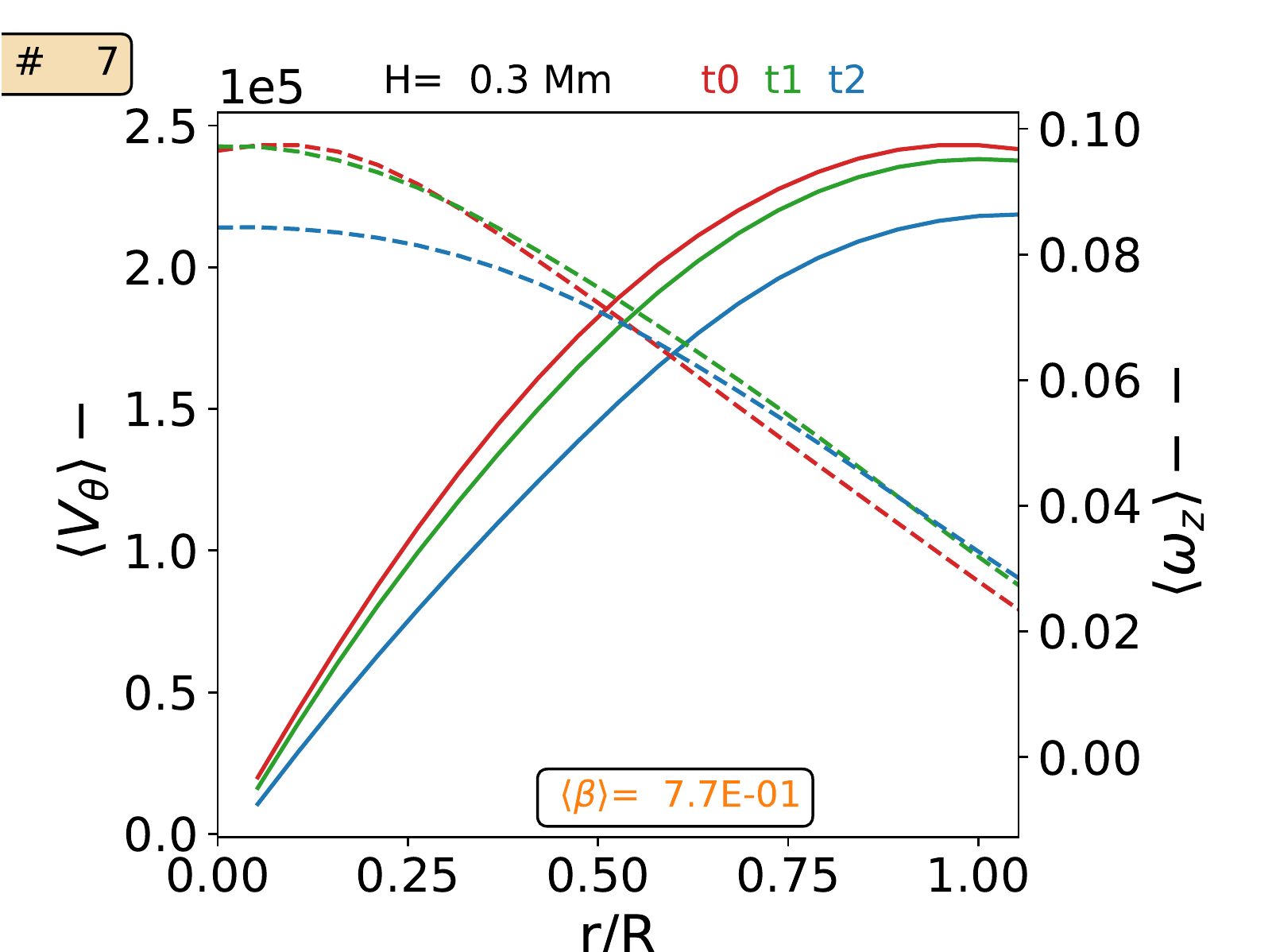}{0.3\textwidth}{(d)V7}
          \fig{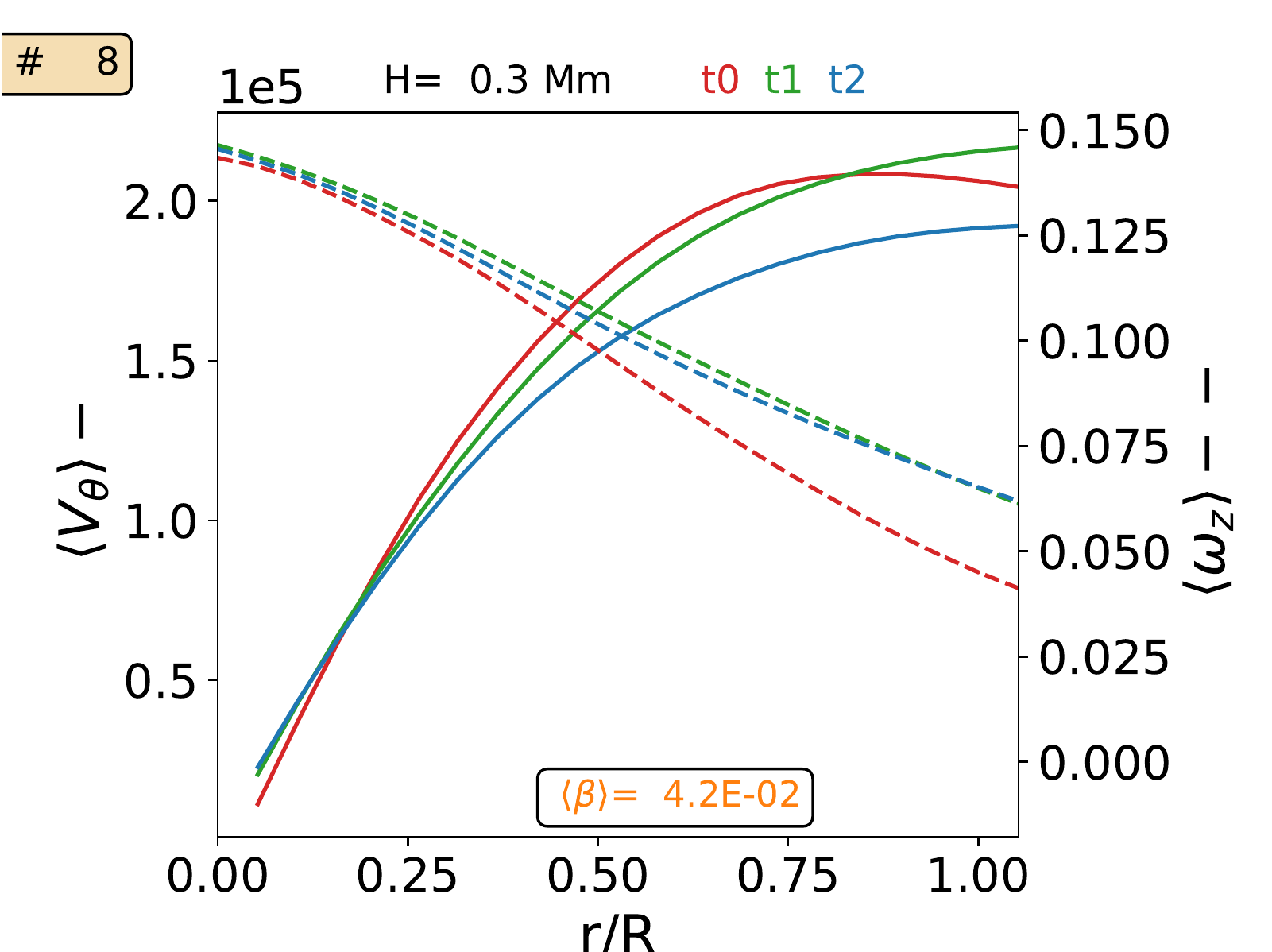}{0.3\textwidth}{(e)V8}
          \fig{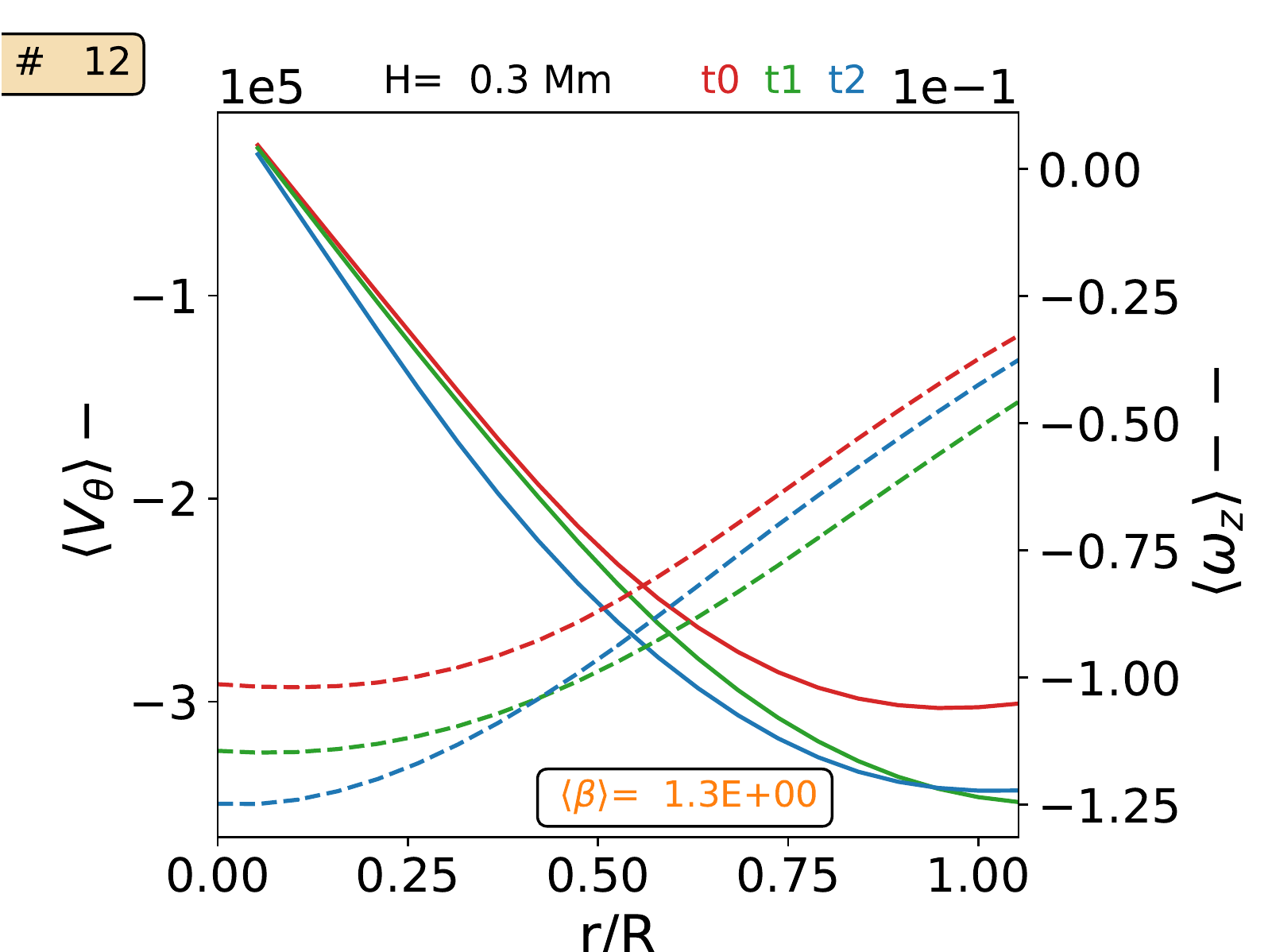}{0.3\textwidth}{(f)V12}
          }
\gridline{\fig{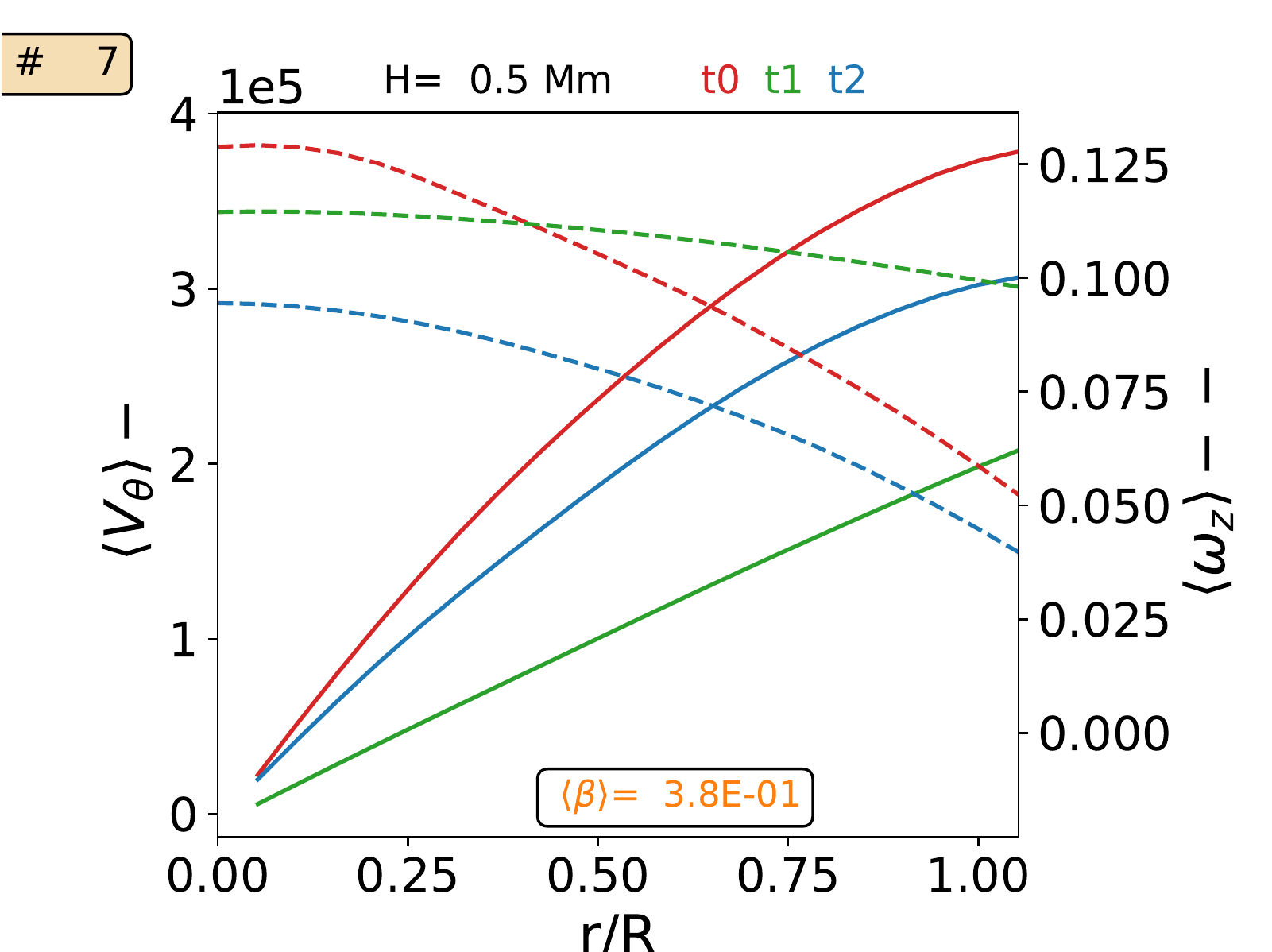}{0.3\textwidth}{(g)V7}
          \fig{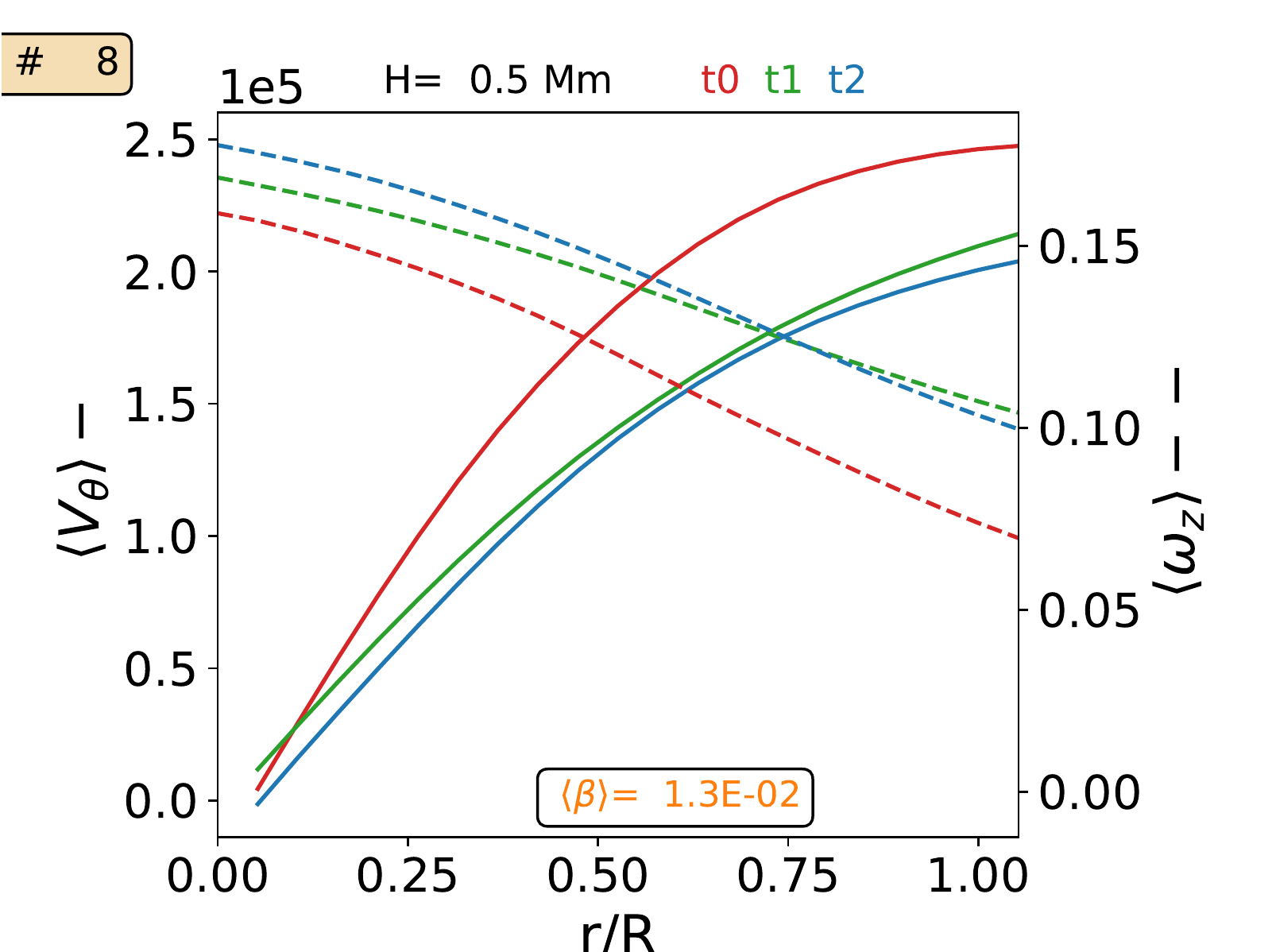}{0.3\textwidth}{(h)V8}
          \fig{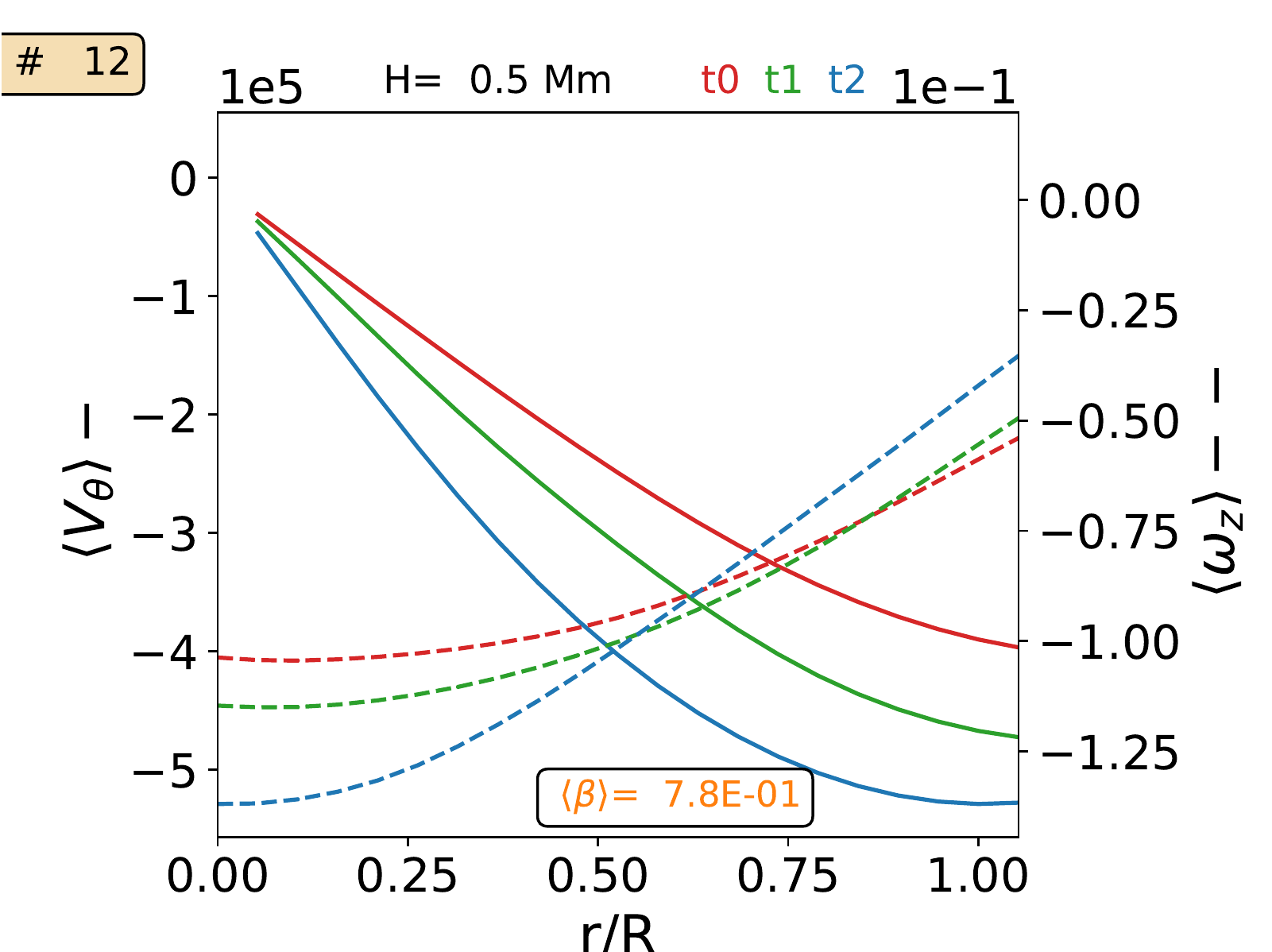}{0.3\textwidth}{(i)V12}
          }
\caption{The average tangential velocity field  (left $y$-axis, solid lines)  along the vortex radius from the center, $r=0$, to the boundary $r=R$ and  the $z$-component of the vorticity vector            (right $y$-axis, dashed lines). The red lines are for the initial time $t_0= 0$, the green lines are for $t_1=25 s$ (green line), and the blue lines are for $t_2=50 s$.  The radial profiles are shown for vortex \#7(a)(d)(g), \#8(b)(e)(h) and \#12(c)(f)(i) at different heights: $H = 0.1$ Mm  (a)(b)(c), $H = 0.3$ Mm (d)(e)(f), $H = 0.5$ Mm (g)(h)(i). The averaged value for plasma $\beta$ at the analyzed times and for the region within the vortex  is displayed in orange \label{fig:velsradial}}
\end{figure*}

For nonmagnetized flows, the main aspects of the vortical flow are drawn from the tangential velocity distribution along the radii. All the detected vortices present the tendency for the tangential velocity profile illustrated by the solid lines in Fig.\ \ref{fig:velsradial} for the selected vortices, \textit{i. e.}, the intensity of tangential velocity decreases around 90\% as one goes away from the vortex boundary, reaching a minimum close to its center. This behavior is in agreement with observational data \citep{Simon_1997} and also MHD simulations \citep{Onishchenko2018}. The differences found among the vortices concern mainly the sign of the averaged tangential velocity as well as the time evolution. The tangential velocities of vortices \#7 and \#8  tend to decrease around 20\% at $H=0.5$ Mm, whereas the variations in the lower part were less than 10\%. The averaged intensity of tangential velocity for vortex \#12  increases more than 30\% in both upper and lower parts of the vortex tube. The negative signal for the averaged tangential velocity of vortex \#12 indicates that it rotates clockwise, as opposed to the rotation direction of vortices \#7 and \#8. Since most of the vortical flow is along the $xy$-plane, we compute the radial distribution of the $z$-component of the vorticity vector, $\omega_z$, which is shown by the dashed lines and right $y$-axis in Fig.\ \ref{fig:velsradial}. The maximum $\omega_z$ is around the vortex center, and it decreases in the radial direction, in agreement with 2D vortices from observational data \citep{Simon_1997}.   The difference found between the vorticity at the center, and the boundary tends to be greater at lower heights. Among the three vortices analyzed, \#12 displays less variation of $\omega_z$ from the center to its boundary. In general, $\omega_z$ tends to vary more overtime at the center than on the vortex boundary and it decreases in time, except for vortex \#12. The averaged plasma $\beta$ within the vortex is also indicated in orange for each height level. We see that, in general, the vortices present low plasma  $\beta$. The lowest value is found for vortex \#8, which has a   plasma   $\beta$ that is 10 times lower than the other vortices, \#7 and \#12.  

\begin{figure*}[htp!]
\gridline{\fig{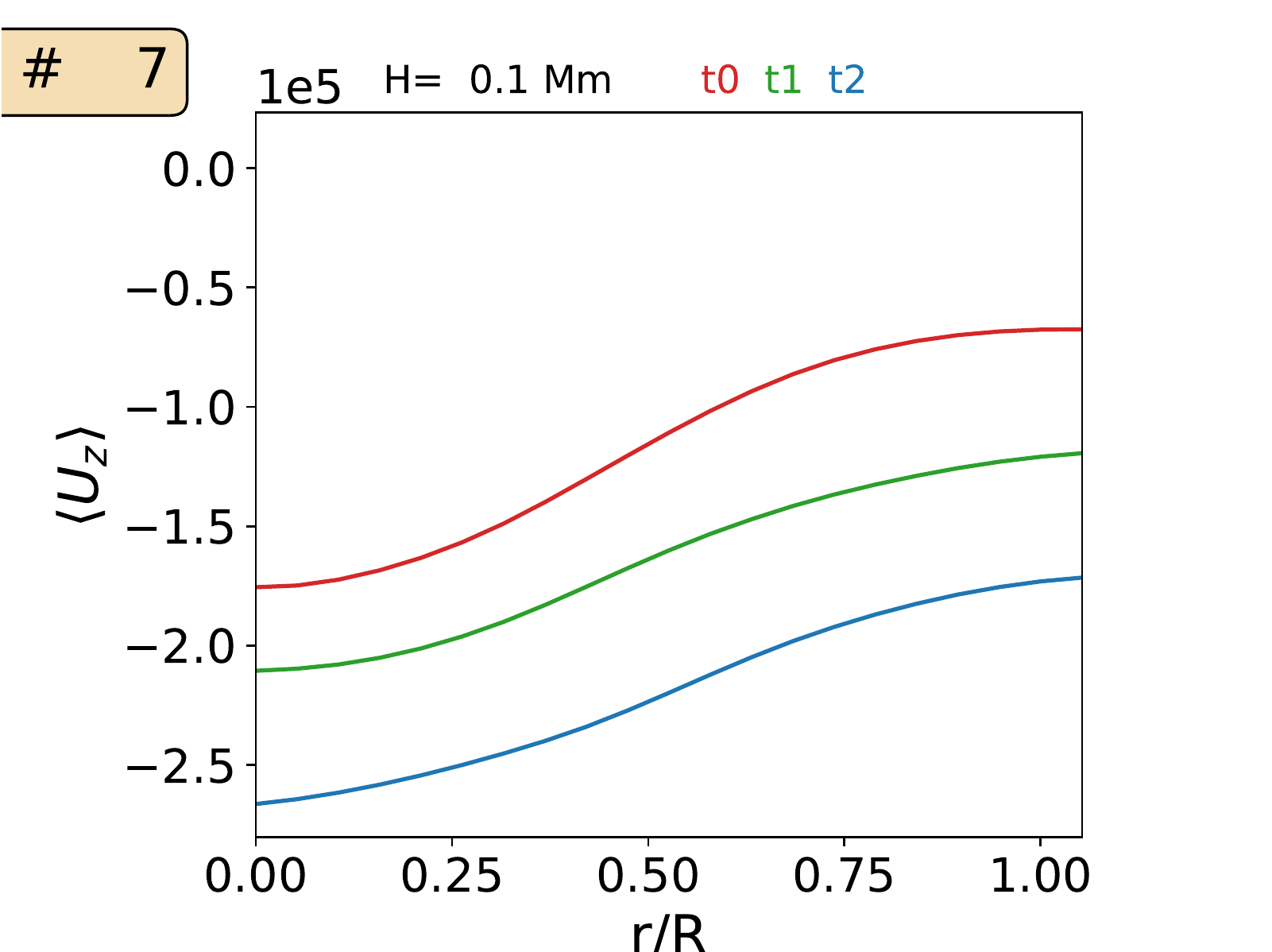}{0.3\textwidth}{(a)V7}
          \fig{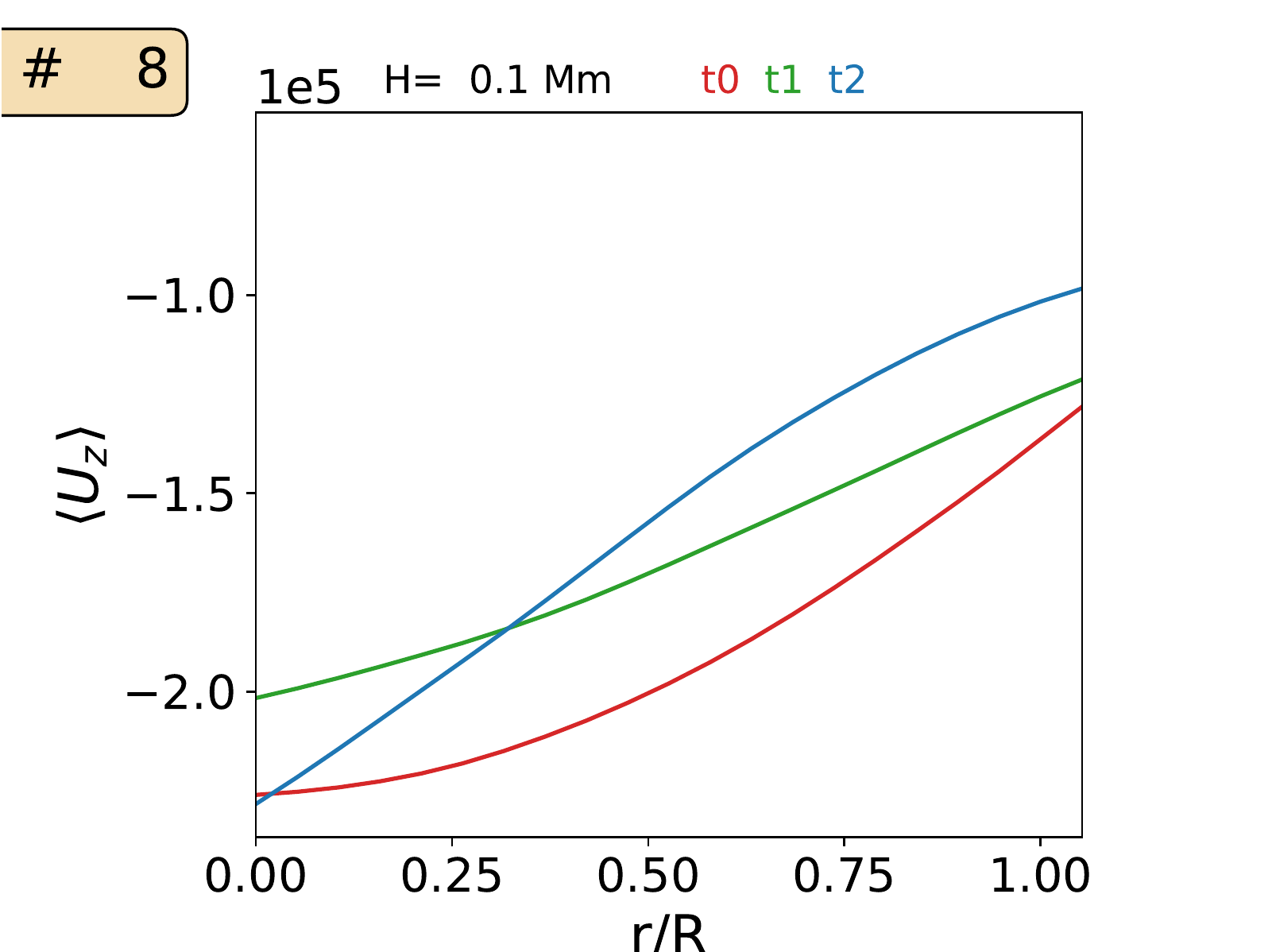}{0.3\textwidth}{(b)V8}
          \fig{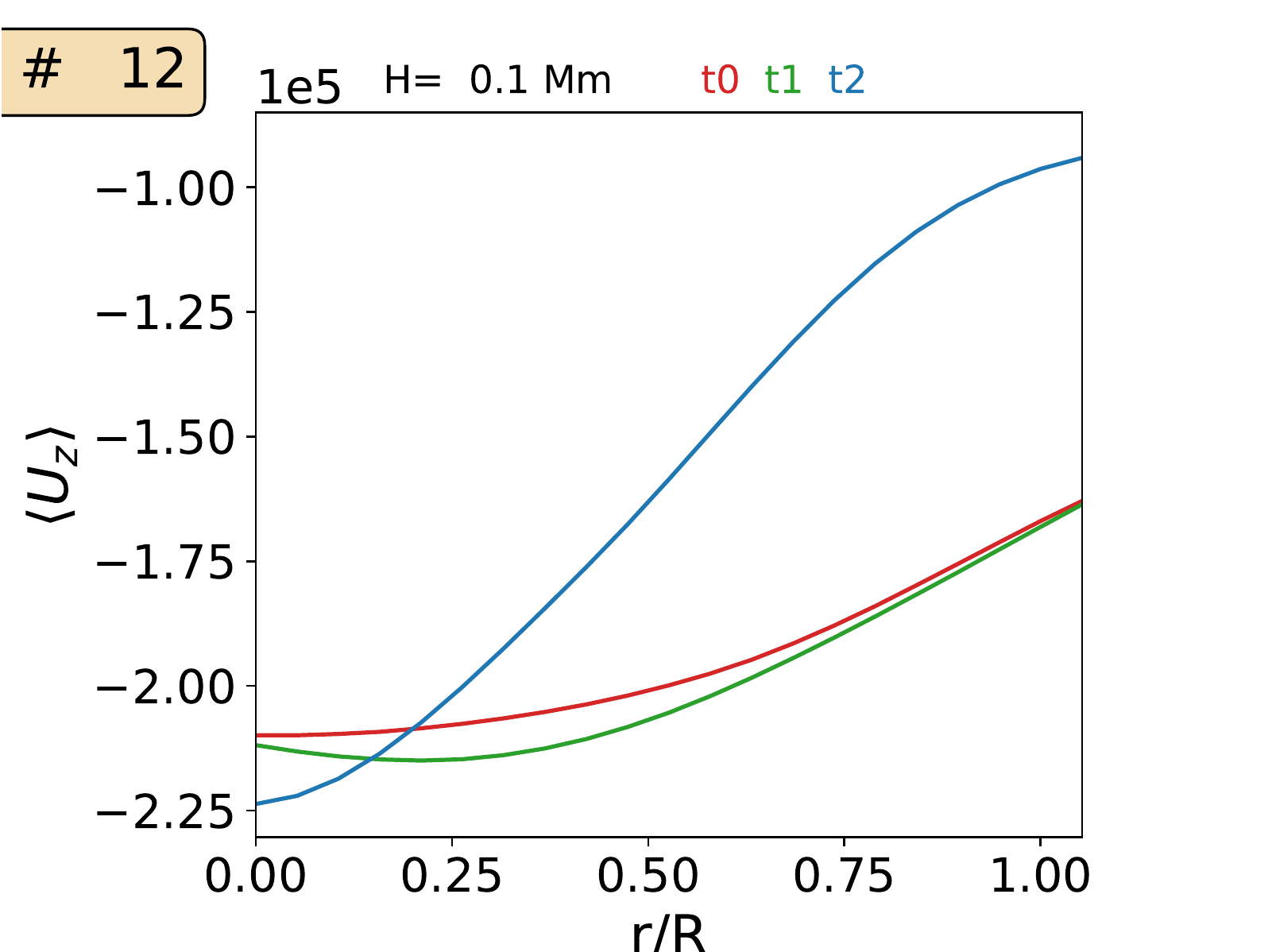}{0.3\textwidth}{(c)V12}
          }
\gridline{\fig{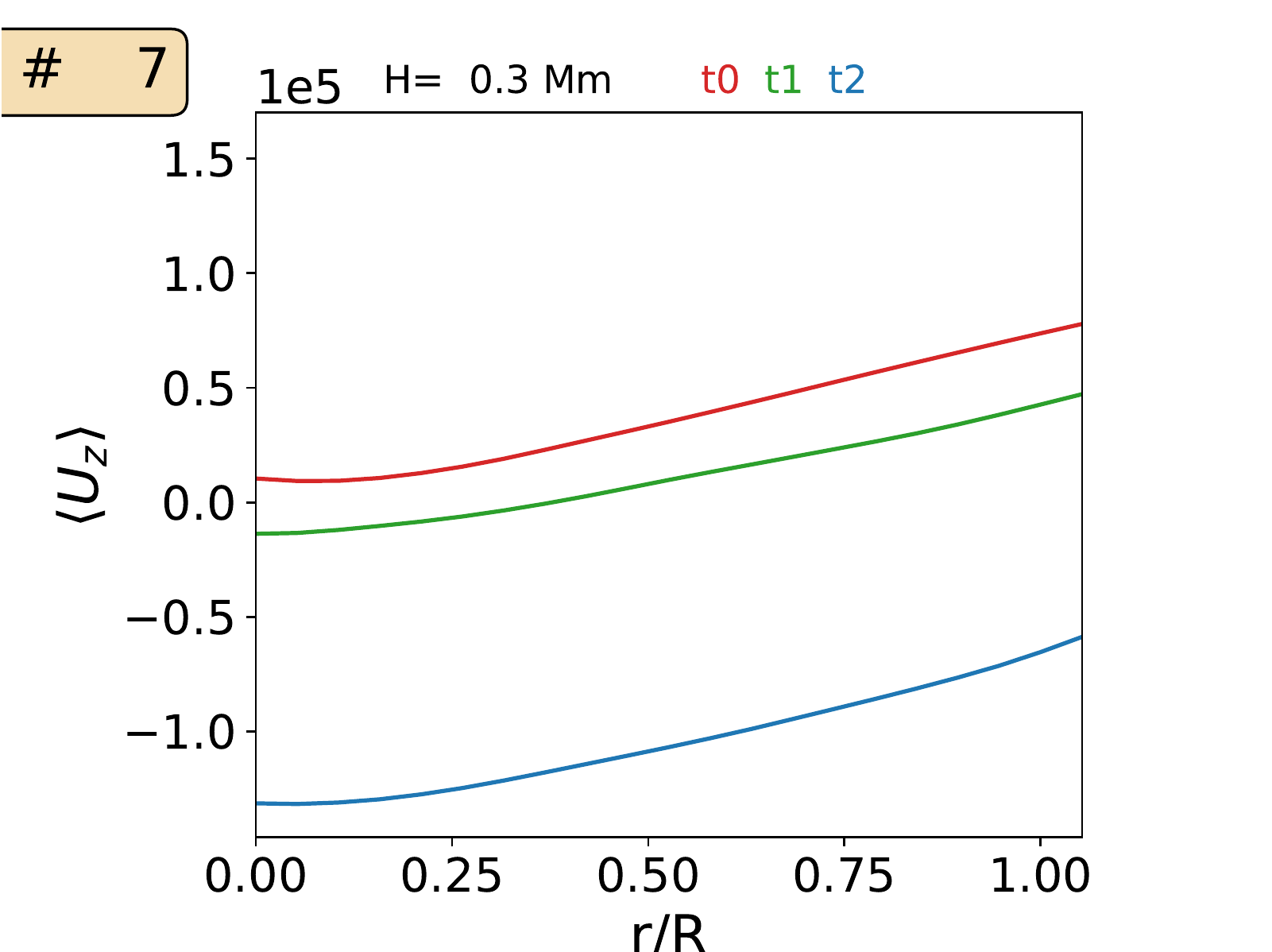}{0.3\textwidth}{(d)V7}
          \fig{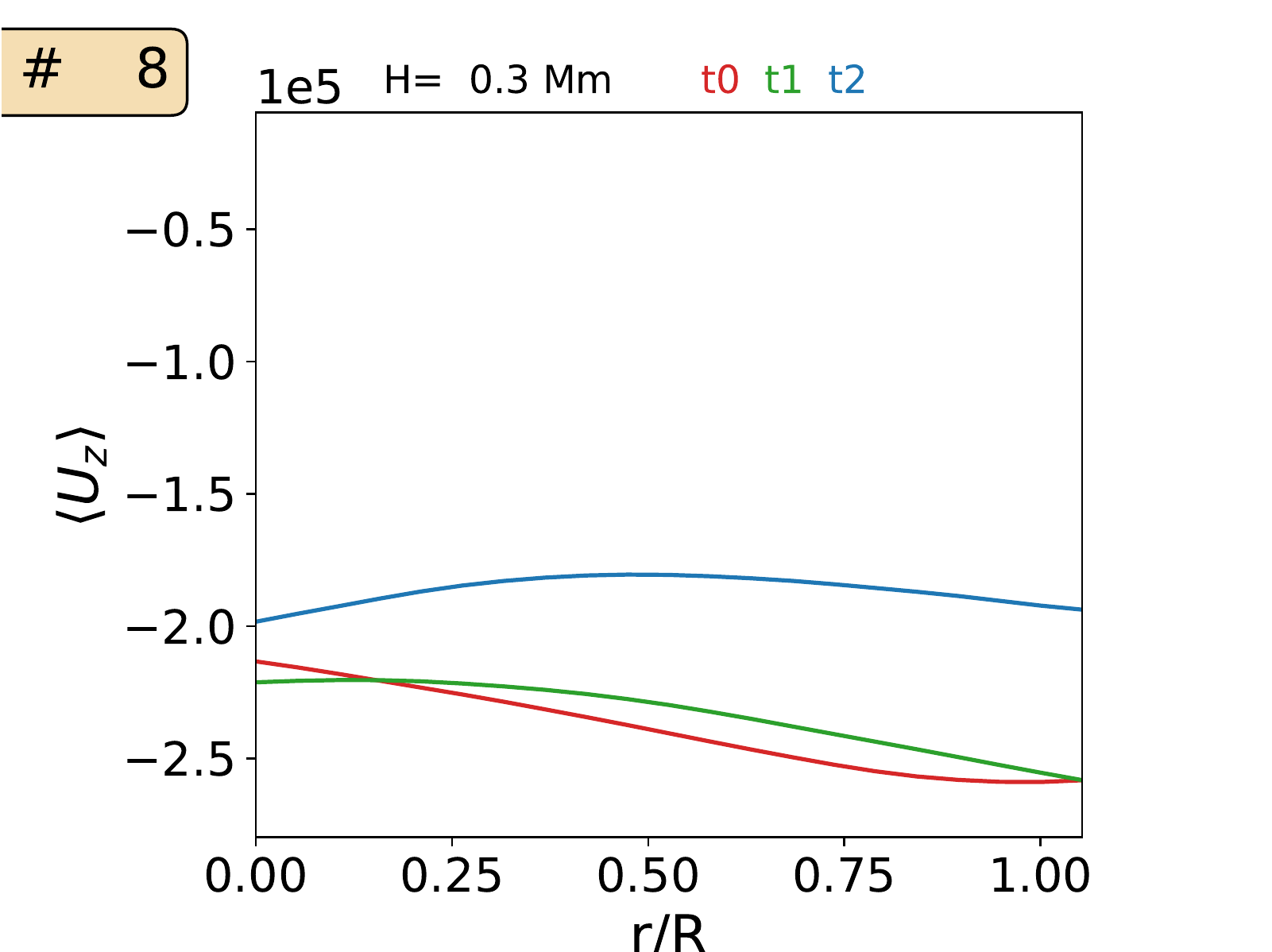}{0.3\textwidth}{(e)V8}
          \fig{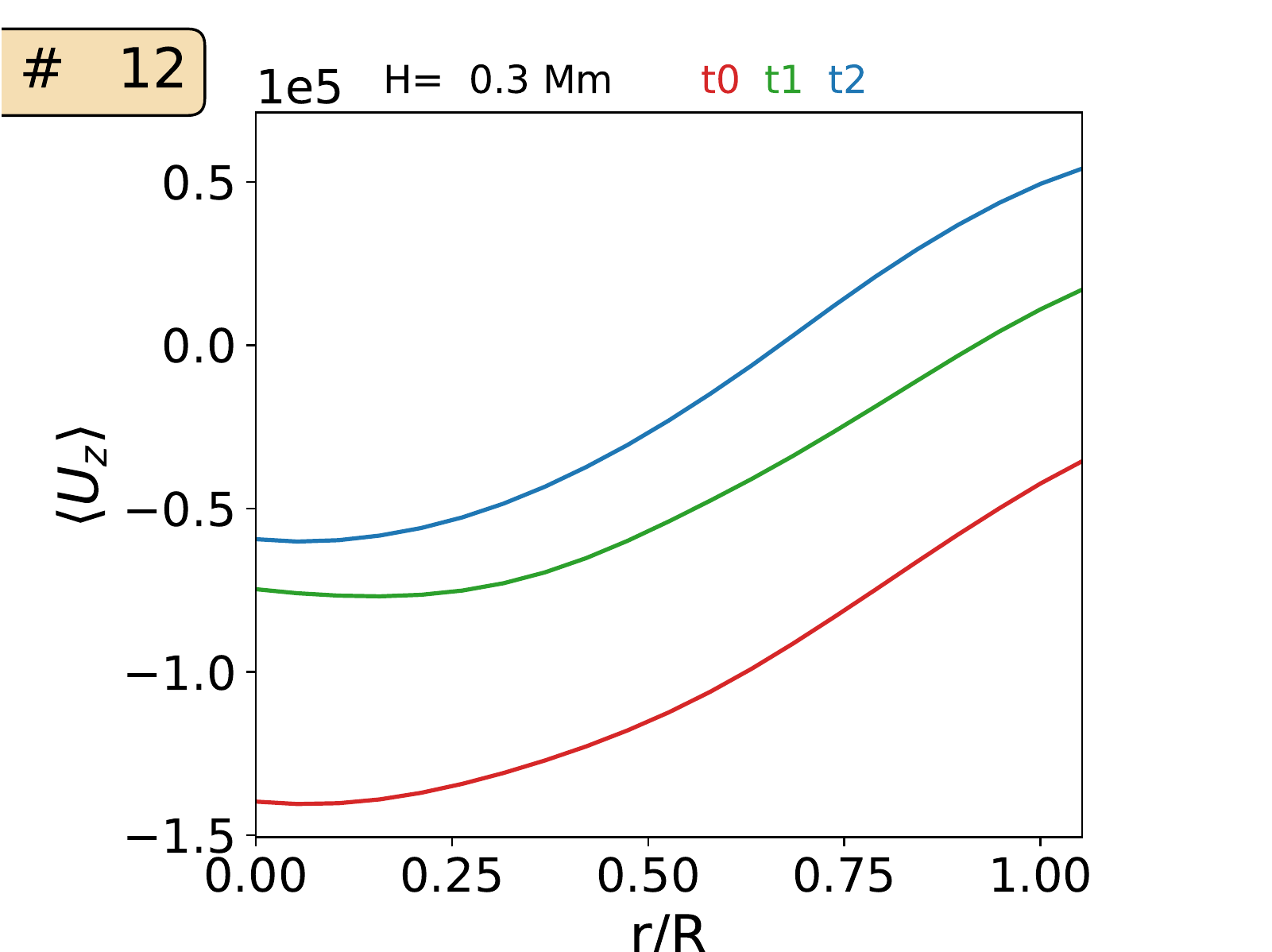}{0.3\textwidth}{(f)V12}
          }
\gridline{\fig{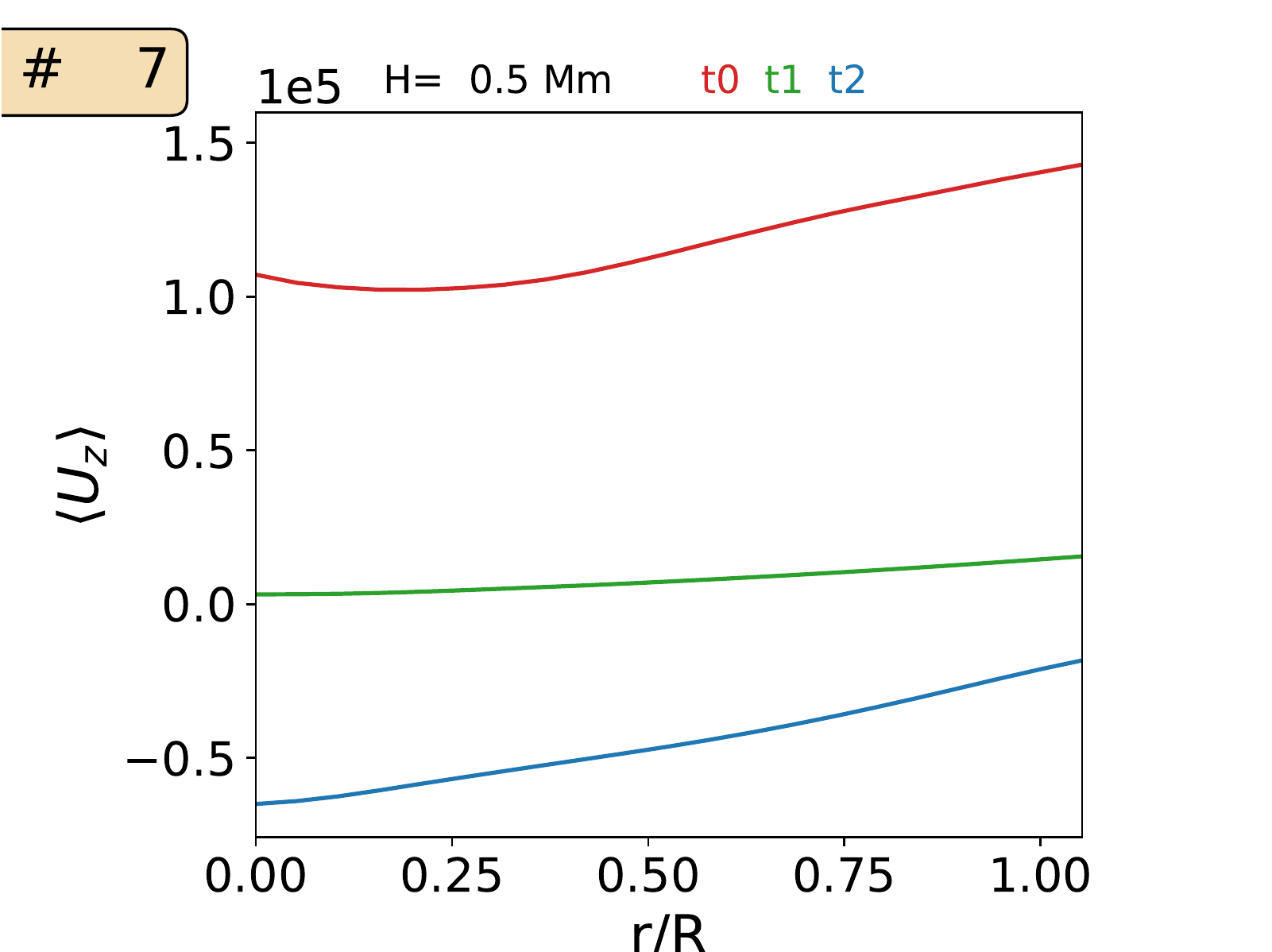}{0.3\textwidth}{(g)V7}
          \fig{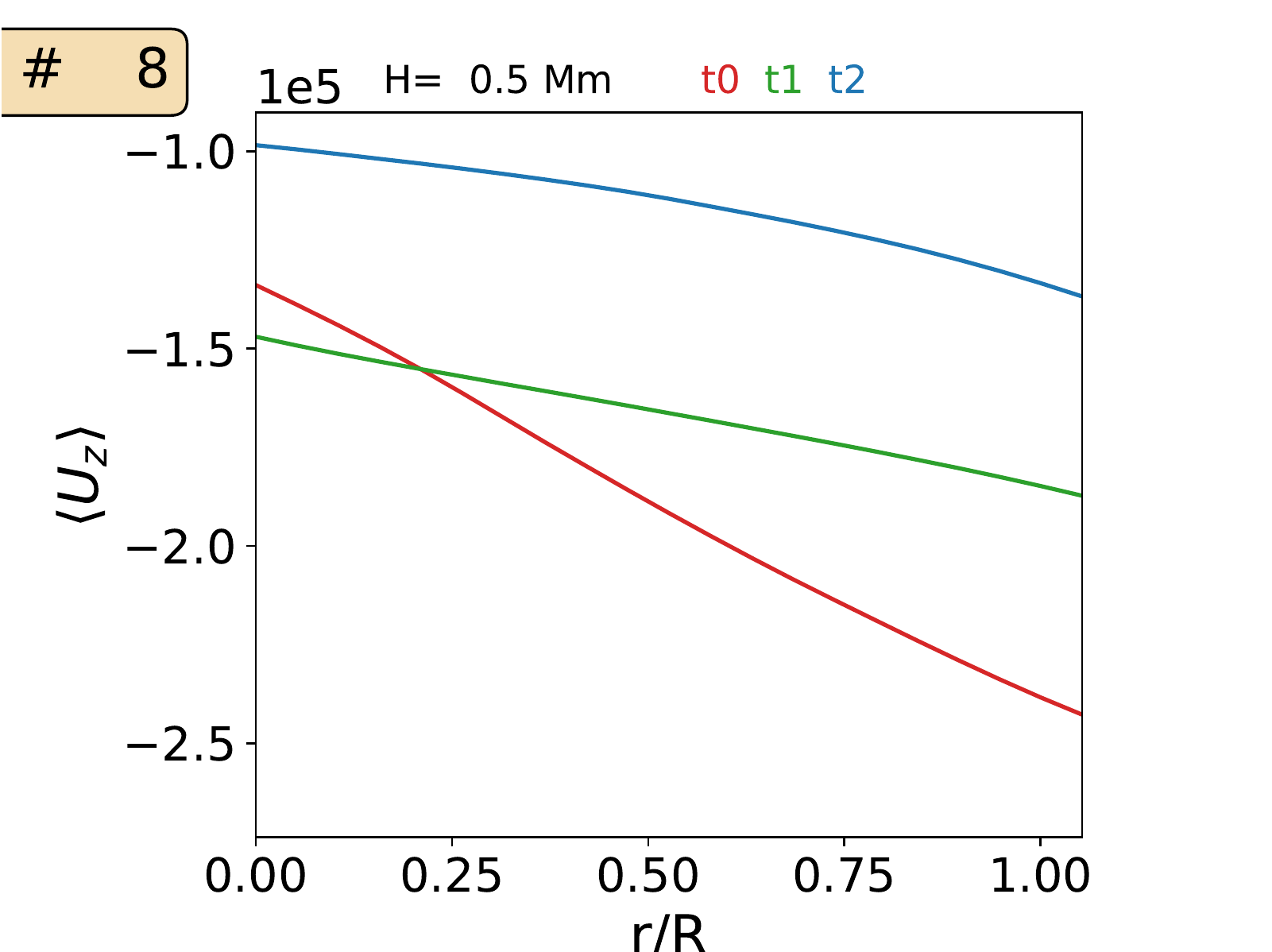}{0.3\textwidth}{(h)V8}
          \fig{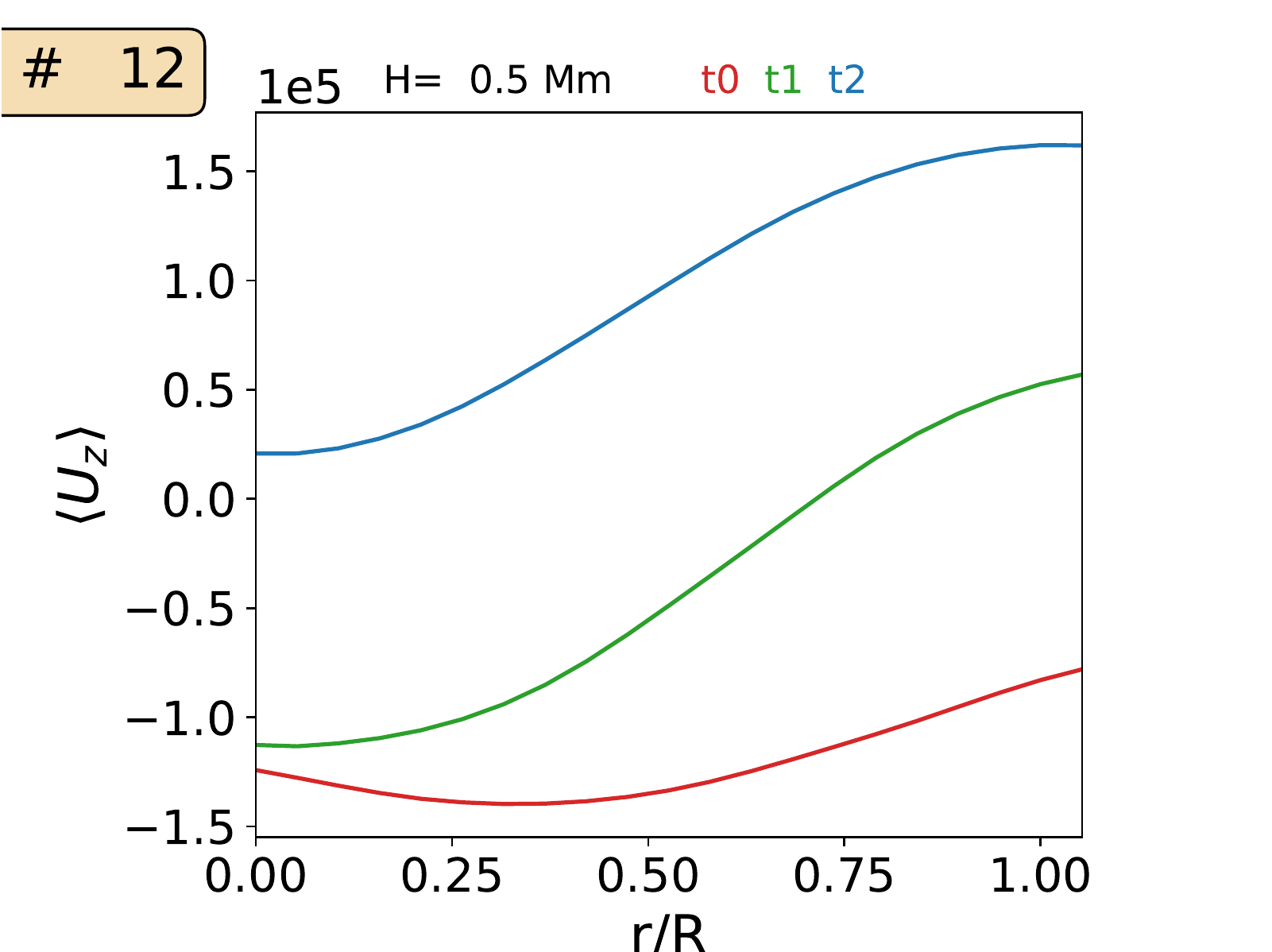}{0.3\textwidth}{(i)V12}
          }
\caption{The average z-component of velocity field  along the vortex radius from the center, $r=0$, to the boundary $r=R$. The red lines are for the initial time $t_0= 0$, the green lines are for  $t_1=25$ s, and the blue lines are for $t_2=50$ s.  The radial profiles are shown for Vortex \#7(a)(d)(g), \#8(b)(e)(h) and \#12(c)(f)(i) at different heights: $H = 0.1$ Mm (a)(b)(c), $H = 0.3$ Mm (d)(e)(f), $H = 0.5$ Mm  (g)(h)(i).  \label{fig:vzradial}}
\end{figure*}

Figure \ref{fig:vzradial} shows the averaged radial profile of the $z$-component of the velocity field. All vortices encompass downflows at $H= 0.1$ Mm. For vortices  \#7 and \#12, one can see a tendency of interchange between up- and downflows at other height levels during their lifetime. As for vortex \#8, there is only downflow, which tends to decrease as a function of time.  The downflows tend to be greater at the center, whereas the upflows are stronger around the vortex boundary. 

In order to obtain a general model for the tangential velocity profile within the vortex region, we have tried four different fitting polynomials:
\begin{itemize}
    \item Linear approximation: $v_{\theta} = ar $.
    \item Quadratic approximation: $v_{\theta} = ar^{2} + br +c $.
    \item Cubic approximation: $v_{\theta} = ar^{3} +br^{2} + cr +d $.
    \item Vortex model approximation : $v_{\theta} = -ar^{3} +br^{5} + cr^{7} $.
    
\end{itemize}
The vortex model approximation is based on the work of \cite{Rodriguez2012}, which established a series expansion to describe the common aspects of existing vortex models in nonmagnetized fluids. Those models are generally based on approximate solutions to the Navier-Stokes equations, which are obtained using different assumptions regarding boundary conditions and viscosity effects. Some of the classical vortex models, like Lamb–Oseen and Burgers vortex, (e.g., \citet{acheson1990}), could not  provide a suitable fit to our data and therefore were left out of this study and replaced by the general model proposed by  \cite{Rodriguez2012}.

\begin{figure*}[htp!]
\gridline{\fig{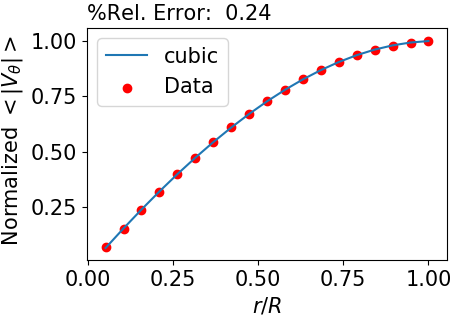}{0.3\textwidth}{(a)}
          \fig{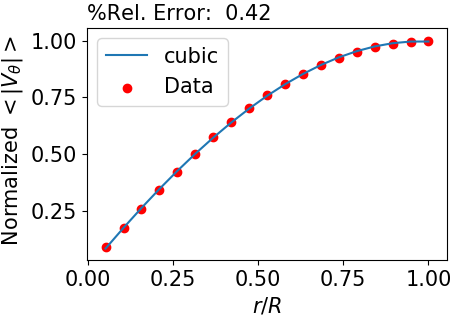}{0.3\textwidth}{(b)}
          \fig{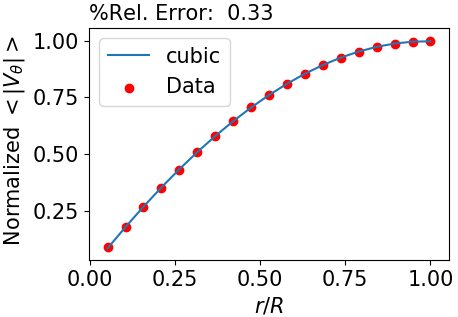}{0.3\textwidth}{(c)}
          }
\caption{ Radial profile of the tangential velocity of vortex \#12 normalized by its maximum value (red dots). The blue curve depicts the cubic function approximation for the profile. \label{fig:vthetacubic}}
\end{figure*}

To evaluate the best approximation, we compute the percentage average relative error, $E(i)$, of each approximation at a given point for each point \textit{i} along the radial direction until the vortex boundary: 
\begin{equation}
    E(i) = 100*\frac{\lambda(i) -\lambda_0(i)}{\lambda_0(i)},
\end{equation}
where $\lambda$ is the value obtained by the approximation given by the polynomial and $\lambda_0$ is the actual value obtained for tangential velocity. We then proceed to averaged $E(i)$ at each height level to obtain $E$ for each vortex. In table  \ref{tab:fit_vtheta}, we display the value of $E$ at different height levels and averaged for all the detected vortices at different times. We see that the cubic approximation gives the best description for the tangential velocity distribution as a function of the vortex radius. Figure \ref{fig:vthetacubic} shows the cubic fit for vortex \#12 at $H=0.1$ Mm, 0.3 Mm, and 0.5 Mm for $t=50$s.
The tangential velocity is normalized by its maximum intensity at the given height level.

\begin{table}[htp!]
    \centering
    \begin{tabular}{|c|c|c|c|c|c}
    \hline
Height (Mm) & Linear & Quadratic & Cubic & Vortex Model    \\ 
\hline
0 &  12.47 & 2.68 & 1.13 & 4.78  \\
0.1  &  8.93 & 1.80 & 1.21 & 2.16\\
0.2 &  8.52 & 1.19 & 0.50 & 1.13 \\
0.3 &  8.05 & 0.97 & 0.29 & 0.99 \\
0.4 &  6.89 & 0.95 & 0.34 & 0.91 \\
0.5 &  6.43 & 1.14 & 0.39 & 1.29 \\
\hline
    \end{tabular}
    \caption{Average Relative Error of Polynomial Fits for the Radial Profiles of the Tangential Velocities of All Detected vortices.}
    \label{tab:fit_vtheta}
\end{table}
The solar vortex tubes present similar curves concerning the description of the rotating flow. A vortex with a solid body rotation has a tangential velocity given by:
\begin{equation}
    V_\theta= \Omega r,
    \label{eq:solid}
\end{equation} 
where $\Omega$ is the angular velocity that is uniform  and $r$ is the vortex radius. Therefore, the tangential velocity would have a linear dependence with $r$ in a rigid body rotation. According to Table \ref{tab:fit_vtheta}, the solar vortices tend to deviate from rigid body rotation as $V_{\theta}$ has a better fit with cubic dependence. We see from Table \ref{tab:fit_vtheta} that closer to the photosphere, the deviation from a solid body rotation is more than 10\%, whereas it decreases for the upper parts of the vortex. The small relative errors found for a cubic approximation is a clear indication that solar vortices do not present a rigid body rotation.

\begin{figure*}
\gridline{\fig{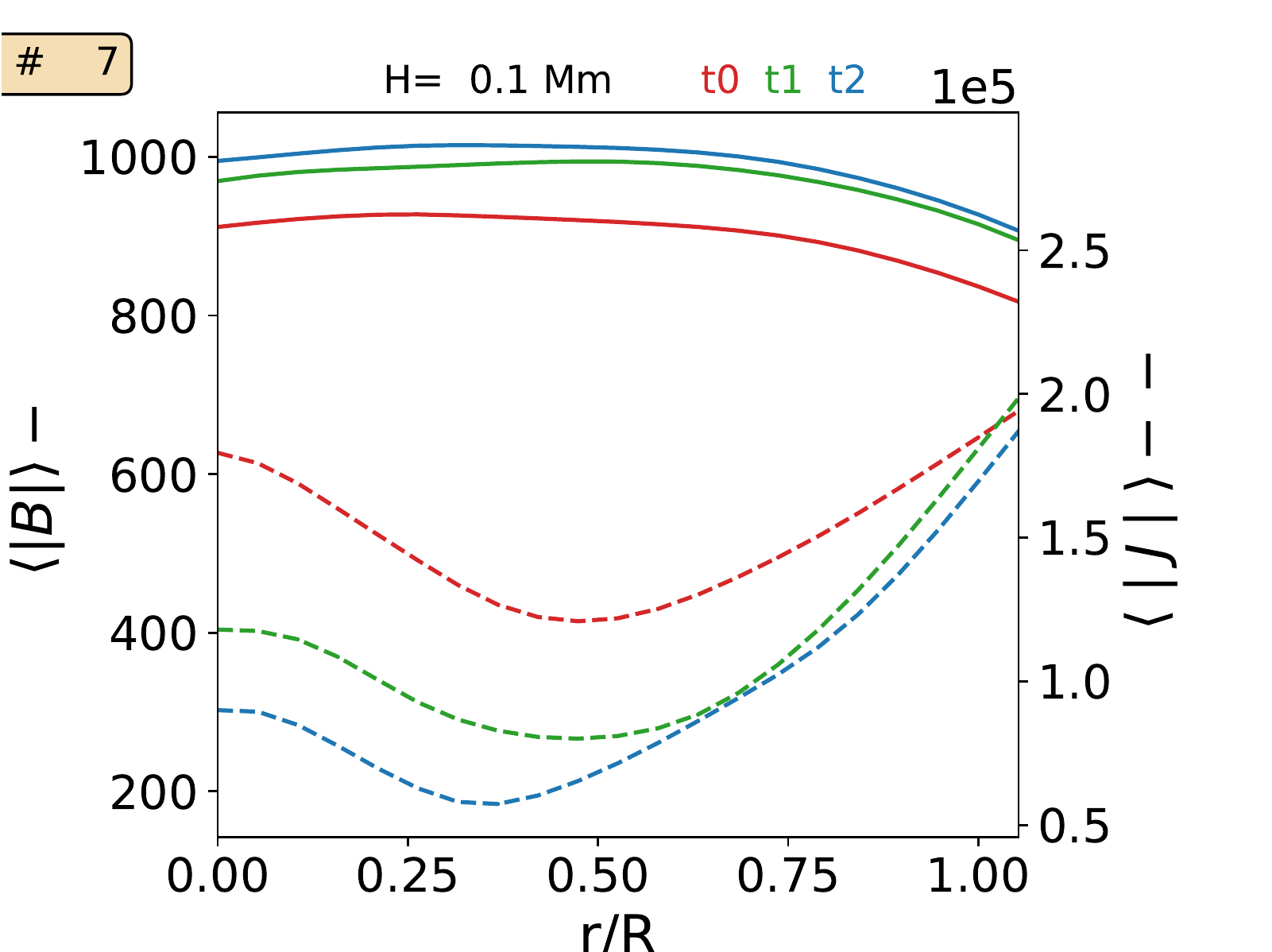}{0.3\textwidth}{(a)V7}
          \fig{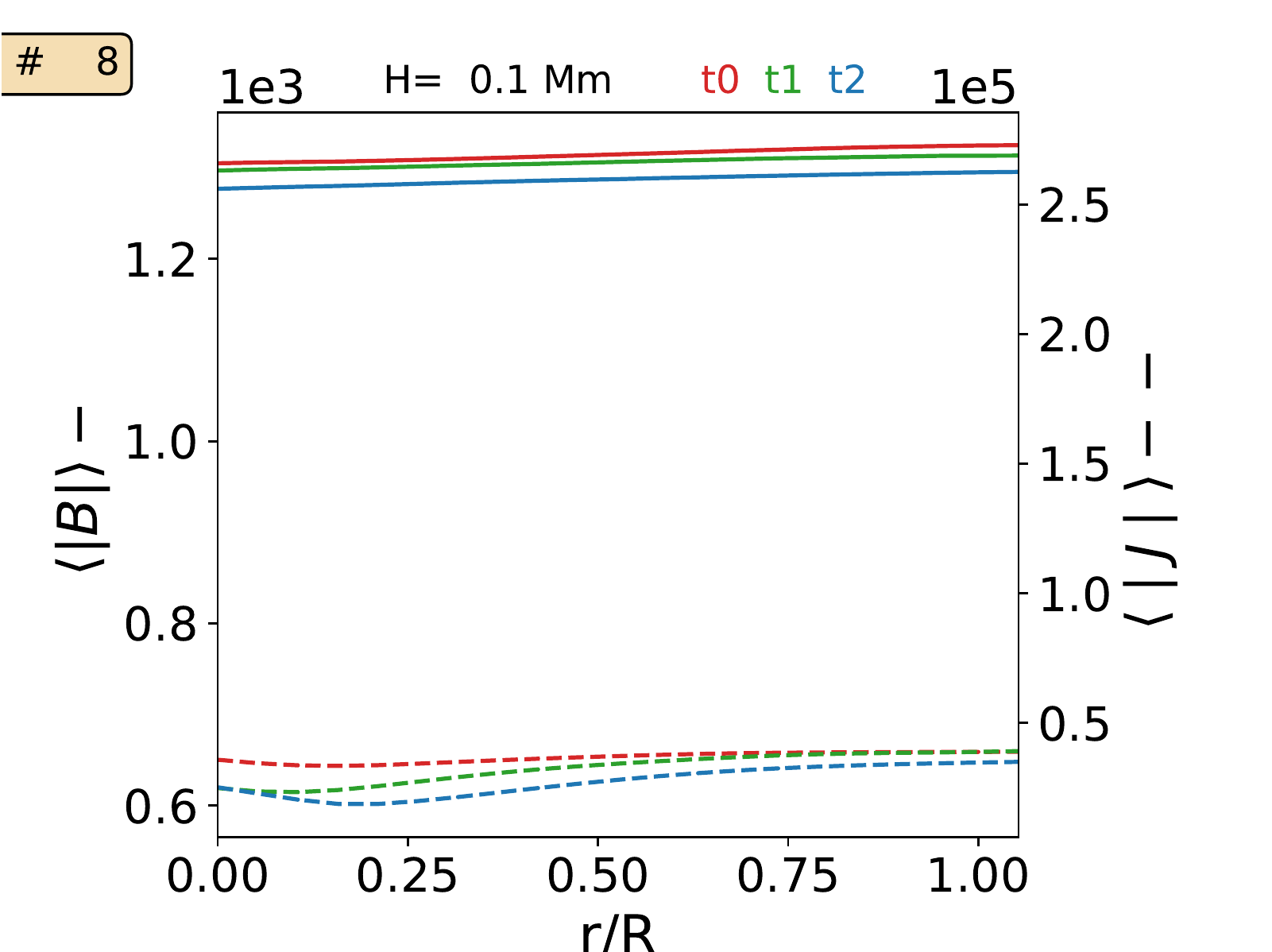}{0.3\textwidth}{(b)V8}
          \fig{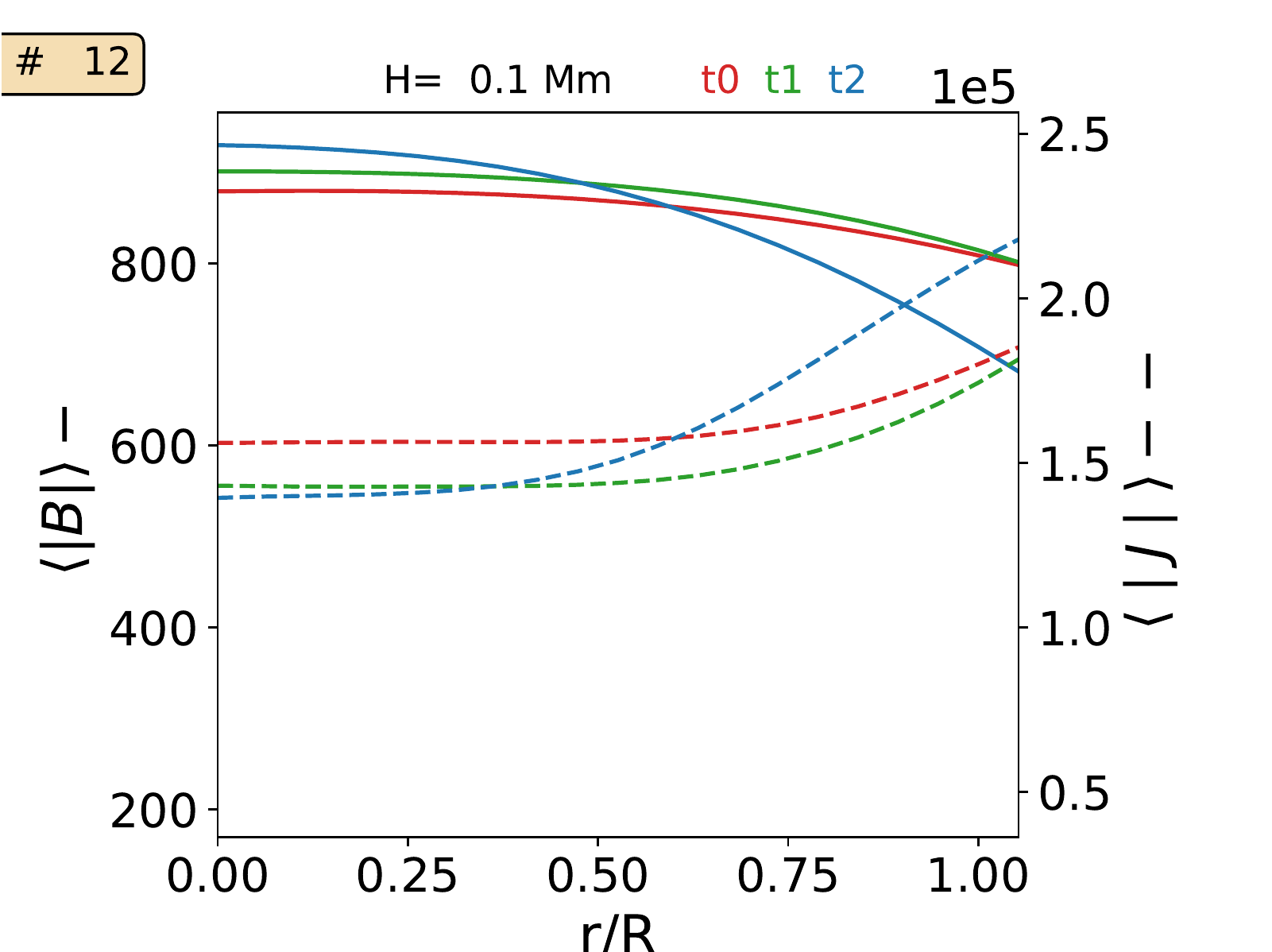}{0.3\textwidth}{(c)V12}
          }
\gridline{\fig{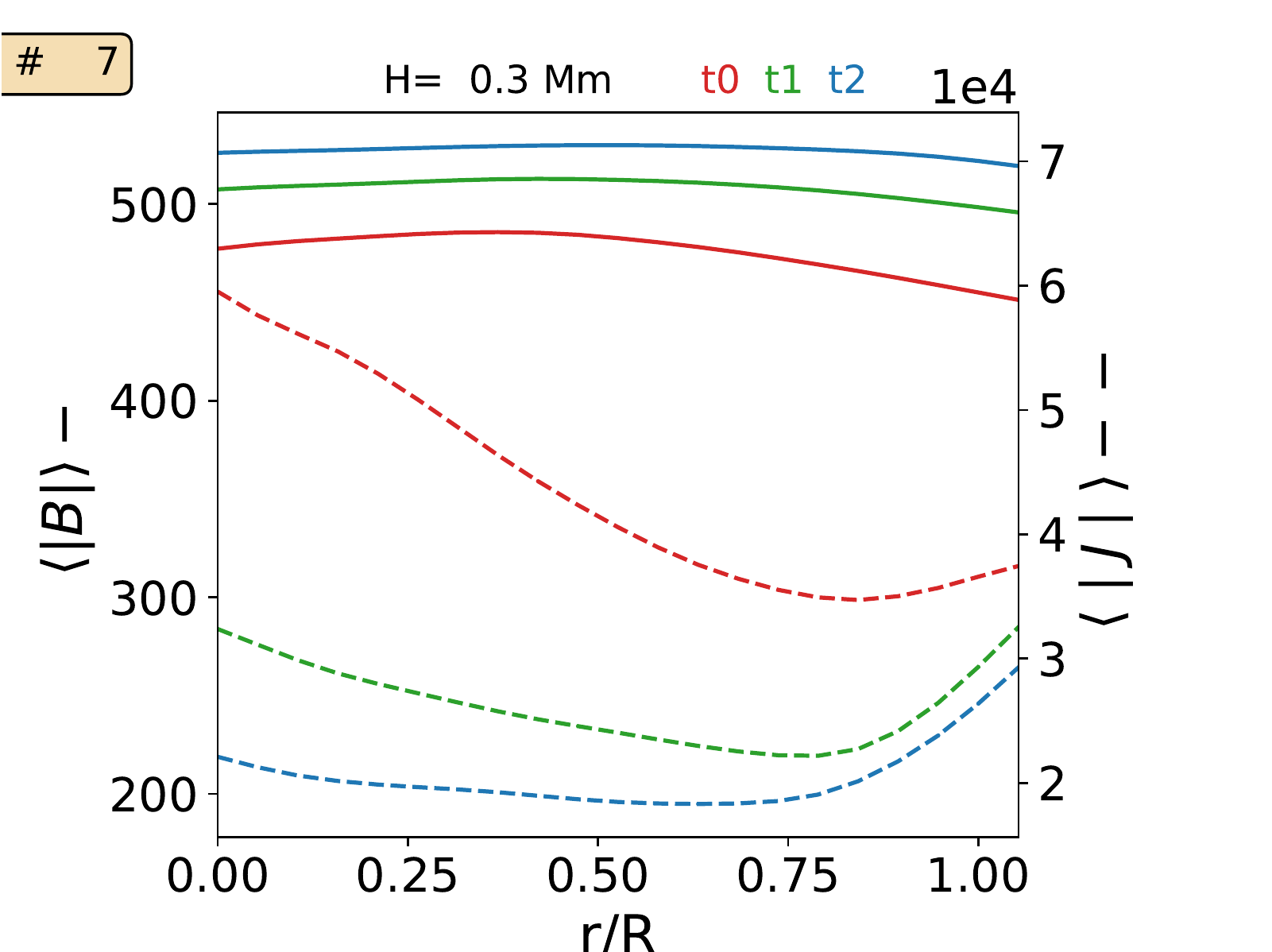}{0.3\textwidth}{(d)V7}
          \fig{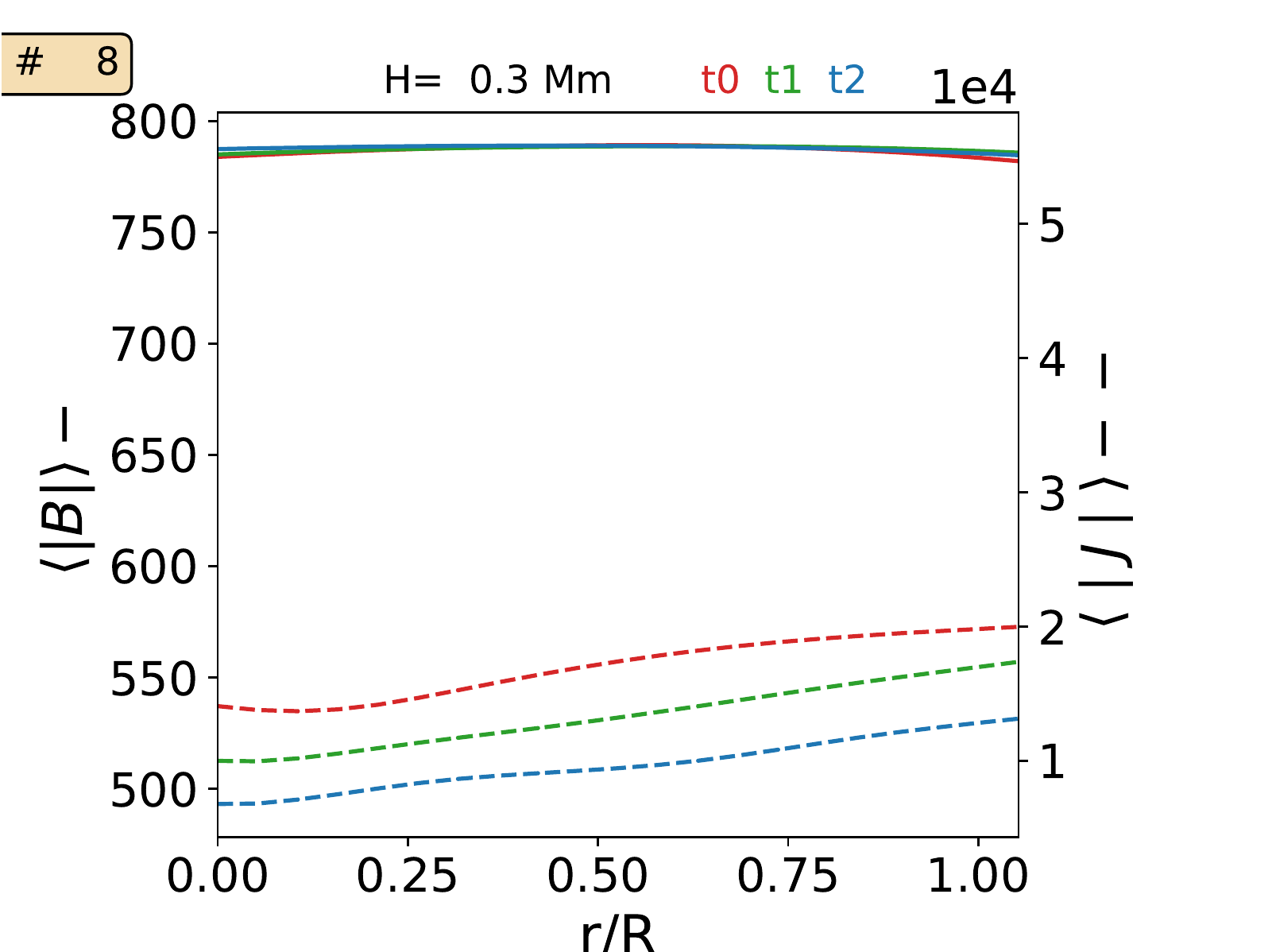}{0.3\textwidth}{(e)V8}
          \fig{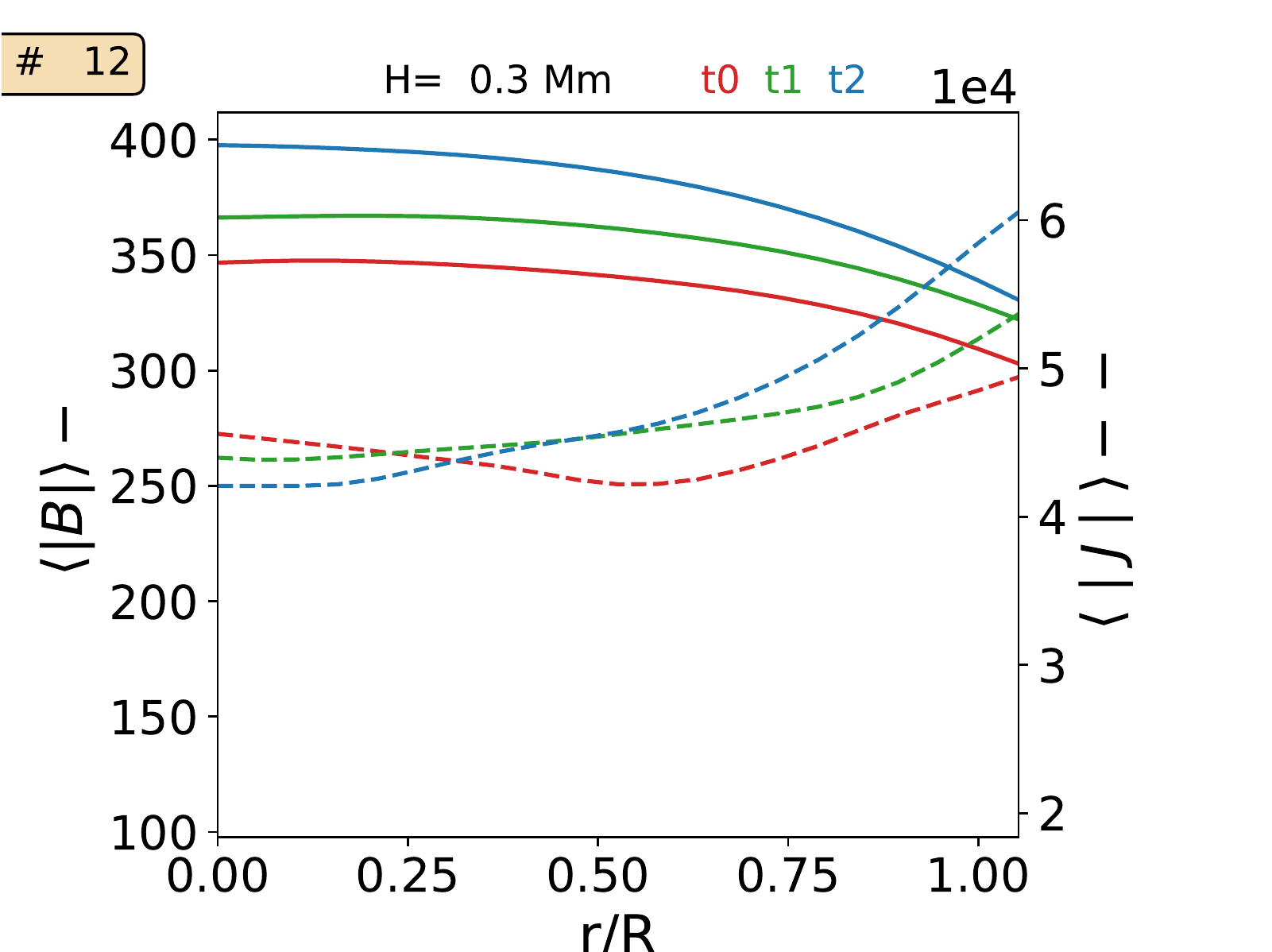}{0.3\textwidth}{(f)V12}
          }
\gridline{\fig{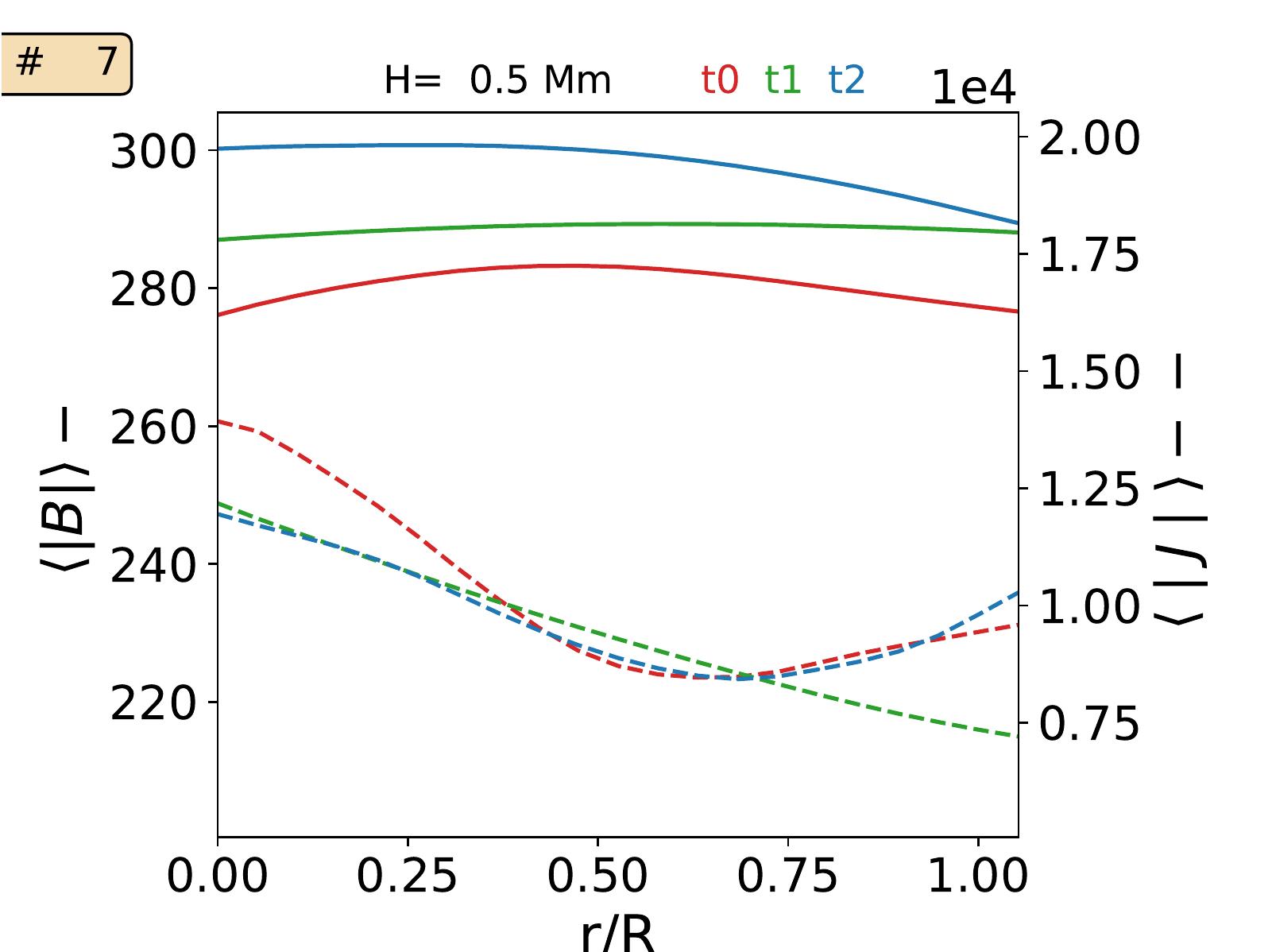}{0.3\textwidth}{(g)V7}
          \fig{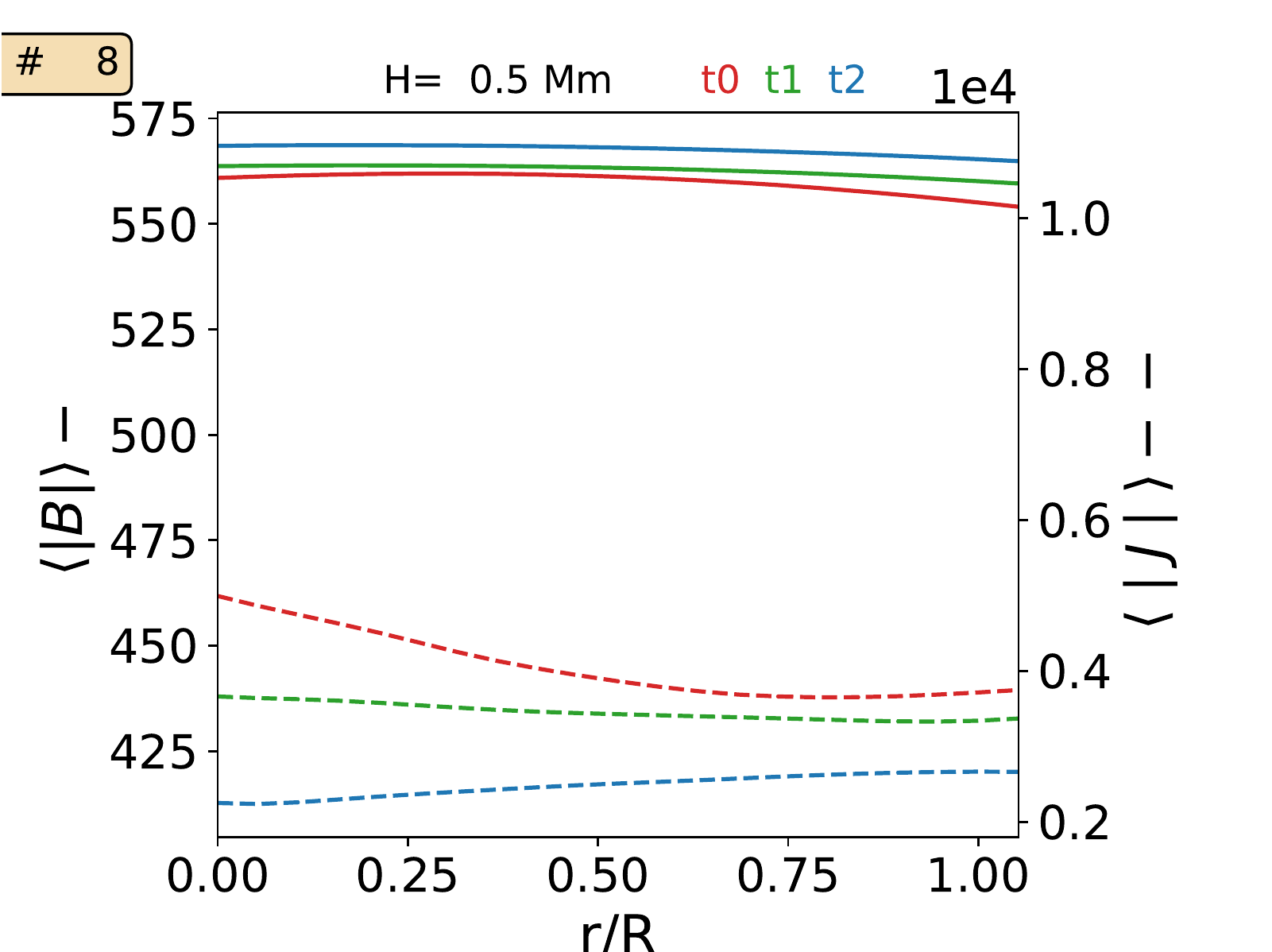}{0.3\textwidth}{(h)V8}
          \fig{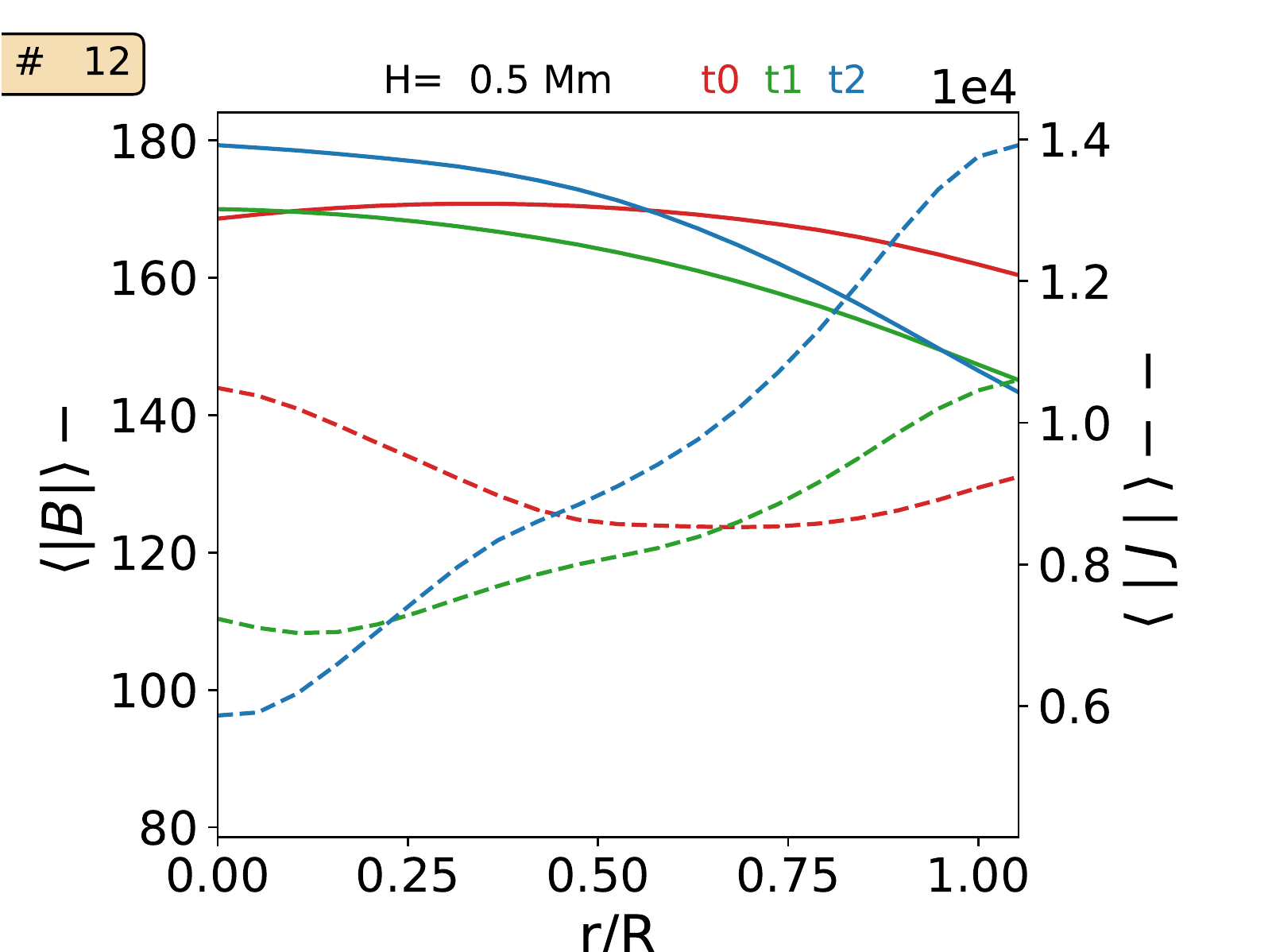}{0.3\textwidth}{(i)V12}
          }
\caption{The average magnetic field intensity (left $y$-axis, solid lines) and current density field intensity (right $y$-axis, dashed lines)  along the vortex radius from the center, $r=0$, to the boundary $r=R$. The red lines are for the initial time $t_0= 0$, the green lines are for $t_1=25$ s (green line), and the blue lines are for $t_2=50$ s.  The radial profiles are shown for vortices \#7 (a)(d)(g), \#8 (b)(e)(h) and \#12 (c)(f)(i) at different heights: $H = 0.1$ Mm (a)(b)(c), $H = 0.3$ Mm (d)(e)(f), $H = 0.5$ Mm(g)(h)(i).  \label{fig:BJradial}
}
\end{figure*}

We now turn to the properties of the magnetic field inside the vortices. Figure \ref{fig:BJradial} shows the averaged radial profile of magnetic field intensity (solid lines, left axis) and the current density intensity (dashed lines, right axis) for vortices \#7, \#8, and \#12.  The magnetic field at the center tends to be between 1\% and 12\% higher than at the boundary.  This difference tends to increase over time, and it is also greater at the parts of the vortices closer to the photosphere. The exception is for vortex \#8, which has stronger magnetic field intensities at its border at $H=0.1$ Mm, where it also loses intensity compared to initial times.  Another interesting feature of vortex \#8  is that it encompasses a magnetic field from $\sim $ 30\% up to $\sim $ 45\% larger than the other selected vortices at any given height level.  
For most of the detected vortices, we found that the magnetic field intensity tends to increase around 1\% $\sim$ 15\% at vortex's center. The current density intensity is within the same range at each height level for all those three vortices, even though there are considerable differences for magnetic field intensities inside each vortex. Initially, at time $t_0$, the current density is higher at the center for the upper part of the vortex, whereas, at $H=0.1$ Mm, the boundary presents greater current density values. Over time, the vortex boundary tends to hold the highest current density within the vortex And, except for vortex \#12, the current density intensity tends to decrease over time.

\begin{figure*}
\gridline{\fig{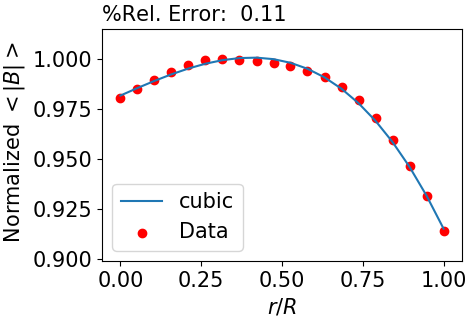}{0.3\textwidth}{(a)}
          \fig{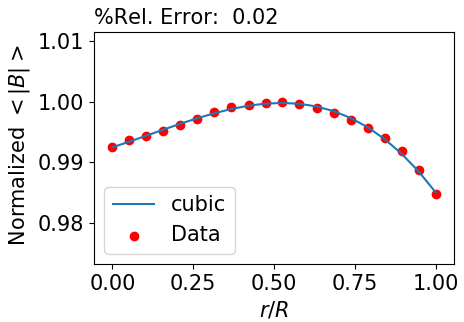}{0.3\textwidth}{(b)}
          \fig{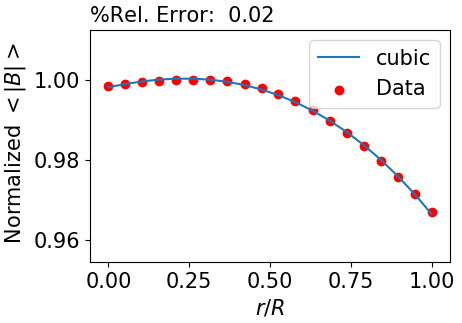}{0.3\textwidth}{(c)}
          }
\caption{Magnetic field intensity radial profile of vortex \#7  for $t= 50$ s normalized by its maximum value (red dots). The blue curve depicts the cubic function approximation for the profile. \label{fig:Bfit}}
\end{figure*}

We also apply the first three fitting polynomials mentioned above in order to fit the magnetic intensity radial profile of the vortices. The average relative errors for each function for different height levels are shown in Table \ref{tab:Bfit}. Again, the best fit is given by the cubic approximation, which is shown for vortex \#7 at different height levels for $t = 50$ s in Fig.\ \ref{fig:Bfit}.
 \begin{table}[htp!]
     \centering
    \begin{tabular}{c|c|c|c}
    \hline
Height (Mm) & Linear & Quadratic & Cubic  \\ 
\hline
0 & 0.65 & 0.062 & 0.035 \\
0.1 & 0.56 & 0.11 & 0.037 \\
0.2 & 0.36 & 0.072 & 0.028 \\
0.3 & 0.34 & 0.071 & 0.024 \\
0.4 & 0.37 & 0.055 & 0.019 \\
0.5 & 0.33 & 0.056 & 0.020
     \end{tabular}
\caption{Average Relative Error of Polynomial Fits for the Magnetic Field Radial Profile of All Detected Vortices.}
     \label{tab:Bfit}
 \end{table}

\section{Discussions} \label{sec:discussions}

\begin{figure*}
\gridline{\fig{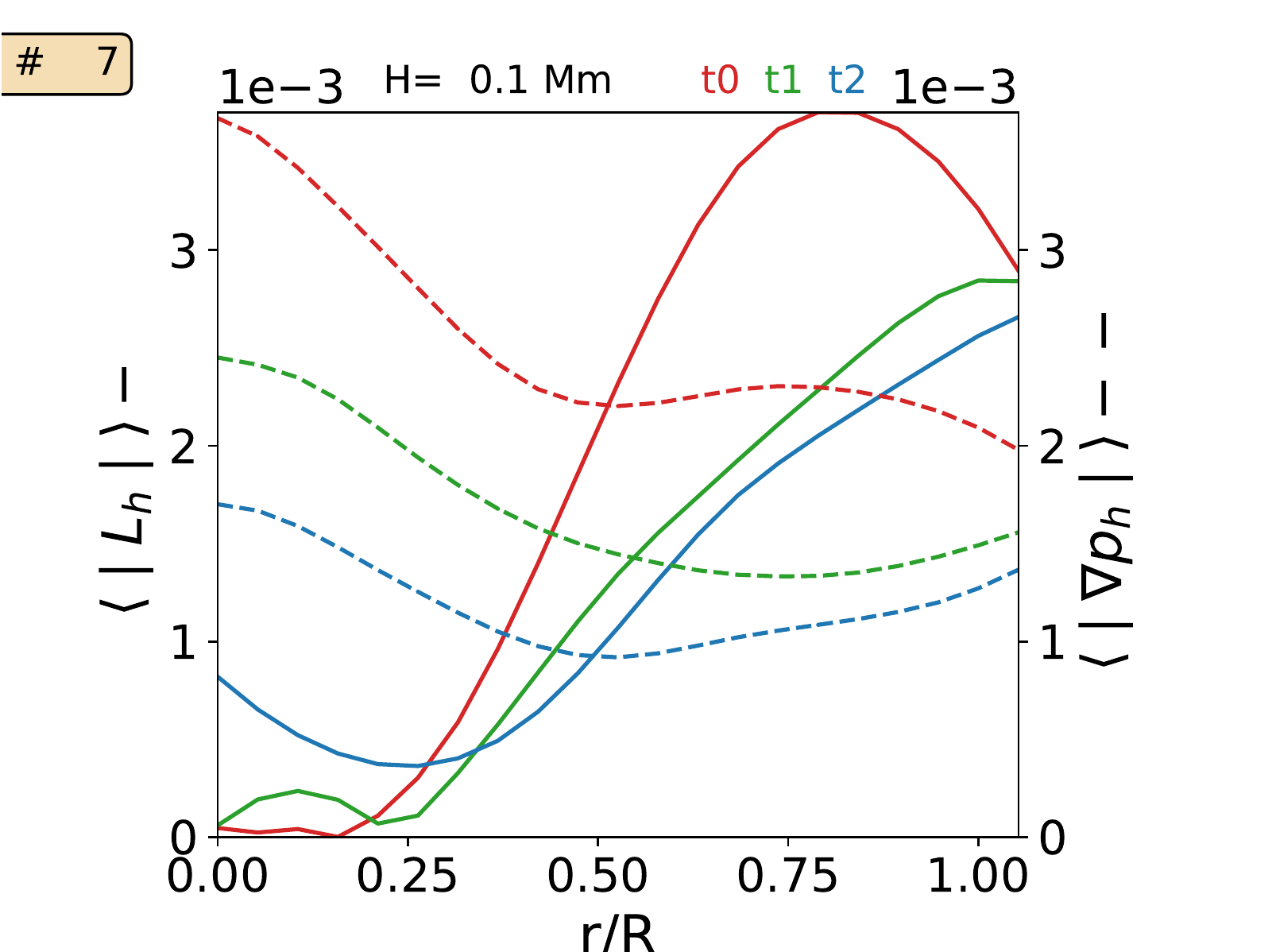}{0.3\textwidth}{(a)V7}
          \fig{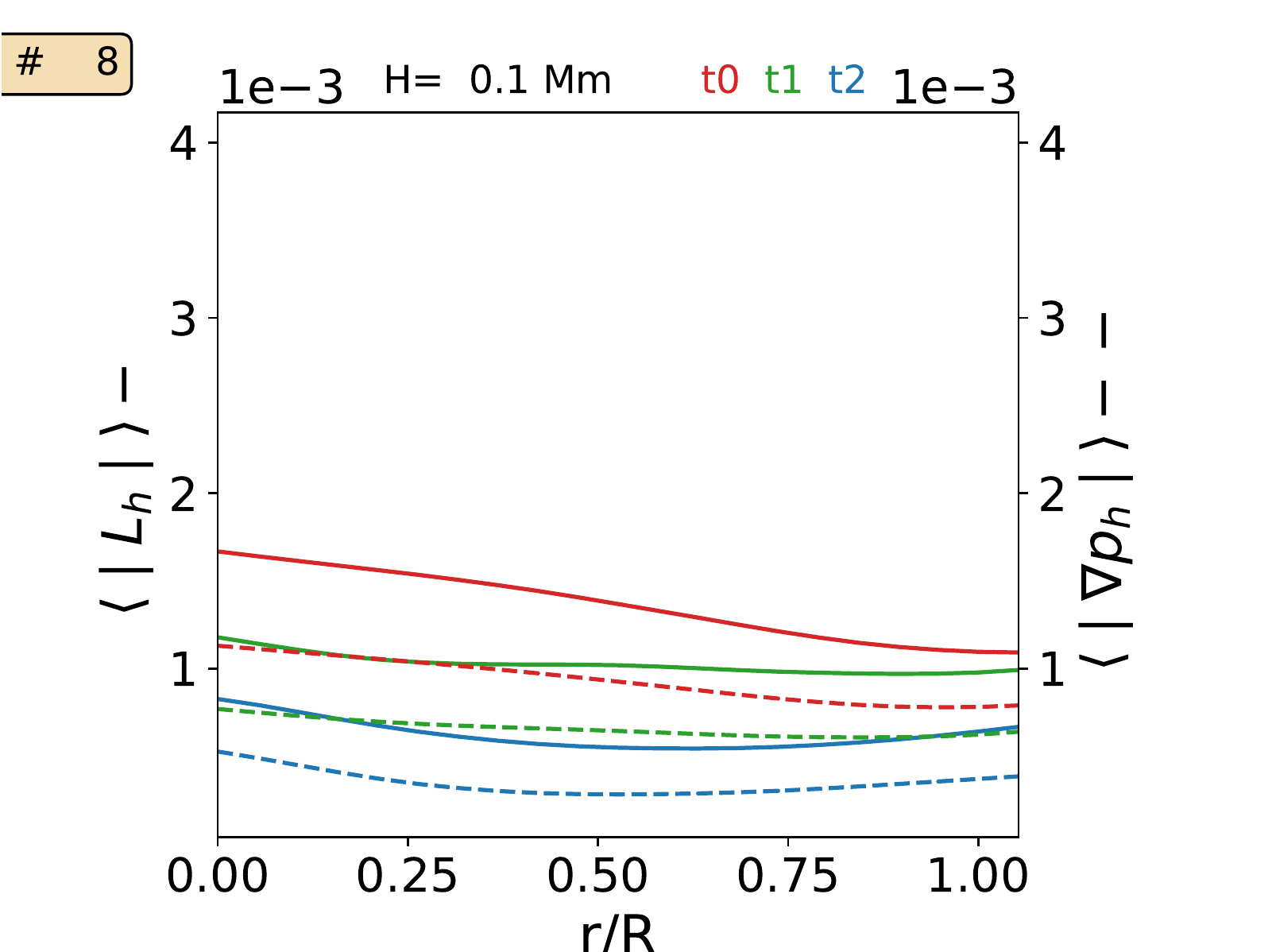}{0.3\textwidth}{(b)V8}
          \fig{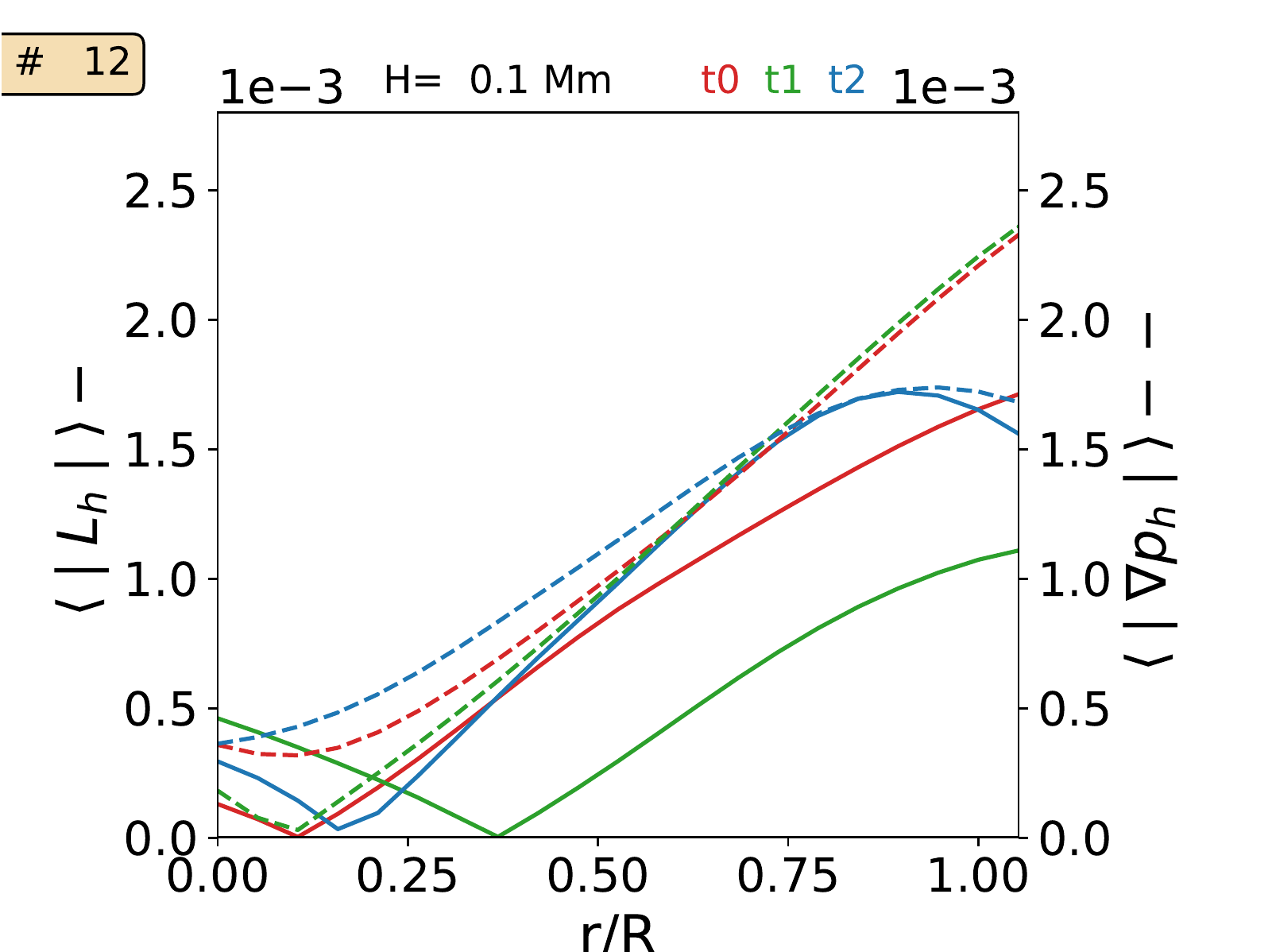}{0.3\textwidth}{(c)V12}
          }
\gridline{\fig{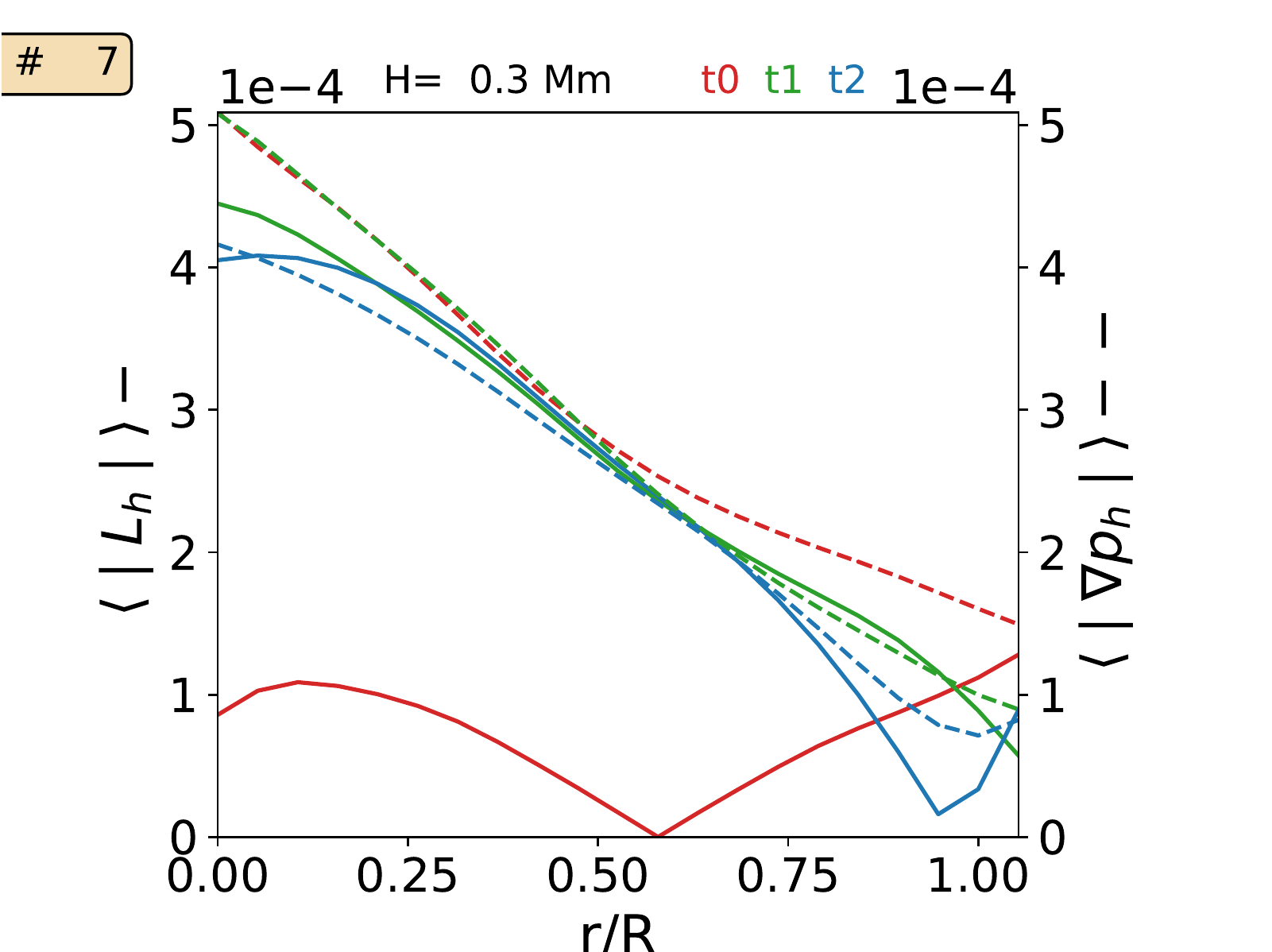}{0.3\textwidth}{(d)V7}
          \fig{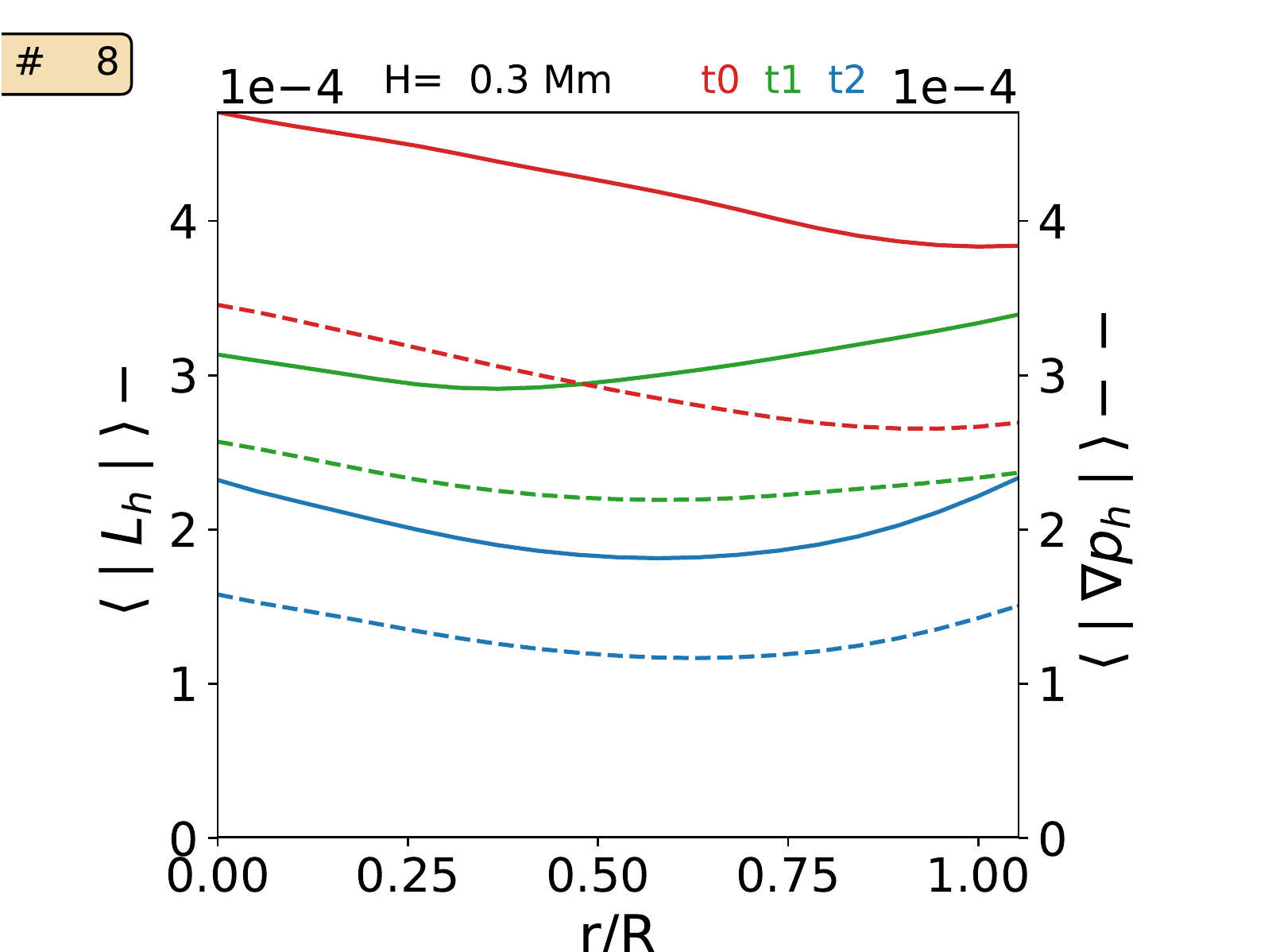}{0.3\textwidth}{(e)V8}
          \fig{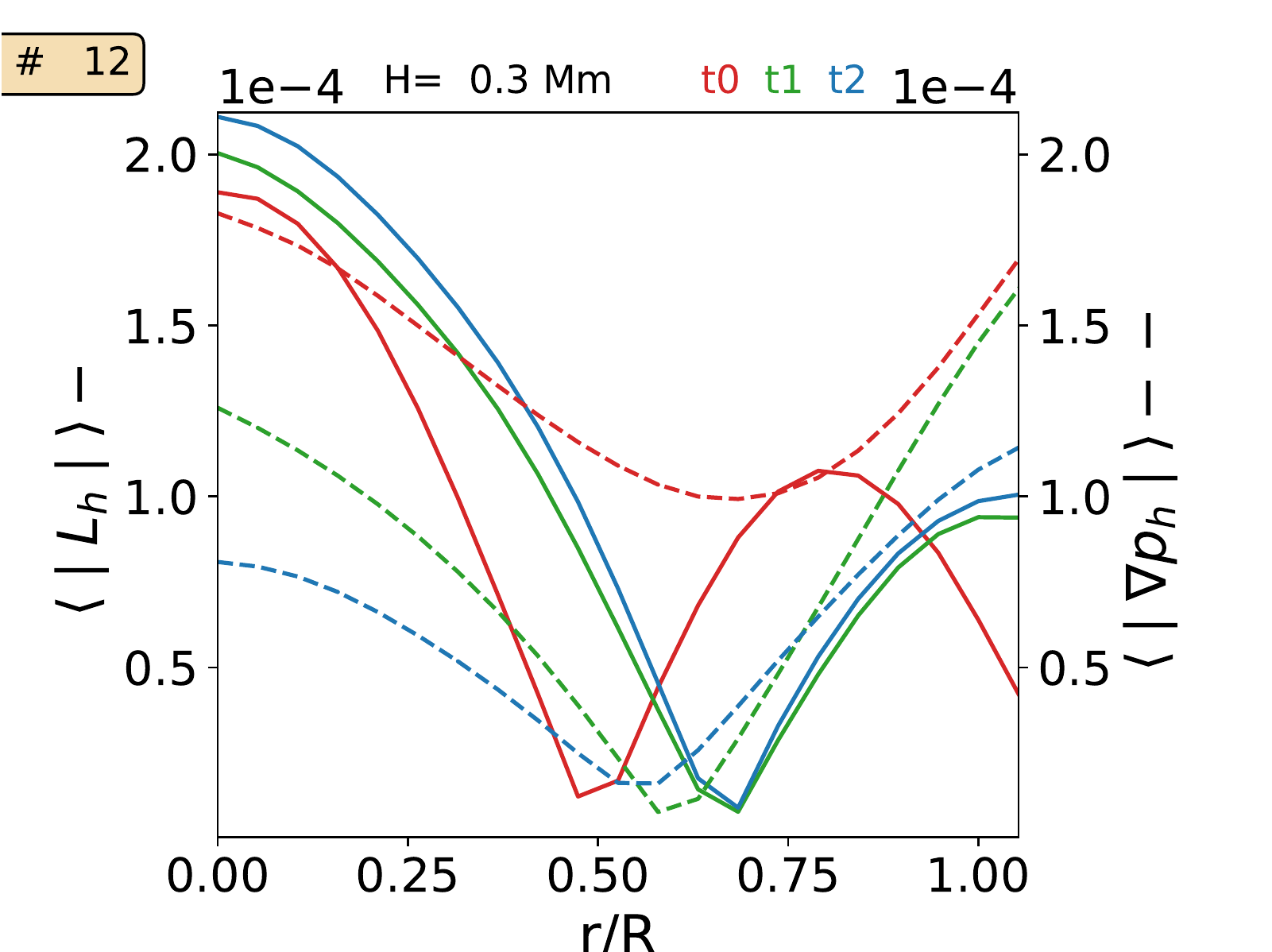}{0.3\textwidth}{(f)V12}
          }
\gridline{\fig{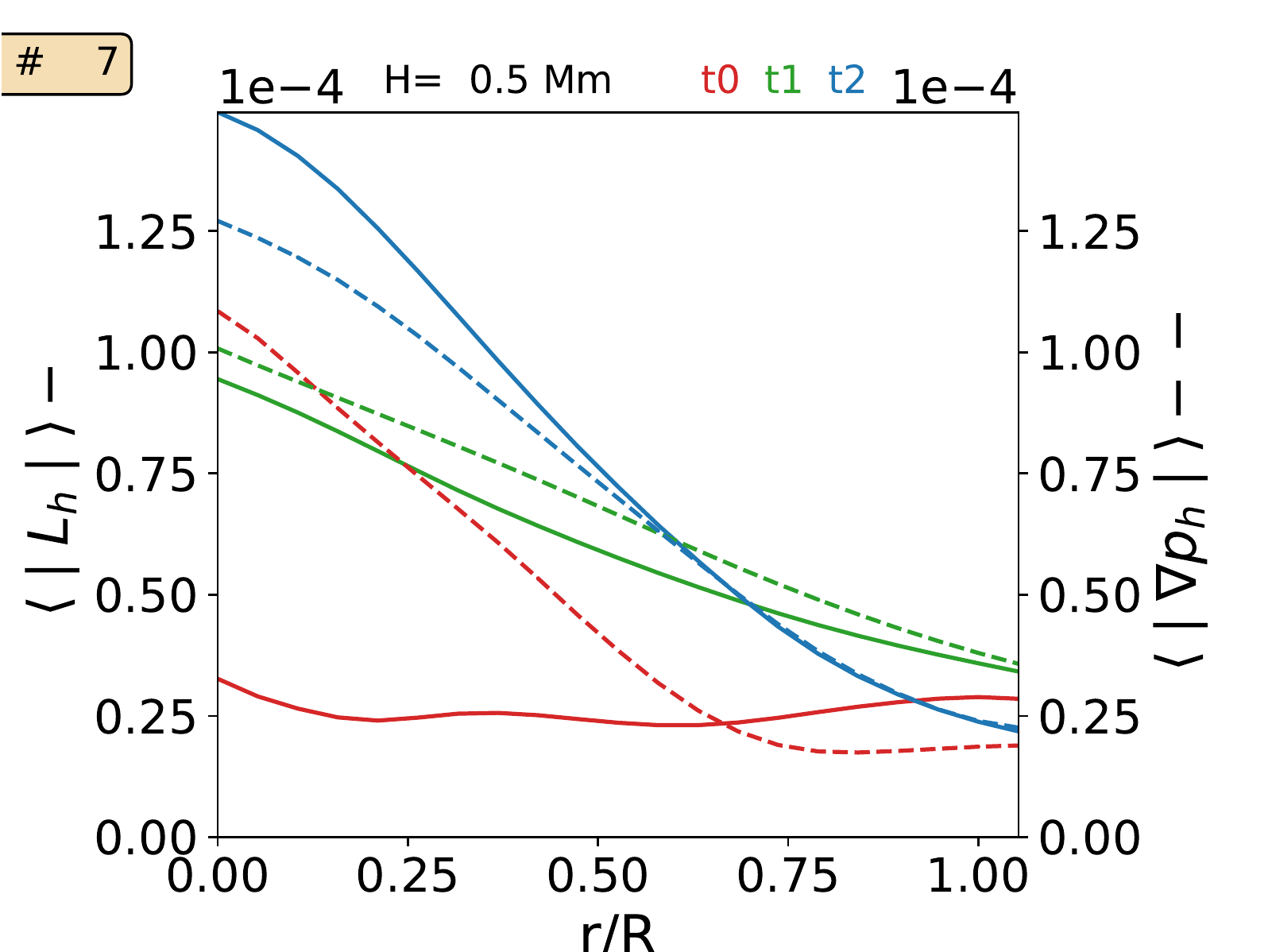}{0.3\textwidth}{(g)V7}
          \fig{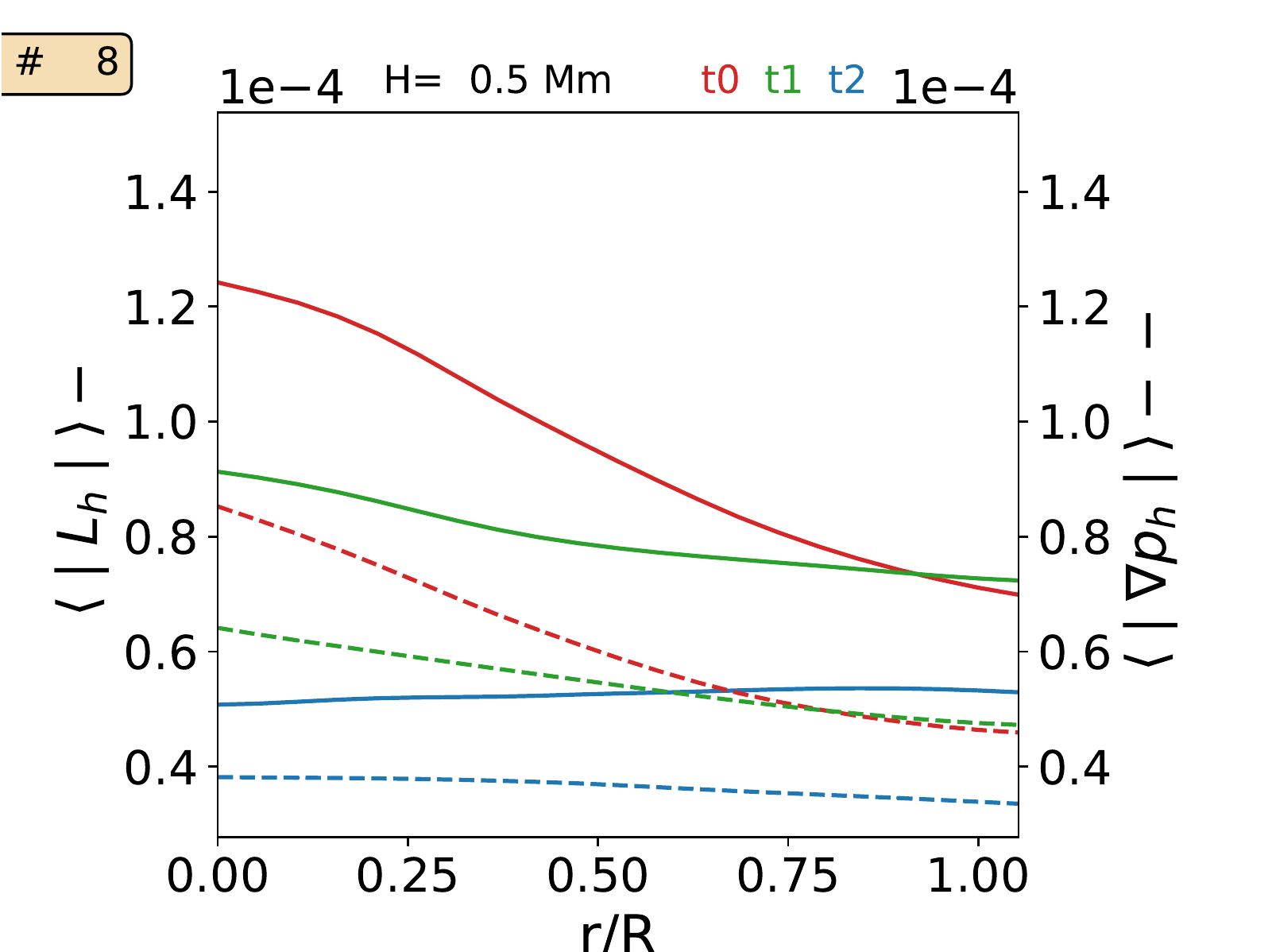}{0.3\textwidth}{(h)V8}
          \fig{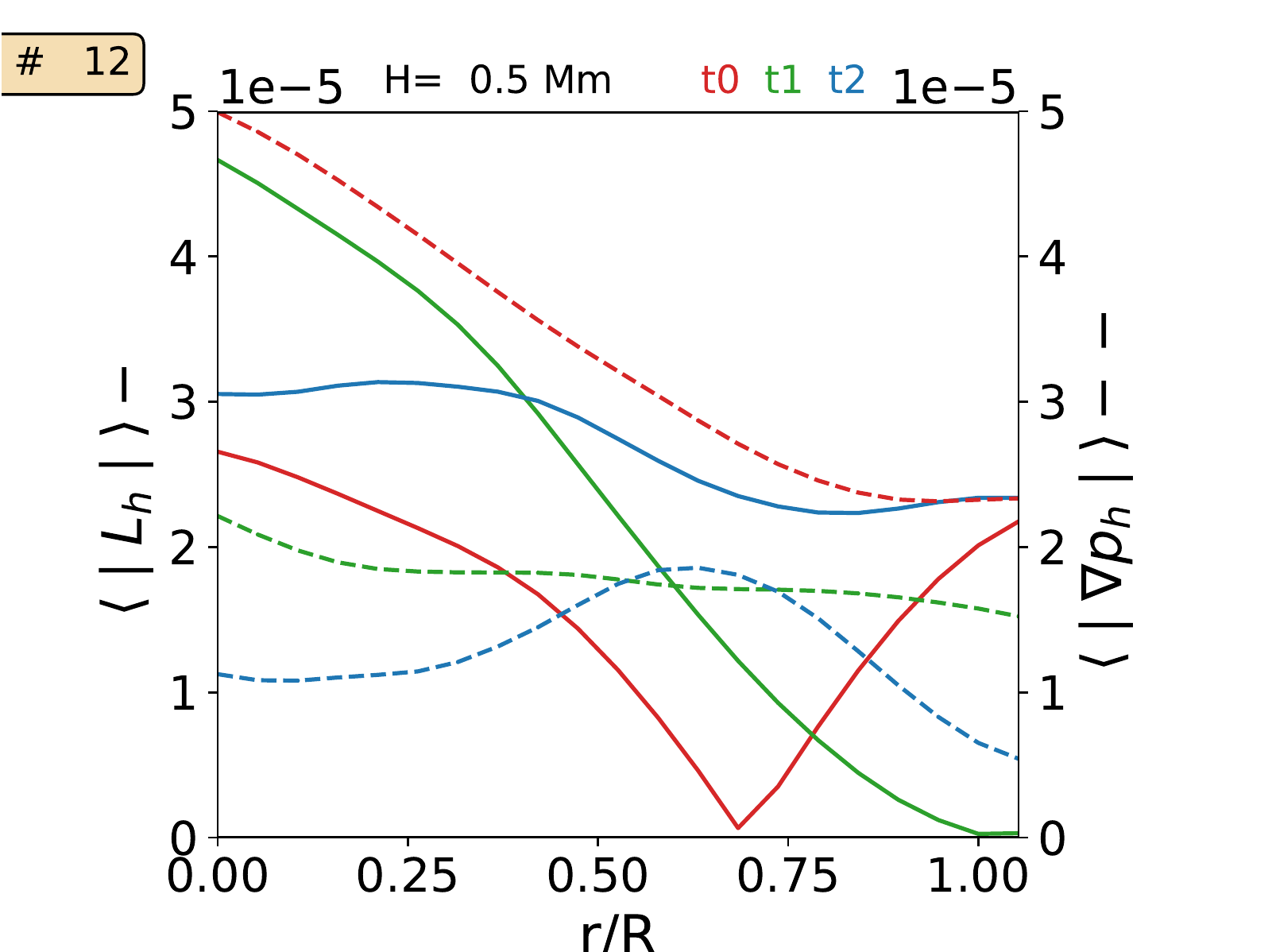}{0.3\textwidth}{(i)V12}
          }
\caption{The average intensity of the horizontal component of the Lorentz force  (left $y$-axis, solid lines) and the average intensity horizontal component of the pressure gradient (right $y$-axis, dashed lines)  along the vortex radii from the center, $r=0$, to the boundary $r=R$. The red lines are for the initial time $t_0= 0$, the green lines are for $t_1=25$ s (green line), and the blue lines are for $t_2=50$ s.  The radial profiles are shown for  vortices \#7 (a)(d)(g), \#8 (b)(e)(h) and \#12 (c)(f)(i) at different heights: $H = 0.1$ Mm (a)(b)(c), $H = 0.3$ Mm (d)(e)(f), $H = 0.5$ Mm (g)(h)(i).  \label{fig:Fgradpradial}}
\end{figure*}

The magnetic field concentration by solar vortices seems to saturate at a given time as the vortex presenting the highest magnetic field intensity, \#8,  displays negligible increments over time. The highest growth of magnetic field concentration is found in vortex \#12, which holds the lowest magnetic field intensity.  The vortices with higher magnetic field concentration also display considerably higher vorticity value, indicating that magnetic field has an important contribution to vorticity evolution. For instance, vortex \#12 presents more significant vorticity increase over time and vortex \#8 the highest  $\omega_z$ among the  three analyzed vortices.  The importance of the magnetic field for vortex dynamics is also suggested by the differences found in tangential velocity radial profiles of solar atmospheric vortices, and vortices models of nonmagnetized fluids. As the pressure gradient, $\nabla P$, is an important force on the dynamics of nonmagnetized vortex flows,  we compare the radial force balance between  $\nabla P$ and the Lorentz force, $L$, in the horizontal plane. Figure \ref{fig:Fgradpradial} shows both horizontal intensities of those forces with the same axis ranges in order to facilitate comparisons. Vortices \#7 and \#12 tend to have their dynamics alternately ruled by the pressure gradient and by the Lorentz force.  The presence of an intense magnetic field in vortex \#8 also leads to the Lorentz force dominating the dynamics in the horizontal plane over gradient pressure forces, which is confirmed by the low plasma $\beta$ found for \#8. 

\begin{figure*}[htp!]
\gridline{\fig{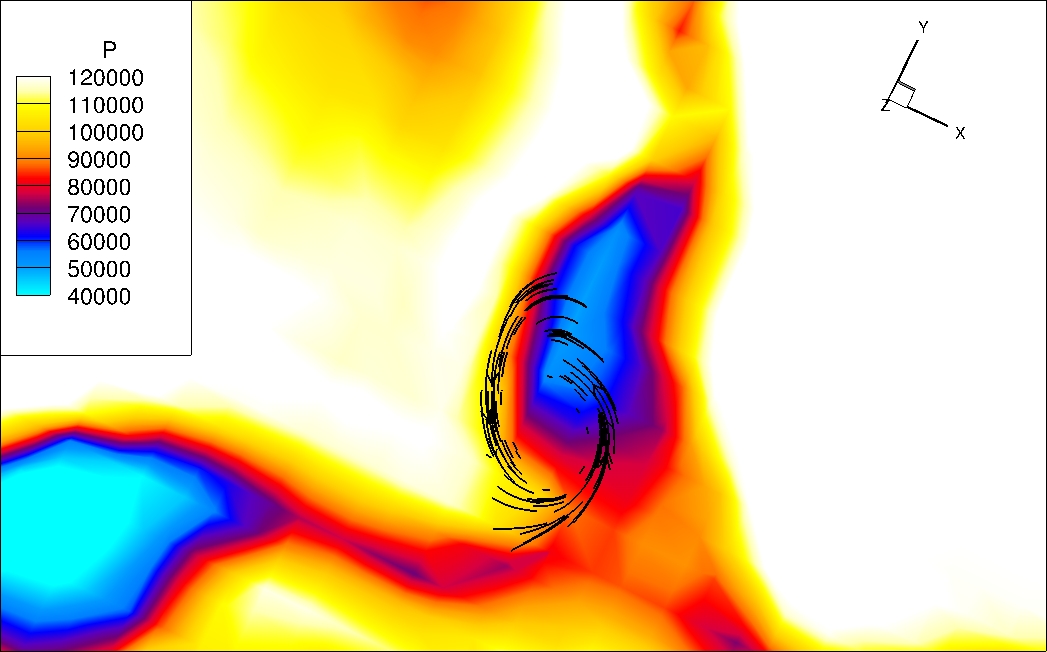}{0.25\textwidth}{(a)Velocity field lines for V7 at $t=t_0$}
          \fig{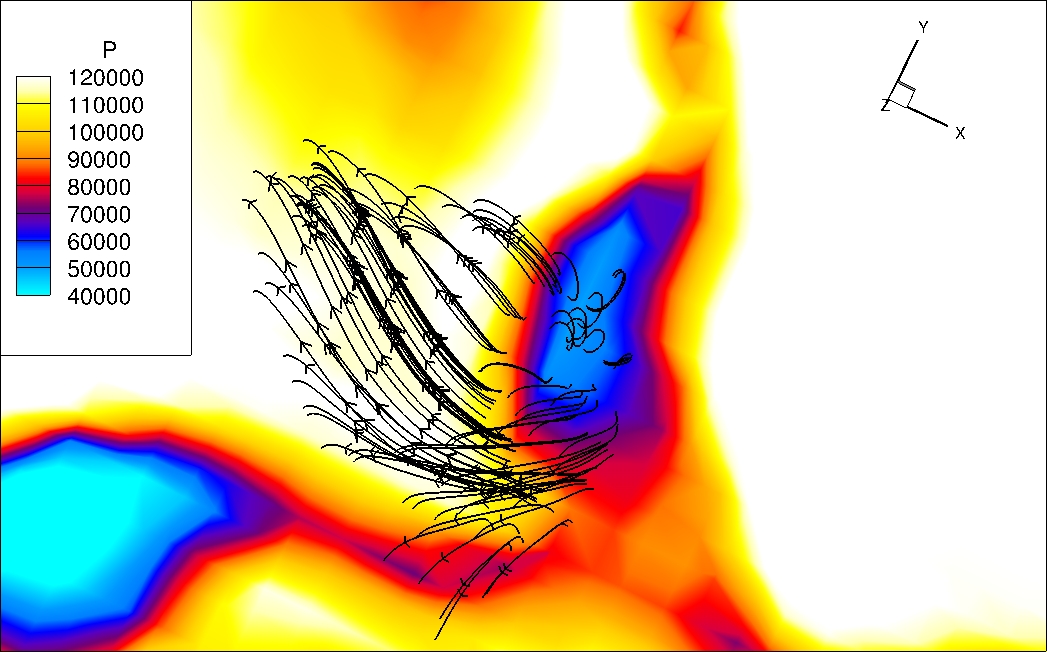}{0.25\textwidth}{(b)Magnetic field lines for V7 at $t=t_0$}
          \fig{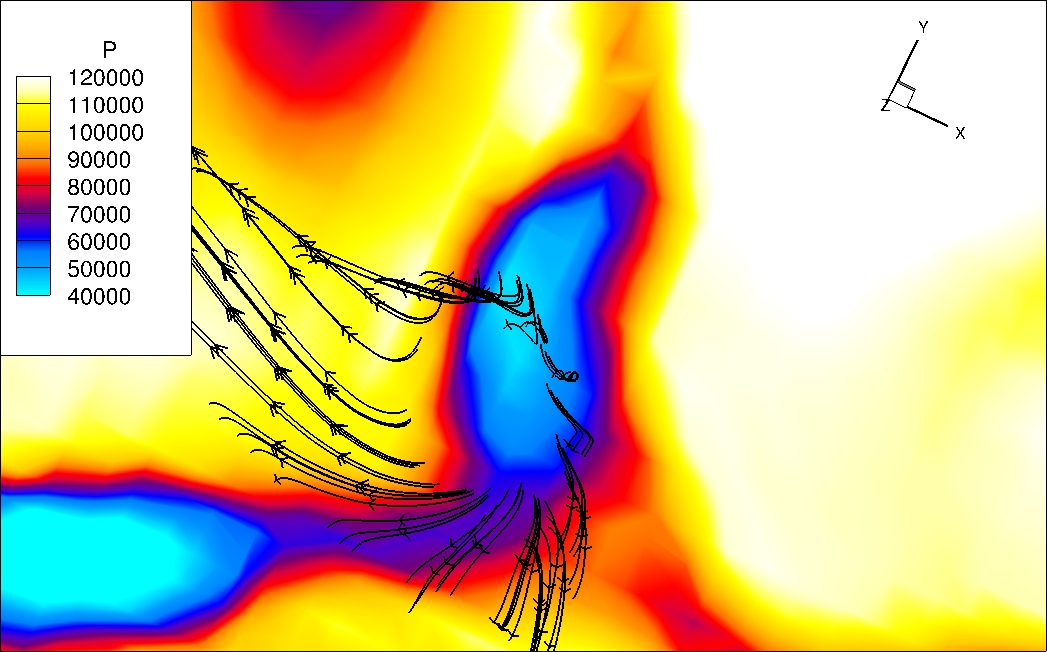}{0.25\textwidth}{(c)Magnetic field lines for V7 at $t=t_f$}
          }
\gridline{\fig{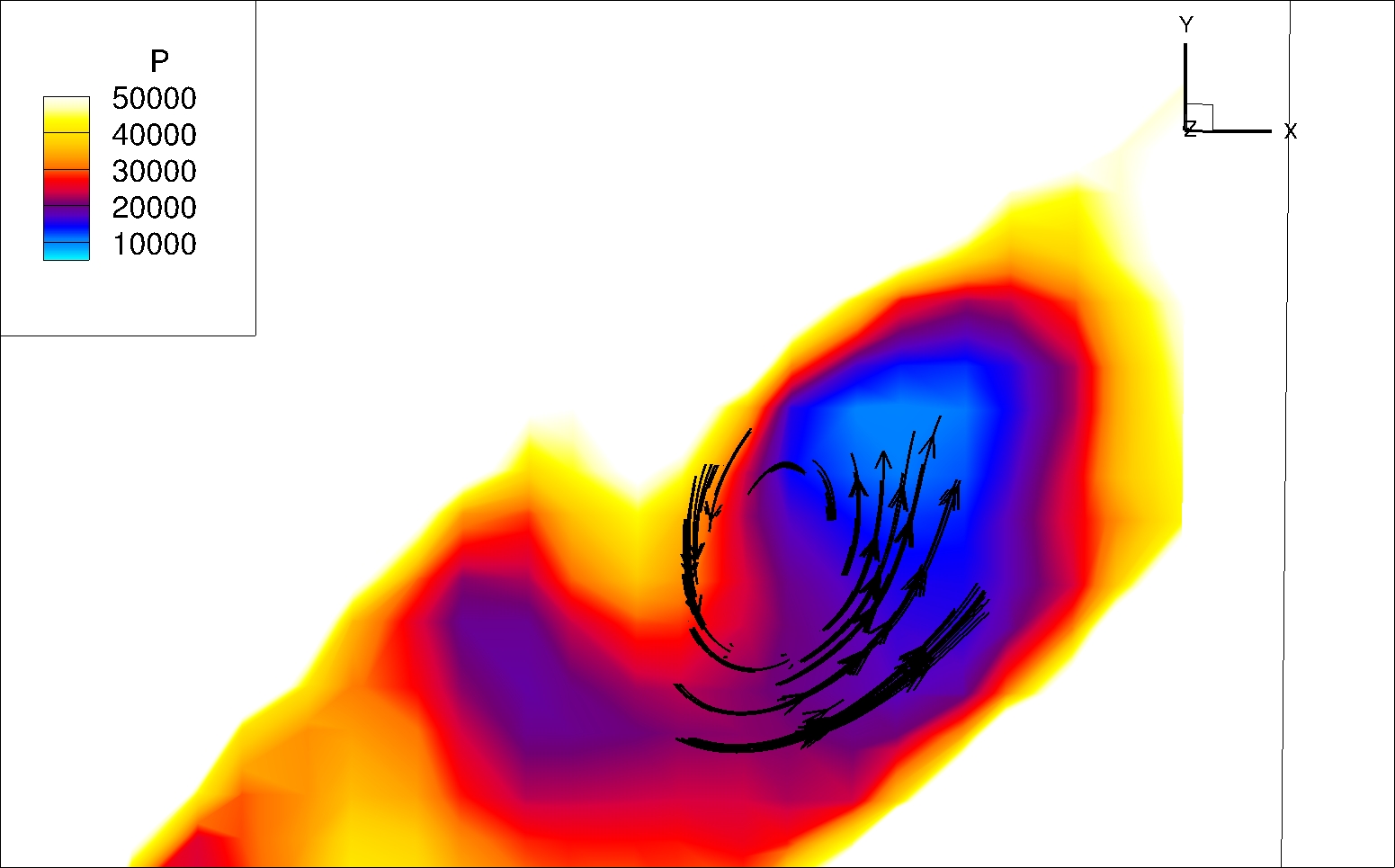}{0.25\textwidth}{(d)Velocity field lines for V8 at $t=t_0$}
          \fig{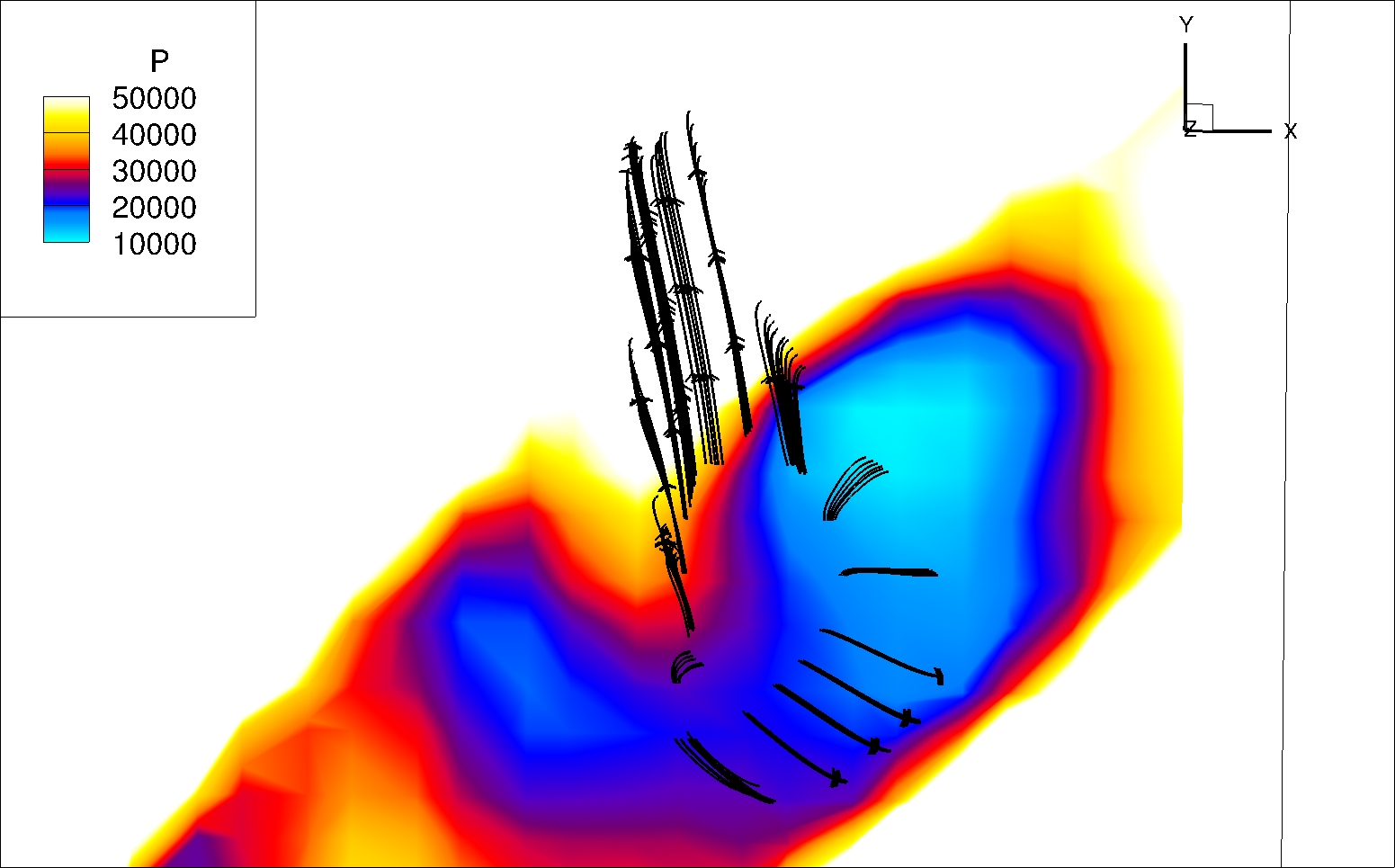}{0.25\textwidth}{(e)Magnetic field lines for V8 at $t=t_0$}
          \fig{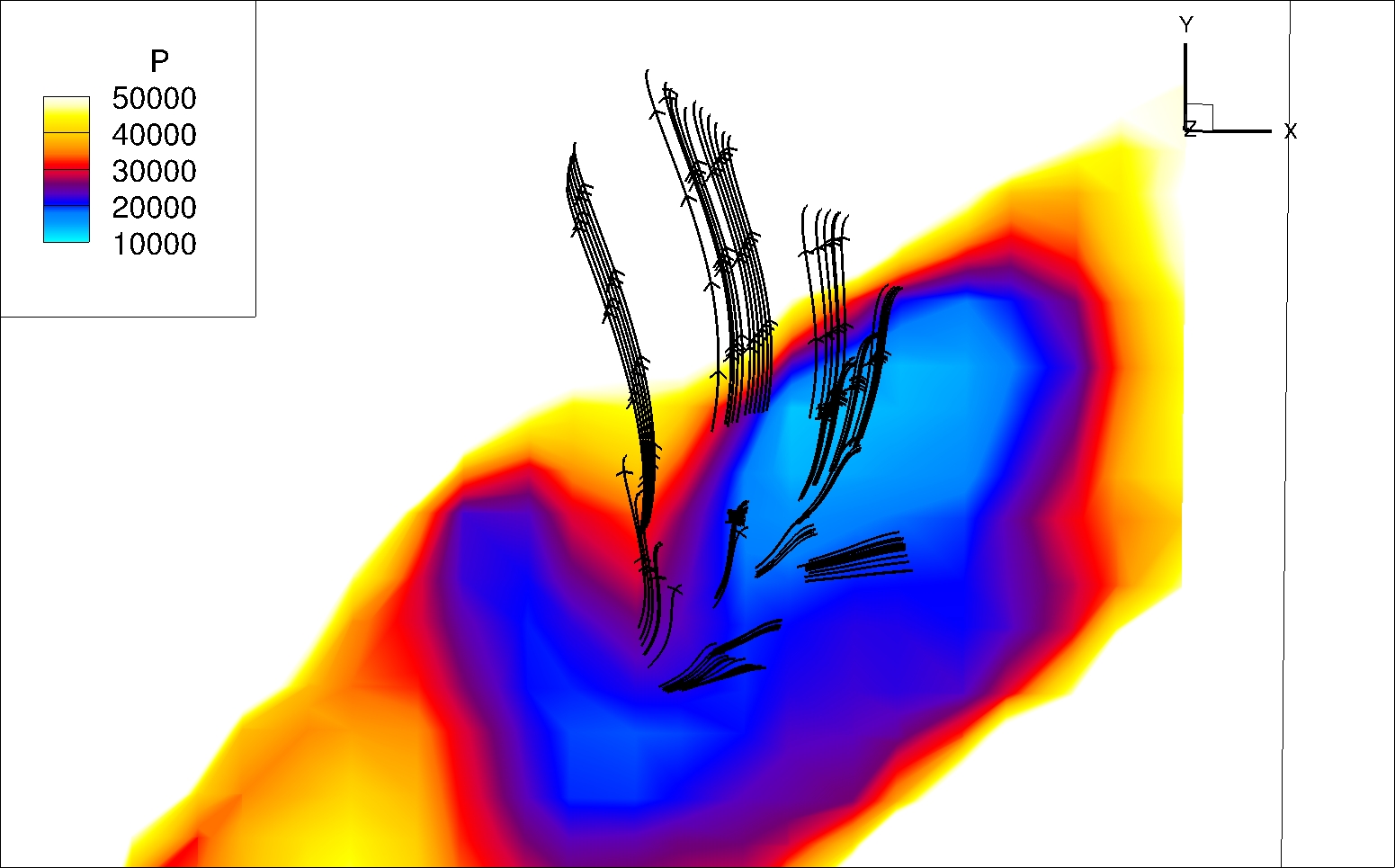}{0.25\textwidth}{(f)Magnetic field lines for V8 at $t=t_f$}
          }
\gridline{ \fig{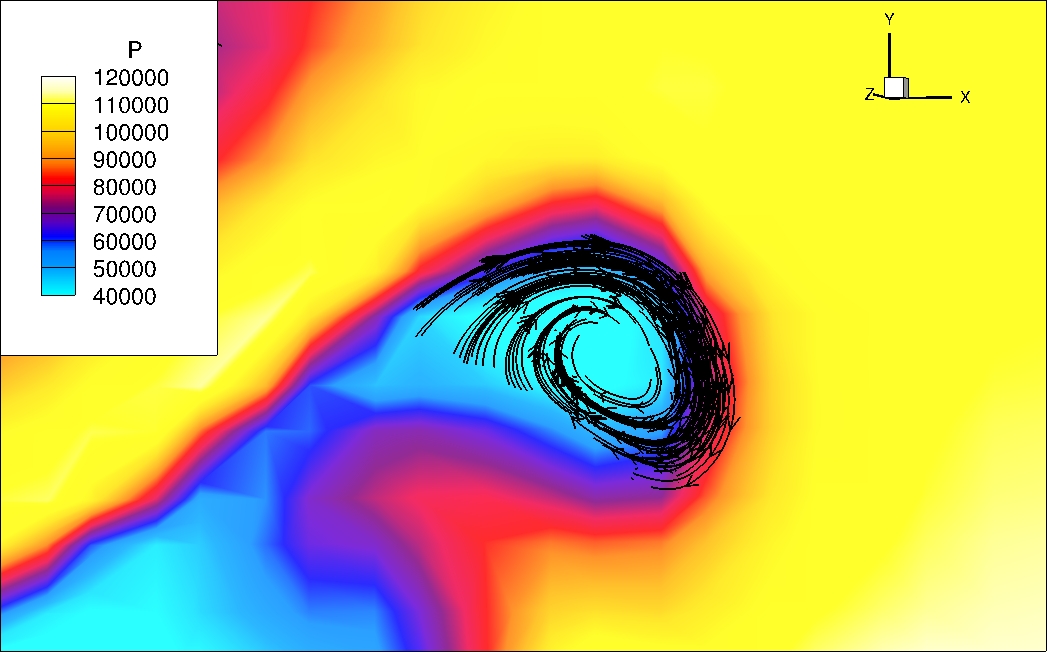}{0.25\textwidth}{(g)Velocity field lines for V12 at $t=t_0$}
          \fig{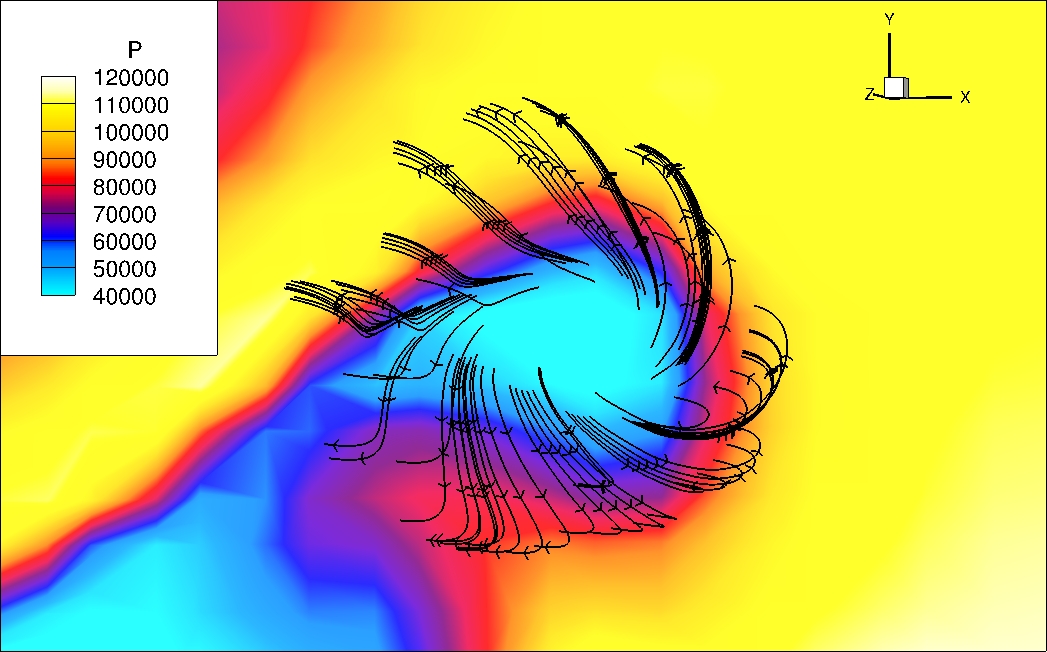}{0.25\textwidth}{(h)Magnetic field lines for V12 at $t=t_0$}
          \fig{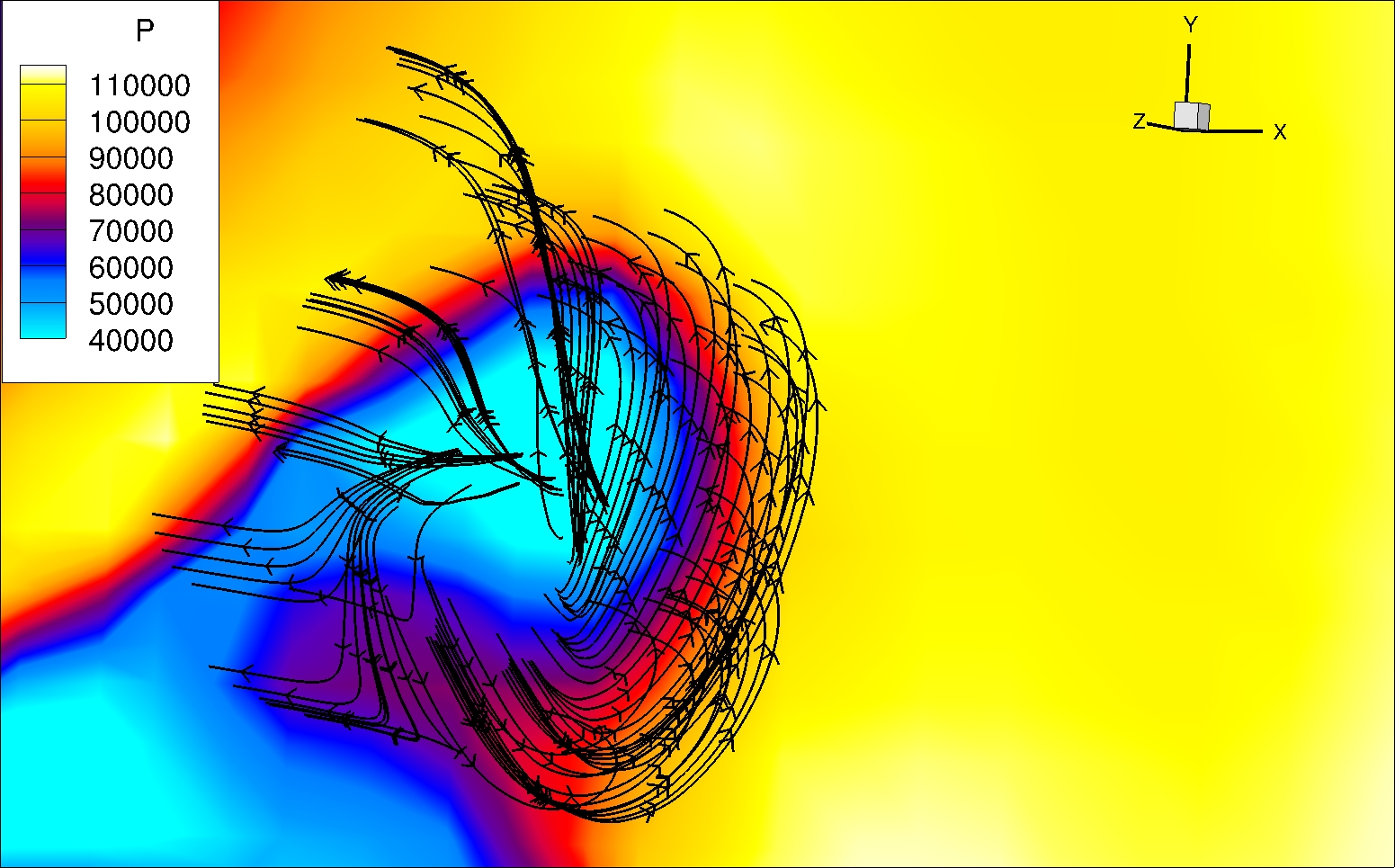}{0.25\textwidth}{(i)Magnetic field lines for V12 at $t=t_f$}
          }
\caption{ Close view of vortex region for vortices \#7 (a,b,c),  \#8 (d,e,f) and  \#12 (g,h,i), where the $xy$-plane is colored by pressure. The panels on the left (middle) display the velocity (magnetic) streamlines traced from the vortex boundary detected by IVD at $t_0=0$ for  $H = 0.0$ Mm. The panels on the right depict the magnetic field lines from points originally at the vortex boundary detected by IVD at $t_0$ for  $H = 0.0$ Mm and advected in time to $t=50$ s. \label{fig:Blinesadv}}
\end{figure*}

 In order to investigate the dynamics imposed by vortical motions on the magnetic field lines, we select a set of points at the boundary of vortices \#7, \#8, \#12 at $H = 0$, and advect them for $\Delta t =50$ s. Figure \ref{fig:Blinesadv} shows the $xy$-plane at $H = 0$ colored by pressure and  (a)  the velocity field streamlines at $t=0$, (b) the magnetic field streamlines for the same points at $t=0$, and (c) the magnetic field lines at $t=50$ s for the advected points. All the vortices shown in Figure \ref{fig:Blinesadv} seem to drag the magnetic field rooted at $H= 0$, leading to some torsion of those lines. Vortex \#8 has the lowest plasma $\beta$ among the analyzed vortices, and it is also the one with less torsion. The vortex with a higher plasma $\beta$,  vortex \#12, was the only one in the whole analyzed domain that was able to create and sustain a twisted magnetic flux. The differences found for vortex \#12 and all the other 17 detected vortices are its higher tangential velocity and plasma $\beta$ values. Among the 17 detected vortices,  some had even higher plasma $\beta$ than \#12, but their $V_\theta$ values were between 50\% and 80\% of the values found for \#12 at different height levels.

\section{Conclusions} \label{sec:conclusions}
The 17 detected vertical vortex tubes reported in this work were found within intergranular regions, in areas of high concentration of magnetic field flux.  The vortex tubes in the lower solar atmosphere were precisely calculated, allowing the study of the plasma dynamics across the vortical flow.   The solar vortex tubes present different shapes as a function of time, being deformed by the forces acting on the flow. The horizontal radii of the vortices are, on average, 40 km at the photosphere and around 80 km at the upper part of the simulation domain. 
As the domain only reaches 600 km above the surface, the upper parts of the vortices are located at the lower part of the chromosphere, but it does not correspond to the observed chromospheric swirls observed in line emissions \citep{Wedemeyer2012, Leenaarts2013, Shetye2019}. Even so, our results indicate that photospheric and chromospheric vortices are part of the same 3D vortex tube.  For the detected vortices, the part laying in the chromosphere tends to cover an area almost twice as large as the part of the vortex at $H= 0.0$ Mm. Another relevant aspect concerning the linking of parts of the vortex at different height levels is the similarity in the radial profiles throughout the solar atmosphere, which confirms that chromospheric and photospheric vortices are not only part of a vortex tube, but they are also under similar dynamics. More specifically, the tangential velocity profiles show that the plasma rotates in the same direction at all height levels, with the chromospheric part of the vortex rotating up to twice as fast as the photospheric part. At all height levels, once the plasma is dragged into the vortex tube, its tangential velocity decreases, indicating eddy viscosity effects. The plasma also carries vorticity, which is, in turn, concentrated in the low-pressure vortex regions as confirmed by the vorticity profile and also matched by observations of mesogranular flows  \citep{Simon_1997}. The in- and outflow of vorticity implies that the vortex is not a conservative system and, therefore, the assumption of conservation of angular momentum as a vortex generation mechanism is misleading. Also, both tangential velocity and vorticity profiles indicate that the ``bathtub effect'' mechanism is likely not responsible for the observed vortices as they do not present in any part of the vortex the expected behavior of a  ``bathtub effect''; \textit{i.e}  a tangential velocity that would initially increase from the boundary toward the center of the vortex and then decay closer to the center. For the solar vortices, there is only decreasing of the tangential velocity from the boundary to the center, indicating the action of another mechanism to vortex creation.  Within the higher parts of the tube vortices, there are intermittent upflows that were described by the investigations of \cite{Kitiashvili2012,Kitiashvili13}, where they used MHD simulations to describe the turbulent convection of quiet Sun regions. Those upflow plasma jets were observed for the upper part of the vortex tube and they are stronger at the vortex boundary and they can become downflows at the center of the vortex as suggested by \cite{Kitiashvili13}.

 The solar vortex tubes also concentrate the magnetic field with a cubic dependence on the radius, leading to the formation of magnetic flux tubes above the solar surface.  The vortices encompass a maximum around 1300 G at the solar surface and up to around 600 G at their upper levels. The  magnetic concentrations found for the detected vortices are similar to the ones detected by other MHD simulations \citep{Kitiashvili2012, Wedemeyer2014}. The magnetic field was found to play an essential role in vortex dynamics. The main forces acting on the vortex, the pressure gradient, and the Lorentz force, have similar intensities at all height levels, which indicates that magnetic effects are as important as hydrodynamic terms for the vortex evolution.  The importance of Lorentz force in vortex dynamics was previously hinted by \cite{Kitiashvili13} for the highest parts of a simulated magnetized solar atmosphere, but our studies suggest that this actually applies to the whole vortex tube. Also, the magnetic field contributions to vorticity seem to be an essential aspect of vorticity evolution, which confirms the findings of  \cite{Shelyag2011}. Besides, another corroboration to the importance of the magnetic field in solar vortex evolution relies on the fact that the tangential velocity profiles in the solar atmosphere have a better fit with cubic approximation instead of a general model for vortices in nonmagnetized fluids.  In turn, the magnetic field is also impacted by the vortices' dynamics, leading to torsion and bending of the magnetic field.  For the 17 analyzed solar vortices, only  \#12 had a magnetic vortex as defined by \cite{Rempel2019} and which is cospatially existing with the kinematic solar vortex tube. Those results are in accordance with the findings of \cite{moll12}, who show that the magnetic field lines tend to expand with height and do not present significant twisting. Our findings also indicate that the generation of twisted magnetic flux tubes by vortical motions in the photosphere can occur only when there are sufficiently high tangential speed vortices at various height levels to overcome the magnetic tension and twist the field lines. Our study hints that most of the detected solar vortices will not have existing cospatially magnetic vortices in the atmosphere, but instead slightly bending magnetic flux tubes. As the rotation of the magnetic field by kinematic vortices is believed to generate the detected chromospheric swirls in line emission observations \citep{Wedemeyer2012,Wedemeyer2014},  our results indicate that the amount of photospheric vortices is likely larger than the number of observed chromospheric swirls. 

\acknowledgments
 SSAS, VF, GV and ER are grateful to The Royal Society, International Exchanges Scheme, collaboration with Brazil (IES$\backslash$R1$\backslash$191114). VF and GV are grateful to Science and Technology Facilities Council (STFC) grant ST/M000826/1 and to The Royal Society, International Exchanges Scheme, collaboration with Chile (IE170301). VF would like to thank the International Space Science Institute (ISSI) in Bern, Switzerland, for the hospitality provided to the members of the team on `The Nature and Physics of Vortex Flows in Solar Plasmas'. E.L.R. acknowledges Brazilian agencies CAPES, CNPq,and FAPESP (Grants No. 88881.309066/2018-01, No 304449/2017-2 and No. 16/24970-7) for their financial support. This research has also received financial support from the European Union’s Horizon 2020 research and innovation program under grant agreement No. 824135 (SOLARNET). This study was financed in part by the Coordenação de Aperfeiçoamento de Pessoal de Nível Superior – Brasil (CAPES) – Finance Code 88882.316962/2019-01.

\bibliography{bibliog}{}
\bibliographystyle{aasjournal}
\end{document}